\documentclass[preprint,12pt]{elsarticle}

\usepackage{graphicx}
\usepackage{amssymb}
\usepackage{arydshln}
\usepackage{wrapfig}
\usepackage{siunitx}
\usepackage{mathtools}
\usepackage{multirow}
\usepackage[table]{xcolor}
\usepackage{array}
\usepackage{appendix}

\usepackage{lineno}
\usepackage{rotating}
\usepackage{supertabular,booktabs}
\usepackage{longtable}
\usepackage{lscape}
\usepackage{geometry}
\usepackage{adjustbox}
\usepackage{natbib}
\usepackage[english]{babel}
\addto\captionsenglish{}
\usepackage{blindtext}
\newcolumntype{P}[1]{>{\centering\arraybackslash}p{#1}}
\definecolor{LightCyan}{rgb}{0.88,1,1}

\newcommand \gray{gamma-ray}
\newcommand \grays{gamma-rays}
\newcommand \cof{$^{14}$C}
\newcommand \cotw{$^{12}$C}
\newcommand \kref{$^{85}$Kr}
\newcommand \acttS{$^{227}$Ac}
\newcommand \luoSs{$^{176}$Lu}
\newcommand \laoTe{$^{138}$La}
\newcommand \gdoFt{$^{152}$Gd}
\newcommand \smofS{$^{147}$Sm}
\newcommand \utTFE{$^{235}$U$_{\mathrm{e}}$}
\newcommand \utTFL{$^{235}$U$_{\mathrm{l}}$}
\newcommand \patTo{$^{231}$Pa}
\newcommand \rnton{$^{219}$Rn}
\newcommand \potoF{$^{215}$Po}
\newcommand \potoe{$^{218}$Po}
\newcommand \potof{$^{214}$Po}
\newcommand \bitof{$^{214}$Bi}

\newcommand \kfz{$^{40}$K}
\newcommand \cosz{$^{60}$Co}
\newcommand \thtTt{$^{232}$Th}
\newcommand \thtTtE{$^{232}$Th$_{\mathrm{e}}$}
\newcommand \thtTtL{$^{232}$Th$_{\mathrm{l}}$}
\newcommand \ratte{$^{228}$Ra}
\newcommand \thtte{$^{228}$Th}

\newcommand \utTe{$^{238}$U}
\newcommand \utTeE{$^{238}$U$_{\mathrm{e}}$}
\newcommand \utTeL{$^{238}$U$_{\mathrm{l}}$}
\newcommand \ratts{$^{226}$Ra}
\newcommand \utTF{$^{235}$U}
\newcommand \pbtos{$^{206}$Pb}
\newcommand \pbtoz{$^{210}$Pb}
\newcommand \pbtof{$^{214}$Pb}

\newcommand \tltze{$^{208}$Tl}
\newcommand \rnttt{$^{222}$Rn}
\newcommand \rnttz{$^{220}$Rn}
\newcommand \csoTS{$^{137}$Cs}
\newcommand \agooz{$^{110\mathrm{m}}$Ag}
\newcommand \coFS{$^{57}$Co}
\newcommand \coFe{$^{58}$Co}
\newcommand \kft{$^{42}$K}
\newcommand \patTT{$^{233}$Pa}
\newcommand \nptTn{$^{239}$Np}
\newcommand \thtTz{$^{230}$Th}
\newcommand \utTT{$^{233}$U}

\newcolumntype{L}[1]{>{\raggedright\let\newline\\\arraybackslash\hspace{0pt}}m{#1}}
\newcolumntype{C}[1]{>{\centering\let\newline\\\arraybackslash\hspace{0pt}}m{#1}}
\newcolumntype{R}[1]{>{\raggedleft\let\newline\\\arraybackslash\hspace{0pt}}m{#1}}
\newcolumntype{H}{>{\setbox0=\hbox\bgroup}c<{\egroup}@{}}

\DeclareSIUnit\mwe{m.w.e.}
\DeclareSIUnit\litre{L}
\DeclareSIUnit\mBqkg{\milli\becquerel\per\kilogram}
\DeclareSIUnit\mBq{\milli\becquerel}
\DeclareSIUnit\GeVcSq{\giga\electronvolt\per c \squared}
\DeclareSIUnit[number-unit-product = ]\percent{\char`\%}
\DeclareSIUnit\year{y}

\DeclarePairedDelimiterXPP\BigOSI[2]%
  {\mathcal{O}}{(}{)}{}%
  {\SI{#1}{#2}}

\newcommand \tableWord{Table}
\newcommand \sectionWord{Section}
\newcommand \figureWord{Figure}

\makeatletter
\def\ps@pprintTitle{%
   \let\@oddhead\@empty
   \let\@evenhead\@empty
   \let\@oddfoot\@empty
   \let\@evenfoot\@oddfoot
}
\makeatother

\begin{document}


\begin{frontmatter}




\title{The LUX-ZEPLIN (LZ) radioactivity and cleanliness control programs}

\author[1,2]{D.S.~Akerib}
\author[3]{C.W.~Akerlof}
\author[4]{D.~Yu.~Akimov}
\author[5]{A.~Alquahtani}
\author[6]{S.K.~Alsum}
\author[1,2]{T.J.~Anderson}
\author[7]{N.~Angelides}
\author[8]{H.M.~Ara\'{u}jo}
\author[6]{A.~Arbuckle}
\author[9]{J.E.~Armstrong}
% ...

\address[1]{SLAC National Accelerator Laboratory, Menlo Park, CA 94025-7015, USA}

\address[2]{Kavli Institute for Particle Astrophysics and Cosmology, Stanford University, Stanford, CA  94305-4085 USA}

\address[3]{University of Michigan, Randall Laboratory of Physics, Ann Arbor, MI 48109-1040, USA}

\address[4]{National Research Nuclear University MEPhI (NRNU MEPhI), Moscow, 115409, RUS}

\address[5]{Brown University, Department of Physics, Providence, RI 02912-9037, USA}

\address[6]{University of Wisconsin-Madison, Department of Physics, Madison, WI 53706-1390, USA}

\address[7]{University College London (UCL), Department of Physics and Astronomy, London WC1E 6BT, UK}

\address[8]{Imperial College London, Physics Department, Blackett Laboratory, London SW7 2AZ, UK}

\address[9]{University of Maryland, Department of Physics, College Park, MD 20742-4111, USA}

% ...

\cortext[cor1]{{akamaha@albany.edu}}
\cortext[cor2]{{umit.utku.12@ucl.ac.uk}}


\begin{abstract}
LUX-ZEPLIN (LZ) is a second-generation direct dark matter experiment with spin-independent WIMP-nucleon scattering sensitivity above \SI{1.4 e-48}{\cm\squared} for a WIMP mass of \SI{40}{\GeVcSq} and a \SI{1000}{\day} exposure. LZ achieves this sensitivity through a combination of a large \SI{5.6}{\tonne} fiducial volume, active inner and outer veto systems, and radio-pure construction using materials with inherently low radioactivity content. The LZ collaboration performed an extensive radioassay campaign over a period of six years to inform material selection for construction and provide an input to the experimental background model against which any possible signal excess may be evaluated.  The campaign and its results are described in this paper.

We present assays of dust and radon daughters depositing on the surface of components as well as cleanliness controls necessary to maintain background expectations through detector construction and assembly. 

Finally, examples from the campaign to highlight fixed contaminant radioassays for the LZ photomultiplier tubes, quality control and quality assurance procedures through fabrication, radon emanation measurements of major sub-systems, and bespoke detector systems to assay scintillator are presented.

\end{abstract}


\end{frontmatter}


\section{Introduction}
\label{S:1}
The LUX-ZEPLIN (LZ) experiment operates a \SI{7}{\tonne} purified liquid-xenon target in a time projection chamber (TPC) and has a design sensitivity capable of excluding at \SI{90}{\percent} confidence level spin-independent WIMP-nucleon cross sections above \SI{1.4 e-48}{\cm\squared} for a \SI{40}{\GeVcSq} mass WIMP, providing excellent discovery potential for WIMPs through nucleon elastic scattering and subsequent detection of light and charge from the collisions~\cite{Akerib:2018lyp}. The detector is currently being installed in the Davis Cavern of the Sanford Underground Research Facility (SURF) at a depth of approximately 4850~ft which is equivalent to approximately 4300 meters of water (henceforth referred to as meters of water equivalent - m w.e.). The detector is described in detail in~\cite{Akerib:2019nima} and~\cite{Mount:2017qzi}. The expected sensitivity of the experiment is achieved through a combination of very low background rates, a large fiducial mass of \SI{5.6}{\tonne}, and a \SI{1000}{\day} WIMP search exposure. The detector backgrounds are predominantly of two kinds: electron recoils (ER) which occur through interaction with the atomic electrons of the target xenon material; and nuclear recoils (NR) which occur through interaction with the nucleus of the xenon. The coincident background levels in LZ are suppressed, tagged and characterized by employing two veto detectors: an instrumented layer of liquid-xenon known as the xenon skin veto, and an outer detector (OD) that uses a Gd-loaded liquid scintillator (GdLS) detection medium. The inner \SI{5.6}{\tonne} fiducial volume further benefits from attenuation of background radioactivity penetrating the TPC.

The construction of LZ requires materials very low in radio-contamination to limit their background contribution in the target, thus maximizing the rare-event search sensitivity. Furthermore, an accurate knowledge of the expected low levels of background from remaining radioactivity and irreducible sources is necessary to ascribe confidence to any possible observation of signal excess. This article presents the results from the radioassay and screening campaign performed to inform the selection of LZ construction materials and to develop the experiment's comprehensive background model.

\sectionWord~\ref{S:2} describes the gamma-ray spectroscopy and mass spectrometry techniques and the facilities used to determine levels of radioactivity in the bulk of the materials, including efforts to cross-calibrate the various detectors deployed. \sectionWord~\ref{S:3} presents the radon emanation facilities available to LZ which were used to inform selection of materials in contact with liquid xenon and to characterize components used for the construction of the LZ experiment. \sectionWord~\ref{S:4} describes our techniques to limit surface depositions of environmental dust and atmospheric radon decay-progeny, with particular emphasis on the period of detector assembly at SURF. 
\sectionWord~\ref{S:5} presents a selection of highlights from the assay program that illustrates our fixed-contaminant radioassays for the LZ PMTs; demonstrates the importance of our quality control (QC) and quality assurance (QA) procedures for construction of the cryostat; revisits a dedicated detector constructed to survey the scintillator for the OD first discussed in~\cite{Haselschwardt:2018vmp}; and presents radon emanation measurements from key components, including \emph{in-situ} assays of the cryostat with the complete detector within. 

Upon completion of any assay, results are uploaded to a dedicated database. This database links assay results to individual components of the LZ detector, greatly simplifying the process of building the expected background model which, in turn, helps to define the expected sensitivity of the detector. The results from the assays performed are tabulated in the Appendix to this article.


\section {Fixed Contamination}
\label{S:2}

\subsection{Origin of Fixed Contamination}
Fixed contaminants are embedded in materials and typically consist of naturally occurring radioactive materials (NORM), the most prevalent being \utTe{}, \utTF{}, \thtTt{}, and their progenies which emit various radioactive species through their decay eventually to stable isotopes of lead; and \gray{} emitting isotopes, \kfz{}, \cosz{} and \csoTS{}. Neutrons are produced through ($\alpha$,n) reactions and through spontaneous fission in the uranium and thorium decay chains. The \utTe{} and \thtTt{} chains are divided into `early' and `late'; for \utTe{}, the early part of the chain (\utTeE{}) contains the isotopes above \ratts{} since chemical processes may induce a break of secular equilibrium at this point, and it will take thousands of years  ($\tau_{1/2}$~1600 years) to be restored. The late part of the chain (\utTeL{}) is counted from \ratts{} and below. Typical p-type high-purity germanium (HPGe) detectors are not sensitive to the low-energy gamma-ray lines from \pbtoz{} at the bottom of the chain but broad energy germanium (BEGe), n-type and well-type detectors available to LZ are.

Secular equilibrium breaking is observed by differences in long-lived isotope concentrations in early and late chain values.
However, it should be noted that the \thtTt{} chain, as defined, includes \ratte{} which has a relatively long half-life of \SI{5.7}{\year}. If \thtTt{} and \thtte{} are depleted in production of a material, it is possible for \thtte{} to grow back in from \ratte{} on a time scale of several years, such that assays may underestimate the ultimate activity. It can be difficult to measure \ratte{} with the same sensitivity as \thtTt{} because HPGe counting used for the former is generally less sensitive than ICP-MS analysis used for the latter, particularly for samples where only small masses are available, but the risk is generally mitigated by using materials where batches of different ages have been measured over the course of a long material selection campaign~\cite{Scovell:2017srl}.

The collaboration also performs assays with in-house inductively-coupled plasma mass spectrometry (ICP-MS) (\sectionWord~\ref{sec:icpms}) and some limited use of glow discharge mass spectromety (GDMS).

\subsection{High-Purity Germanium Screening}
\label{sec:HPGe}

Gamma-ray spectroscopy was used to identify the bulk of the radio-isotopes contributing to neutron and \gray{} emission. In order to achieve sensitivity to the required low levels, these measurements were typically of 1 to 2 week duration. Assays were made using 12 HPGe detectors, or variants of HPGe detectors, available to the LZ collaboration across four sites, described in the following subsections and with key parameters summarized in \tableWord~\ref{tab:GeDetInf} and performance characteristics summarized in \tableWord~\ref{tab:GeDetPerf}. In the early stages of the LZ screening program, a campaign of blind cross-calibration across all detectors was undertaken to verify the consistency of analysis and interpretation across the different sites. The cross-calibration campaign is described in \sectionWord~\ref{sec:cross-cal}. HPGe assay sensitivity to both early and late chain activities was critical to the comprehensive modeling of backgrounds.

\subsubsection{BHUC}
The Black Hills Underground Campus (BHUC)~\cite{Mount:2017iam} is a facility located at the 4850~ft level of SURF which hosts a class 2000 cleanroom containing six low- and ultra-low background HPGe detectors: \textsc{Maeve, Morgan, Mordred}, SOLO, and the TWINS. \textsc{Maeve}, an Ortec \SI{85}{\percent} relative efficiency p-type detector (where the efficiency is defined as relative to that of a 3-inch $\times$ 3-inch NaI detector exposed to 1332~keV \cosz{} \grays{} with a source-detector distance of \SI{25}{\cm}) was previously situated in the Davis campus at SURF and, before that, at LBNL's Oroville site for 15 years. \textsc{Morgan}, an Ortec \SI{85}{\percent} relative efficiency p-type detector, is effectively identical to \textsc{Maeve} in performance. \textsc{Mordred}, an Ortec \SI{60}{\percent} relative efficiency n-type detector, was retrofitted with ultra-low background electronics to improve its performance for low-background assay. \textsc{Mordred} has good sensitivity to low-energy gamma rays and is therefore particularly well-suited for U early chain assays. SOLO, a \SI{30}{\percent} relative efficiency p-type detector, was previously sited in the Soudan Underground Laboratory and played an important role in the LUX experiment's material assay campaign~\cite{akerib:2014rda,akerib:2013tjd,Akerib:2016vxi}. While the crystal is small, it has exceptionally low backgrounds. The newest detectors in the BHUC are referred to as the TWINS, a pair of Ortec \SI{90}{\percent} relative efficiency p-type detectors in a common shield. The TWINS can operate in coincidence or anti-coincidence and in combined singles data acquisition mode where spectra from each detector are combined without any regard for events which are detected in coincidence.  

In the BHUC, \textsc{Maeve, Morgan, Mordred}, and SOLO are situated in separate graded shields. The TWINS are installed in a common shield. Each shield provides at least \SI{20}{\cm} of low-activity lead shielding with \SI{2.5}{\cm} of oxygen-free, low conductivity copper within the lead. The shield surrounding \textsc{Maeve} has an inner layer of \SI{2.5}{\cm} of ultra-low activity lead and the shield surrounding SOLO has an inner layer of \SI{5}{\cm} of ancient lead. The study of low-background lead for detector shielding is discussed in detail in~\cite{ALESSANDRELLO1991106}. All HPGe detectors are constructed using low-background designs and include remote preamplifiers. All detectors are cooled using liquid nitrogen from a fully automated filling system. The background radon in the BHUC counting room varies between 500-1000~Bq/kg. In order to suppress the background in the detectors caused by radon, a dedicated gas generator was installed which produces low activity nitrogen gas from a liquid nitrogen dewar at a rate of approximately 1.4 litres per minute. The gas purge flushes the detector counting cavities as well as the lead and copper shields. An additional detector, Ge-IV, operated by the University of Alabama, is installed outside the cleanroom, although this has not been used for assays discussed in this paper. 

\subsubsection{BUGS}
The Boulby Underground Germanium Suite (BUGS) hosts seven gamma spectroscopy detectors \SI{1.1}{\km} underground at the Boulby Underground Laboratory in a class 1000 cleanroom. Since 2013, the majority of screening efforts for the LZ experiment were performed using the Chaloner, Lunehead, and Lumpsey detectors. These detectors are, respectively, a Mirion (formerly Canberra) BE5030 broad-energy ultra-low background (ULB) HPGe detector\footnote{\label{note}It is inappropriate to classify BEGe and SAGe well detectors by their relative efficiency as they are designed to maximize efficiency to low-energy gamma-rays rather than to maximize efficiency to a \SI{1332}{\kilo\electronvolt} \cosz{} gamma-ray. For the BEGe type detector, the model number signifies the area of the front face and the thickness of the crystal. In the case of the BE5030 detector, it has a \SI{50}{\cm\squared} front face and a thickness of \SI{30}{\mm}.}, a Mirion ULB SAGe well-detector, and a refurbished \SI{100}{\percent} relative efficiency Ortec p-type detector used previously for the ZEPLIN--II and ZEPLIN--III experiment's low background counting~\cite{alner:2007ja,araujo:2011as,akimov:2011tj}. The BUGS detectors are housed in custom shields designed and built by Lead Shield Engineering Ltd. The shields comprise \SI{9}{\cm} thickness of lead and an inner layer of \SI{9}{\cm} thickness of copper with interlocking retractable roofs to simplify sample loading. The lead used in these shields has mostly been recycled from lead used to shield previous low-background experiments hosted at the Boulby Underground Laboratory. The characterizations and sensitivities of these detectors are discussed in \cite{Scovell:2017srl}.

In addition to these detectors, BUGS has installed additional Mirion ``specialty ultra-low background'' (S-ULB) detectors which have been used to screen later LZ samples since 2017. These comprise two p-type detectors, Belmont and Merrybent, with relative efficiencies of \SI{160}{\percent} and \SI{100}{\percent}, respectively, and Roseberry, a BE6530 BEGe type detector. For uniformity, these detectors are housed in similar shields to those used for the ULB standard Mirion detectors with the exception of the shield used for the Belmont detector which includes an inner liner of very low-background copper. These three detectors display substantially lower backgrounds than those of the ULB standard to significantly enhance the throughput rate of assays for LZ. The Belmont detector in particular, was used to further lower the \utTeL{} upper limits for titanium reported in~\cite{Akerib:2017iwt}.

The shields used for all detectors are purged using nitrogen from a Wirac NG6 gas generator. The Boulby Underground Laboratory benefits from a low baseline radon level (averaging \SI{\sim 2.5}{\becquerel\per\m\cubed}). To remove residual radon in the nitrogen purge gas, charcoal traps containing approximately \SI{6}{\kilogram} of Carboact activated charcoal are deployed in a Labcold ULTF416 $-80$~$^{\circ}$C chest freezer. This radon reduction system is based on the design of a radon emanation detector developed at the Centre de Physique des Particules de Marseille (CPPM)~\cite{Noel:2015nla}. As an example, the use of this purge system reduces the count rate in the \SI{609}{\kilo\electronvolt} line by at least a factor of 40 (from 16.4 counts/kg/day to less than 0.4 counts/kg/day at \SI{90}{\percent} confidence level).

\subsubsection{LBNL}
Lawrence Berkeley National Laboratory (LBNL) has two HPGe detectors devoted to assay~\cite{Smith:2015aoa}. These are housed in a near-surface room shielded with over \SI{1.5}{\m} of low radioactivity concrete. The \textsc{Merlin} detector is an Ortec \SI{115}{\percent} n-type detector. \textsc{Merlin} is shielded by \SI{20}{\cm} of lead with an inner layer of \SI{2.5}{\cm} of copper and is equipped with a plastic scintillator cosmic-ray veto. The BIG-8 detector is an \SI{85}{\percent} p-type detector shielded with \SI{10}{\cm} of lead and \SI{1.2}{\cm} of copper. It is equipped with a NaI veto that encloses the Ge crystal. Both detectors are flushed with nitrogen boil-off gas. The cosmic-ray vetos and local shielding result in detector performance equivalent to being sited at a depth of approximately \SI{500} m w.e.

\subsubsection{Alabama}
The University of Alabama operates two above-ground Canberra p-type low-background HPGe detectors~\cite{Tsang:2019apx}. These are Ge-II and Ge-III which have relative efficiencies of \SI{60}{\percent} and \SI{105}{\percent}, respectively. Each of these detectors is housed in shielding comprising \SI{20}{\cm} of lead with an inner layer of \SI{5}{\cm} of copper, instrumented with \SI{5}{\cm} thick plastic scintillator cosmic-ray veto detectors. The sample chambers are continuously flushed with nitrogen boil-off gas to displace radon. Despite their above-ground location, the background rates achieved this way are comparable to some of the underground devices, as reported in \tableWord~\ref{tab:GeDetPerf}. Ge-II and Ge-III have been used for items assayed using Neutron Activation Analysis (NAA), described in \sectionWord~\ref{sec:naa}. Ge-III was further utilized extensively for studies of $^{210}$Pb surface activities, their deposition through radon exposure, and their removal. 

\begin{table}
\caption{Key characteristics of the 12 detectors used in the LZ HPGe screening campaign. Crystal mass and volume is included to give an idea of the relative sizes of the crystal. In addition the relative efficiency is given for the p-type detectors and the area of the front face is given for the BEGe detectors.\label{tab:GeDetInf}}
    \tabcolsep=4pt
    \centering
    \begin{tabular}{ lllcccc }
    \toprule
    \multirow{3}{*}{\textbf{Location}} & 
    \multirow{3}{*}{\textbf{Detector}} & 
    \multirow{3}{*}{\textbf{Type}} & 
    \multirow{2}{*}{\textbf{V}} & 
    \multirow{2}{*}{\textbf{M}} & 
    \textbf{Relative} & 
    \textbf{Face} 
    \\
    &
    & 
    & 
    \multirow{2}{*}{\textbf{[cm$^{3}$]}} & 
    \multirow{2}{*}{\textbf{[kg]}} & 
    \textbf{Efficiency} & 
    \textbf{Area} 
    \\
    &
    & 
    & 
    & 
    & 
    \textbf{(\%)} & 
    \textbf{[cm$^{2}$]} 
    \\
    \hline
    \hline
    \\
    \multirow{6}{*}{BUGS} & Belmont & p-type & 600 & 3.2 & 160 & - \\
    & Merrybent & p-type & 375 & 2.0 & 100 & - \\
    & Lunehead & p-type & 375 & 2.0 & 100 & - \\
    & Roseberry & BEGe & 195 & 1.0 & - & 65 \\
    & Chaloner & BEGe & 150 & 0.8 & - & 50 \\
    & Lumpsey & SAGe well & 263 & 1.4 & - & - \\
    \hline
    LBNL & \textsc{Merlin} & n-type & 430 & 2.2 & 115 & - \\
    \hline
    \multirow{4}{*}{BHUC}& \textsc{Maeve} & p-type & 375 & 2.0 &85 & - \\
    & \textsc{Morgan} & p-type & 375 & 2.0 & 85 & - \\
    & \textsc{Mordred} & n-type & 253 & 1.3 &60 & - \\
    & SOLO & p-type & 113 & 0.6 & 30 & - \\
    \hline
    \multirow{2}{*}{Alabama} & Ge-II  & p-type & 260 & 1.4 & 60  & - \\
    & Ge-III & p-type & 406 & 2.2 & 105 & - \\
    \\
\bottomrule
\\

\end{tabular}
\end{table}
\begin{table}
\caption{Performance characteristics for each detector used in the LZ fixed contaminant screening campaign. For each detector, as is standard, the full-width at half maximum (FWHM) of the 1332~keV \cosz{} gamma-ray is shown. It can be seen that BEGe and SAGe-well type detectors (Roseberry, Chaloner \& Lumpsey) typically achieve the best resolution. A comparison of the integral counts between 60-2700~keV is given scaled to the mass of the detector crystal. Finally, a comparison between count rates for several peaks of interest are given corresponding to standard NORM (or in the case of \cosz{}, anthropogenic) isotopes. The gaps in this table reflect the absence of a measurable peak for the given energy either because of a lack of sensitivity (in the case of \SI{46.5}{\kilo\electronvolt} for p-type detectors) or due to very low background levels. The \textsc{Merlin} detector is dedicated to pre-screening NORM contamination in advance of samples being shipped to be assayed using a BHUC detector. This being the case, \cosz{} levels are not routinely quoted and, therefore, a background level does not appear in this table.\label{tab:GeDetPerf}}
\begin{adjustbox}{width=\textwidth,center}
    \tabcolsep=4pt
    \centering
    \begin{tabular}{ lccccccc }
    \toprule
    \multirow{3}{*}{\textbf{Detector}} & 
    \textbf{FWHM}  &  
    \textbf{Integral (60 -}  & 
    \textbf{\tltze{}}   &  
    \textbf{\bitof{}}  &  
    \textbf{\cosz{}} & 
    \textbf{\kfz{}} &
    \textbf{\pbtoz{}}\\ 
    & 
    \textbf{1332 keV}  &  
    \textbf{2700) keV}  & 
    \textbf{2614.5 keV}   &  
    \textbf{609.3 keV}  &  
    \textbf{1332.5 keV} & 
    \textbf{1460.8 keV} &
    \textbf{46.5 keV}\\ 
    & 
    \textbf{[keV]}  &  
    \textbf{[kg$^{-1}$$\cdot$day$^{-1}$]}  & 
    \textbf{[kg$^{-1}$$\cdot$day$^{-1}$]}   &  
    \textbf{[kg$^{-1}$$\cdot$day$^{-1}$]}  &  
    \textbf{[kg$^{-1}$$\cdot$day$^{-1}$]} & 
    \textbf{[kg$^{-1}$$\cdot$day$^{-1}$]} &
    \textbf{[kg$^{-1}$$\cdot$day$^{-1}$]}\\
    \hline
    \hline
    \\
    Belmont & 1.92 & 135.0 & 0.3 & 1.4 & 1.6 & 1.0 & -\\
    Merrybent & 1.87 & 167.4 & 0.4 & 1.8 & 0.6 & 1.9 & -\\
    Lunehead & 1.86 & 582.4 & 2.0 & 4.7 & 1.5 & 9.2 & -\\
    Roseberry & 1.58 & 181.1 & - & - & 0.6 & 0.7 & 0.3\\
    Chaloner & 1.56 & 1053.0 & 1.7 & 9.5 & 1.2 & 8.3 & 1.7\\
    Lumpsey & 1.66 & 4256.8 & 12.2 & 60.3 & 1.6 & 7.0 & 13.7\\
    \textsc{Merlin} &3.59 &68868.3 & 9.7 & 7.5 &- &20.0 &80.2\\
    \textsc{Maeve} &3.19 &956.1 & 1.8 & 1.4 & 0.5 &3.5 & 49.6\\
    \textsc{Morgan} & 2.68 & 1338.8 & 3.2 & 8.8 &3.8 &4.8 & 4.6\\
    \textsc{Mordred} &2.44 & 2103.2 & 2.1 & 3.9 &1.6 &7.4& 29.0\\
    SOLO & 5.52 &786.9 & - & 3.3 &- &- &-\\
    Ge-II & 3.6  & 9600 & - & 3.6 & 10.3 & 2.3 & -\\
    Ge-III & 2.71 & 8600 & - & 9.6 & 1.7 & 2.5 & 1.6 \\
    \\
\bottomrule
\\

\end{tabular}
\end{adjustbox}
\end{table}

\subsection{HPGe Cross-Calibration\label{sec:cross-cal}}
Early in the LZ assay efforts, it was recognized that samples would be distributed amongst a large variety of detectors with different backgrounds, shielding arrangements, and histories. To cross-calibrate and evaluate the systematic uncertainties in assays performed with a number of the detectors listed in \tableWord~\ref{tab:GeDetPerf}, a sample of latite with well-characterized uranium, thorium and potassium content was prepared. This material has been used by LBNL for more than 30 years to characterize its detectors. The uniformity of the radioactivity has been studied and is confirmed to be flat across a variety of sample sizes from \SI{\sim 1}{\mm} up to several~cm. An S5 Marinelli beaker of this mineral was prepared and sealed. The content and activity was not known by the rest of the collaboration and the same beaker was subsequently analyzed by all groups on all their detectors. The analyses were sent to a central site, amassed and compared by one individual who had knowledge of the true contamination of the calibration source material. 

\begin{figure}
\centering
\includegraphics[width=0.8\textwidth]{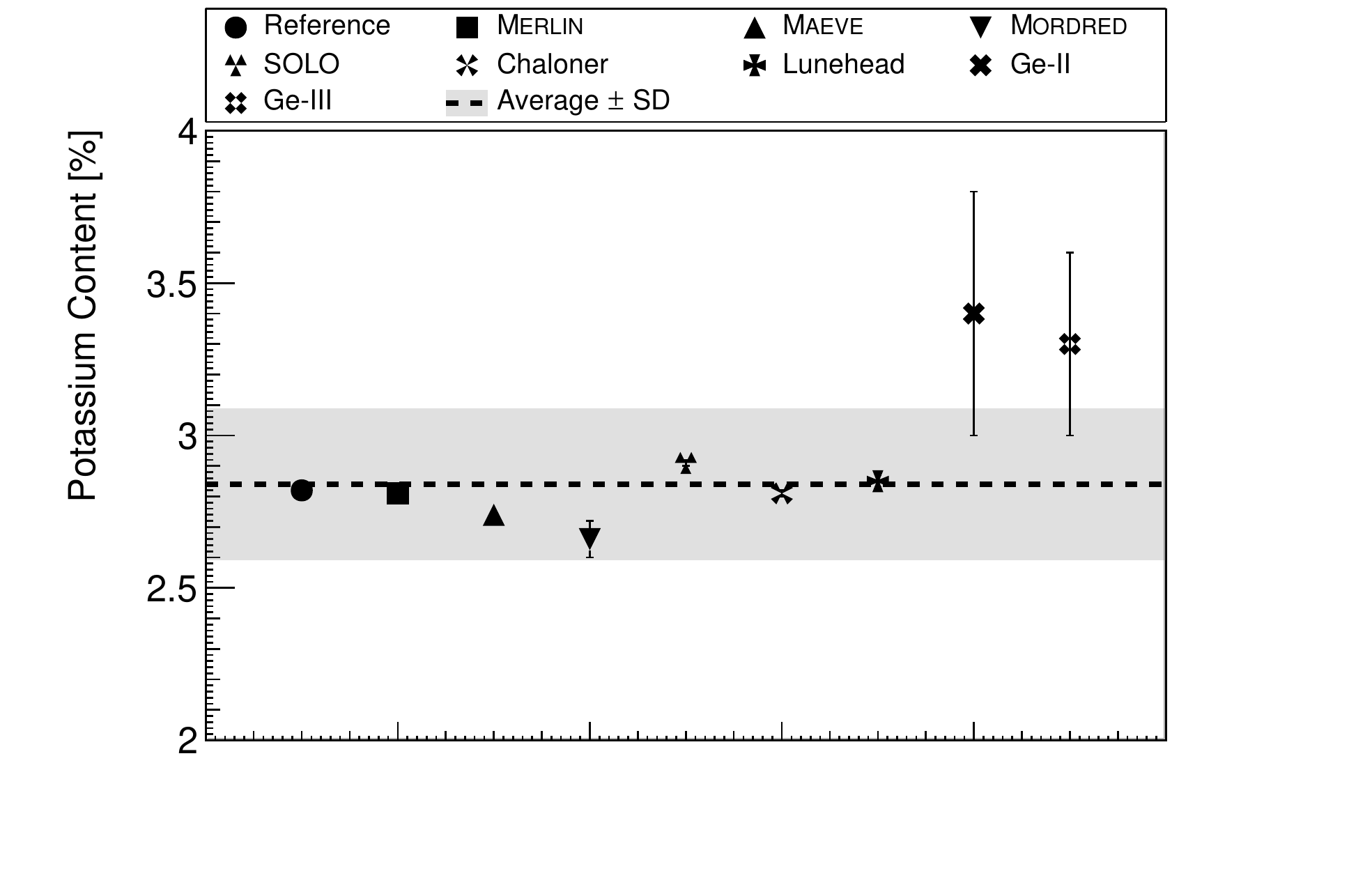}
\caption{Cross-calibration results for potassium concentration for the detectors used in the LZ HPGe radioassay campaign. The reference value for potassium concentration in this sample is \SI{2.82(1)}{\percent}. Excluding this, an error-weighted mean of \SI{2.84(2)}{\percent} was derived from the measurements of \kfz{}. In this figure, the gray band represents the standard deviation of the measurements with a value of \SI{0.25}{\percent}.}\label{crosscal}
\end{figure}

This comparison uncovered some issues with several analyses, mostly due to problems with the Monte Carlo simulations of the detectors. After discrepancies were highlighted by the individual who amassed the results, these issues were identified and corrected. The results were again compared across all the detectors. \tableWord~\ref{tab:GeCrossCal} lists reference values for each isotope; compares results from each detector; and gives their combined average and standard deviation. A visual comparison between detectors used in the calibration program for potassium is shown in \figureWord~\ref{crosscal}. It is important to note at this point that when a concentration is reported, in parts per value $e.g.$ ppm, ppb, ppt (\si{\gram} of U/Th per \si{\gram} of material), it is no longer pertinent to refer to late chain or early chain values as the concentration defines the concentration of the progenitor isotope (\utTe{ }, \utTF{}, \thtTt{}) assuming secular equilibrium~\cite{malling:2013jya}. There is some residual disagreement between Ge-II, Ge-III and the other detectors. These two surface detectors were used primarily for neutron activation analysis and pre-screening of samples before sending them for assay on an underground detector. We noted the discrepancies but because of their limited use, we chose to accept this in a larger systematic for these two detectors. The vast majority of the assays performed as part of the LZ campaign were carried out on the detectors shown in \figureWord~\ref{crosscal}. The cross-calibration effort confirmed that the modeling of detector geometries and efficiencies were correctly handled and provides a reasonable estimate on the systematic variation among the assays of \SI{\sim 10}{\percent} thus giving the collaboration confidence that each individual facility is able to produce consistent and accurate assay results. This being the case, newer detectors that were used later in the campaign (such as the S-ULB detectors added to BUGS) were not required to be characterized using the latite sample. Each facility was able to implement their own calibration and characterization methods and the subsequent assay results were accepted to be accurate within statistical errors and within the systematic errors of the detector used. Agreement between the detectors used in the cross-calibration campaign and newer detectors was also informally verified by assaying identical samples on different detectors.

For some materials, such as the titanium, additional steps were taken to increase our confidence in the assay precision. These included assaying the same sample in multiple locations and at different times (to allow for the decay of cosmogenically activated isotopes of scandium) and assaying samples using mass spectrometry. This cross-calibration also verified that all counters had effective Rn-reducing purge systems. Periodically, the LZ assay campaign screening of selected samples was repeated on different detectors to monitor for stability of assays. These ongoing comparisons spanned a variety of source materials and a wide range of contamination levels, while also probing for Th in-growth in particular LZ components.
For many materials we complemented the HPGe assays with Inductively-Coupled Plasma Mass Spectrometry (ICP-MS) and, after their installation at Boulby, with the new S-ULB detectors to further verify our measured concentrations of \utTeE{} and \thtTtE{}.

Results from the assays deploying gamma spectroscopy are presented in \tableWord~\ref{tab:HPGeBig}.

\begin{table}
\caption{Results from the HPGe cross-calibration performed using a sample of latite. For the \utTeE{} and \utTeL{} columns, the contamination reported is that of the progenitor isotope \utTe{} assuming secular equilibrium and for the \thtTtE{} and \thtTtL{} columns, the contamination reported is that of \thtTt{} assuming secular equilibrium. \label{tab:GeCrossCal}}
\begin{adjustbox}{width=\textwidth,center}
    \tabcolsep=4pt
    \centering
    \begin{tabular}{ lccccc }
    \toprule
    \textbf{Detector} & 
    \textbf{\utTeE{} (ppm)}  & 
    \textbf{\utTeL{} (ppm)} &  
    \textbf{\thtTtE{} (ppm)} &  
    \textbf{\thtTtL{} (ppm)}  & 
    \textbf{K ($\%$)}   \\  
    \hline
    \hline
    \\
Reference & 8.5(1) & 8.87(4) & 12.1(1) & 12.1(1) & 2.82(1) \\
\hline
\textsc{Merlin} & - & 8.92(9) & 12.4(1) & 12.4(1) & 2.81(3) \\
\textsc{Maeve} & 8.6(1) & 8.6(1) & 11.9(1) & 11.9(1) & 2.74(3) \\
\textsc{Mordred} & 10.2(1) & 7.92(5) & 11.3(1) & 11.3(2) & 2.66(6) \\
SOLO & - & 6.16(1) & 9.94(1) & 12.5(7) & 2.91(1) \\
Chaloner & 7.9(2) & 8.73(5) & 11.1(1) & 11.1(1) & 2.81(1) \\
Lunehead & - & 8.5(1) & 11.8(1) & 11.8(1) & 2.85(1) \\
Ge-II & 11.4(15) & 9.6(13) & 12.2(17) & 12(16) & 3.4(4) \\
Ge-III & 10.3(10) & 9.2(9) & 12.8(13) & 12.1(12) & 3.3(3) \\
\hline
Average & 9.2(2) & 7.61(3) & 10.54(5) & 11.9(1) & 2.84(2) \\
Std. Dev.  & 1.26 & 0.98 & 0.84 & 0.46 & 0.25 \\
    \\
    \bottomrule
    \\
\end{tabular}
\end{adjustbox}
\end{table}

\subsection{Mass Spectrometry}
\label{sec:icpms}

Inductively-Coupled Plasma Mass Spectrometry (ICP-MS) allows very precise direct measurement of the elemental abundances of uranium and thorium in small samples. The assays can be very quick, taking hours to days depending on requisite sensitivity down to sub-ppt levels and depending on related sample preparation protocols. ICP-MS has been used extensively in LZ to quickly measure \utTe{} and \thtTt{} in small samples to either reject or clear materials for use, or to pre-screen materials prior to  assay with gamma spectroscopy which can determine the complete activity through the \utTe{} and \thtTt{} decay chains. 
The speed of ICP-MS allowed rapid analysis of test pieces provided by manufacturers at specified points in the production processes to detect potential issues and to ensure radioactivity and cleanliness compliance. The manufacture of the Ti cryostat is one such example, highlighted in \sectionWord~\ref{sec:NSTi}.  

The majority of ICP-MS assays for LZ were performed using a dedicated mass spectrometry laboratory at UCL, housed in a class 1000 cleanroom facility and operating an Agilent 7900 spectrometer installed in 2015 exclusively for LZ~\cite{Dobson:2017esw}. Sample preparation and analysis procedures have been developed for materials with U/Th concentrations in the \si{ppt} to \SI{1}{ppb} range: Samples are microwave-digested in pre-cleaned modified-PTFE vessels using ultra-high purity acids. They are then diluted, without further chemical treatment, into disposable \SI{50}{\milli\litre} polypropylene (PP) vessels ready for ICP-MS analysis. Fractional recoveries of \thtTz{} and \utTT{} spikes added prior to digestion are used to correct for \thtTt{} and \utTe{} signal loss from a range of sources. In particular, this enables accurate analysis of samples with high total dissolved solids (TDS) where the instrument response degrades throughout the run. A full assay including digestion, ICP-MS measurement and analysis can be completed in a single day. The UCL facility was upgraded in 2019 with an Agilent 8900 ICP-MS. 

In addition to the ICP-MS system at UCL, some material samples were assayed using facilities at the University of Alabama, the  Centre for Underground Physics in Korea, and the Black Hills State University.  At the University of Alabama, the LZ group set up a sample preparation laboratory in a class 500 cleanroom equipped with a cryogenic mill, microwave digestion system, and digestion bomb.  Further processing of samples, including spiking and resin-based extraction of U/Th isotopes was carried out in a separate cleanroom.  The samples were then given to the Department of Geological Sciences which processed the samples using a Perkin-Elmer SCIEX-ELAN 6000 system.  In Korea and at Black Hills State University, samples were measured using Agilent 7900 spectrometer, as was used at UCL. Results from ICP-MS assays for LZ are presented in \tableWord~\ref{tab:ICPMSBig}.

Finally, in the early days of the LZ assay program, a small number of items were assayed using Glow Discharge Mass Spectrometry (GDMS). These assays were performed using a Thermo-Fisher VG 9000 GDMS instrument operated by the National Research Council of Canada (NRC). GDMS can achieve sensitivities of around \SI{10}{ppt} for conductive materials. For this reason, GDMS was used for some assays of titanium. Results from the GDMS assays for LZ are presented in \tableWord~\ref{tab:GDMSBig}.

\subsection{Neutron activation Analysis (NAA)}
\label{sec:naa}

As with ICP-MS, NAA allows sensitive analysis of small concentrations of Th and U in small samples. It only constrains the early decay series. It can be utilized for materials where the matrix does not form long-lived radioactivity after neutron capture. As such, it is well-suited for plastics.

For NAA assay in LZ, the University of Alabama group utilizes the \SI{6}{MW_{th}} tank-type MIT Reactor II (MITR-II, located at the Massachusetts Institute of Technology) to activate samples. This technique is typically used for small size samples of a few mm in diameter and several cm in length.
LZ samples were prepared at the University of Alabama in a cleanroom prior to their shipping to the reactor for activation. Surfaces were leached extensively in high-grade acids to distinguish bulk from surface activities. The polyethylene vials used for irradiation are cleaned, welded shut, and leak tested. Samples are typically irradiated for \SI{10}{\hour} in the high-flux sample insertion facility of MITR-II before being returned for counting analysis. Storage within sealed polyethylene vials serves to protect the samples from cross-contamination during transport and activation. There is a typical delay of around \SI{24}{\hour} between activation being completed and samples being assayed using Ge-II or Ge-III, but this is acceptable when compared to the half-lives of the activation products used for NAA ($e.g.$ \kft{} - $\tau_{1/2}$ = \SI{22.3}{\hour}, \patTT{} - $\tau_{1/2}$ = \SI{26.97}{days}, and \nptTn{} - $\tau_{1/2}$ = \SI{2.36}{days}). Assays typically lasted 2 to 4 weeks and a double-differential time-energy analysis is used to determine elemental concentrations of samples. Neutron capture cross sections, averaged over the energy distributions of the three reactor neutron flux components, taken from the JENDL4.0 database are utilized in the data analysis. For each of the LZ activation campaigns the neutron fluxes were determined by activating the NIST reference material, fly ash, immediately following the sample. This allows to reference the elements of interest directly to a standard. This method is discussed in depth in~\cite{Leonard:2017okt}.

The techniques employed by the University of Alabama group routinely achieve a sensitivity of $10^{-12}$~g U/Th per g of material and, indeed, sensitivities as good as $10^{-14}$~g of U/Th per g of material have been reported by the same group for assays related to the EXO-200 experiment \cite{Leonard:2017okt,leonard:2007uv}. Such sensitivity has been key for assays and selection of raw materials not readily amenable to direct HPGe due to sample mass or minimal detectable activity requirements, or to ICP-MS due to difficulties in sample digestion and preparation. Selection of PTFE to manufacture the LZ TPC reflectors are one such example where NAA has been deployed, with results presented in \tableWord~\ref{tab:NAABig}. 


\section{Radon Emanation}
\label{S:3}

\subsection{Origin of Radon Emanation}\label{S:3_1}

All isotopes of radon are radioactive and only five are naturally found in minute quantities in nature. Those of interest for the LZ background model and often other experiments in search for WIMP dark matter are \rnttt{} ($\tau_{1/2}=$ \SI{3.82}{\day}) from the \utTe{} decay chain and \rnttz{} ($\tau_{1/2}=$ \SI{55.8}{\second}) from the \thtTt{} decay chain; hereafter called radon and thoron, respectively. Due to the long lifetime of their progenitor isotopes, radon and thoron are produced at a near-constant rate within detector material over the lifetime of an experiment. The emanation rate of a material can be broken down into two parts: emanation due to recoiling radon atoms and emanation due to diffusion. Emanation due to diffusion can vary drastically depending on chemical and lattice structures of a material, density, surface roughness, and temperature. The diffusion length, $L(m)$, of radon for a given material can be represented as $L(m) = \sqrt{D/\lambda}$, where $D$ is the diffusion coefficient and $\lambda{}$ the decay constant.

The background from radon emanation in LZ is dominated by the ground-state to ground-state or ``naked'' $\beta$-emission from the \pbtof{} progeny of the \rnttt{} sub-chain as it decays to \bitof. The relatively long half-life of \rnttt{} leads to a homogeneous mixing within the target volume, resulting in a uniform ER background with a $\beta$-spectrum up to 1019 keV. The background from \rnttz{} is expected to be significantly suppressed due a much smaller diffusion length as a result of its shorter half-life, hence most of it is expected to decay within the material volume in comparison to \rnttt{}  before diffusing out, or before maximally mixing with the active xenon volume.

Radon emanation accounts for \SI{\approx 66}{\percent} of the projected ER background in the WIMP search region of interest in LZ~\cite{Akerib:2018lyp}, predominantly from a projected \rnttt{} specific activity of \SI{2}{\micro\becquerel\per\kilogram} that corresponds to approximately \SI{20}{\mBq} in the 10 tonnes of xenon, from which \SI{11}{\mBq} is in the fiducial volume. As demonstrated by previous LXe-based rare-event search experiments, including LUX and ZEPLIN--III, the amount of radon in the target can be deduced through analysis of the \pbtof{} $\beta$-spectrum and from coincidence tagging of decaying radon daughter $^{214}$Bi and $^{214}$Po~\cite{akerib:2014rda,araujo:2011as}. While such \emph{in-situ} techniques are powerful in constraining the observed radon emanation background once the detector is closed and operational, a screening campaign akin to that for fixed contaminants is required to inform material selection for detector and sub-system construction, and for the experiment background prediction.  

\subsection{Radon Emanation Screening}
\subsubsection{Screening Techniques}

Radon screening typically involves reconstructing the radon emanation rate by measuring the radon sub-chain daughter isotopes. An approximate way of achieving this uses gamma spectroscopy to measure the \bitof{} ($\tau_{1/2}=$ \SI{19.9}{\minute}) and \pbtof{} ($\tau_{1/2}=$ \SI{26.8}{\minute}) decay rates, from which the radon activity is inferred. Although this method provides useful limits for emanation rates, it is extremely difficult to distinguish radon daughters decaying in the bulk of the material to those that decay outside. Thus, a precise emanation rate cannot be deduced without a material-specific diffusion model. 

A more direct and precise approach, one that has been utilised in four of the LZ facilities, is to directly measure the activity of radon that has emanated out from the material. The sample material is initially enclosed in an air-tight chamber that is filled with a low-radon carrier gas, typically helium or nitrogen. This carrier gas prevents recoiling radon atoms from embedding into the chamber walls.After an emanation period that allows the radon concentration in the chamber to approach equilibrium (\SI{\sim 1}{week}), the emanated radon atoms are transferred into a detector that measures the rate of \potoe{} ($\tau_{1/2}=$ \SI{3.1}{m}) and \potof{} ($\tau_{1/2}=$ \SI{164}{\micro\second}), with a mean decay time of $\sim$\SI{71}{\minute} after the initial \rnttt{} decay. A second approach of measuring the emanation rate is by identifying the delayed \bitof{}Po coincidence. In both cases, the radon emanation rate is reconstructed by correcting for the detection and transfer efficiencies, measured during dedicated calibration runs with radon sources of known activity.

The first of these reconstruction techniques determines the radon emanation rate by detecting the $\alpha$-particles emitted from the \potoe{} and \potof{} isotopes. These systems use electrostatic silicon PIN-diodes to attract and capture the predominantly positively charged ions \SI{87.3\pm1.6}{\percent} of radon daughter nuclei by using an electric field that is generated from the negative voltage applied on the PIN-diode~\cite{Pagelkopf:2003wtf}. The $\alpha$-particles emitted from the \potoe{} and \potof{} ions are detected by a PIN-diode as they undergo $\alpha$-decay and are distinguished by the energies they deposit; \SI{6.1}{\mega\electronvolt} and \SI{7.9}{\mega\electronvolt}, respectively. Of the four radon-emanation screening facilities used by LZ, three use electrostatic PIN-diode detectors for radon emanation. The fourth facility collects the harvested radon by dissolving it in organic liquid scintillator by means of a carrier gas. The delayed \bitof{}-\potof{} coincidences are then counted to infer the corresponding \rnttt{} decay rate. 
All facilities operate at room temperature such that the expected suppression of diffusion-dominated radon emanation at low temperature is not probed.

The LZ collaboration performed cross-calibrations for the four radon facilities deployed as part of our assay program. A rubber sample  previously screened by the EXO collaboration~\cite{Albert:2015nta, Miller:2017tpl} was assayed at each of the radon emanation facilities. Prior to the emanation period, the sample was prepared under the same conditions to reduce the chances of environmental contamination. The surface of the sample was scrubbed with isopropyl alcohol-soaked lint-free wipes and inspected with UV-light to ensure no presence of surface contamination. The activity of the sample was $\BigOSI{10}{\mBq}$ and was thus well above the minimal detectable activities of the radon systems. \tableWord~\ref{tab:ReDet} presents the results of the cross-calibration and a summary of key details of the LZ radon screening facilities.

\begin{table}
\caption{Comparison of the key highlights of the four radon emanation facilities used by LZ. The chambers detailed are those used in containing the sample material, where radon is collected. Some facilities operate two chambers as detailed below. Chamber blank rates detail the emanation rate from the chambers alone and are background subtracted for sample measurements. Detector efficiency represents the fraction of activity measured from the total radon inside the detecting volume; independent of chamber usage and transfer efficiency. The cross-calibration figures represent the reconstructed emanation rate of a standard rubber sample previously used by other collaborations. When not stated, overall uncertainties are estimated to be 10-20\%. \label{tab:ReDet}}
\begin{adjustbox}{width=\textwidth,center}
    \tabcolsep=4pt
    \centering
    \begin{tabular}{ lcccccc }
    \toprule
    \multicolumn{1}{c}{\textbf{Detector}} & 
    \textbf{Type} & 
    \multirow{1}{*}{\textbf{Chamber Volumes}} &  
    \multirow{1}{*}{\textbf{Chamber Blank Rates}} &  
    \multirow{1}{*}{\textbf{Transfer Efficiency}} & 
    \multirow{1}{*}{\textbf{Detector Efficiency}} & 
    \multirow{1}{*}{\textbf{Cross-Calibration}}\\ 
    & 
    & 
    \multirow{1}{*}{\textbf{[L]}} & 
    \multirow{1}{*}{\textbf{[mBq]}} & 
    \multirow{1}{*}{\textbf{[\%]}} & 
    \multirow{1}{*}{\textbf{[\%]}} & 
    \multirow{1}{*}{\textbf{[Measured/EXO--activity]}}\\ 
    \hline
    \hline
    \vspace{-4mm}
    \\ 
    SDSM\&T & 
    PIN-diode & 
    \multicolumn{1}{c}{\begin{tabular}[c]{@{}c@{}}13\\ 300\end{tabular}} & \multicolumn{1}{c}{\begin{tabular}[c]{@{}c@{}}0.2\\ 0.2\end{tabular}} & 
    \multicolumn{1}{c}{\begin{tabular}[c]{@{}c@{}}94\\ 80\end{tabular}} & 
    25 & 
    \multicolumn{1}{c}{\begin{tabular}[c]{@{}c@{}}0.89$\pm$0.15\\ 1.11$\pm$0.28\end{tabular}}
    \vspace{2mm}
    \\ 
    Maryland & 
    PIN-diode & 
    4.7 & 
    0.2 & 
    96 & 
    24 & 
    1.13 $\pm$ 0.19
    \vspace{2mm} \\
    UCL & 
    PIN-diode & 
    \multicolumn{1}{c}{\begin{tabular}[c]{@{}c@{}}2.6\\ 2.6\end{tabular}} &
    \multicolumn{1}{c}{\begin{tabular}[c]{@{}c@{}}0.2\\ 0.4\end{tabular}} & 
    \multicolumn{1}{c}{\begin{tabular}[c]{@{}c@{}}97\\ 97\end{tabular}} & 
    30 & 
    1.49$\pm$0.15
    \vspace{2mm} \\ 
    Alabama & 
    Liquid Scint. & 
    \multicolumn{1}{c}{\begin{tabular}[c]{@{}c@{}}2.6\\ 2.6\end{tabular}} &
    \textless{}0.4 & 
    34 &
    36 & 
    0.83$\pm$0.17
    \\
    \bottomrule
\end{tabular}
\end{adjustbox}
\end{table}
 
LZ makes use of two portable radon collection systems for equipment that is too large or delicate to move to the radon emanation facilities or for assays of systems under construction in the SURF Surface Assembly Laboratory (SAL). Emanated radon is transferred to a cold trap consisting of copper beads or wool that is double-sealed and then transported by car or overnight shipping to the radon facility at South Dakota School of Mines and Technology (SDSM\&T) or University of Maryland. The collected radon would then be transferred over into the respective radon detector with transfer efficiencies taken into account from portable-system specific calibrations. The activity is then reconstructed by correcting for the transportation time and detector efficiency. These portable systems were critical for measurements of radon emanation from the assembled LZ detector and from large instrumentation used in the circulation path. Results from the radon emanation assay campaign are presented in \tableWord~\ref{tab:RnBig}. For majority of the measurements, smaller samples are sent to the facilities detailed below and radon is often collected in emanation chambers and measured with their respective detectors.

\subsubsection{SDSM\&T}

The SDSM\&T system uses two electropolished stainless steel chambers as the radon collection media: a \SI{13}{\litre} vessel for smaller components and a \SI{300}{\litre} vessel for larger components. Emanation samples are placed in the chambers with care taken not to introduce dust into the chambers or onto the samples. The chambers are then filled with nitrogen gas that has been scrubbed of radon by an activated charcoal trap cooled to \SI{196}{\kelvin} by a mixture of dry ice and isopropyl alcohol.

After the sample has emanated, the radon is concentrated and transferred to the \SI{1.7}{\litre} detection chamber in a multi-stage process. In the first stage, the radon is transferred from the emanation chamber to a large brass wool trap cooled to \SI{77}{\kelvin} by liquid nitrogen (LN$_{2}$). A high transfer efficiency is achieved even for the \SI{300}{\litre} vessel by repeatedly pumping the chamber out through the trap and refilling. 
The radon is then transferred to a small brass wool trap by warming the large trap and cooling the small trap and flowing clean nitrogen through first the large trap and then the small trap. Due to the volume of gas allowed to flow, the transfer efficiency from the large trap to the small trap is \SI{\approx 100}{\percent}. Then with the small trap and detection chamber at low pressure, the small trap is warmed and nitrogen is allowed to flow through the small trap into the detection chamber to raise the pressure in the detection chamber to \SI{100}{Torr}. This process transfers \SI{\sim 95}{\percent} of the radon to the detection chamber, for overall transfer efficiencies of \SI{80}{\percent} (\SI{94}{\percent}) for the \SI{300}{\litre} (\SI{13}{\litre}) chamber. The detector is an electrostatic silicon PIN-diode detector as described above. The detector efficiency was determined to be $23 \pm 2 \, \%$ for \potoe{} and $25 \pm 2 \, \%$ for \potof{} under standard operating conditions. A system for performing emanations at LN temperatures is under construction.

\subsubsection{University of Maryland}

The Maryland system's primary focus was to measure emanation rates from volumes that act as their own emanation chambers, such as the LZ compressor system. The Maryland system also contains a \SI{4.6}{\litre} stainless steel vessel, used to calibrate the system and to perform measurements on smaller samples. The system uses an activated charcoal trap operated at LN$_{2}$ temperature to initially scrub the radon from the helium carrier gas. The radon from the emanation volume is captured by a copper pellet trap also cooled to LN$_{2}$ temperature. The copper trap is a 0.5" electropolished stainless steel tube bent into a U-shape and containing 180 g of copper pellets (Atlantic Equipment Engineers CU-131). The pellets range in size from 1/16" to 3/32". The radon is released from the copper at room temperature and transferred to a \SI{1.7}{\litre} chamber containing an electrostatic PIN diode detector. 
The efficiency of the copper pellet trap was determined to be near 100\% by repeated trapping and counting of a radon sample. The absolute efficiency of the detector was determined to be 24\% by measuring a calibrated natural rock source purchased from Durridge. Ion drift simulations were carried out to study the performance of the detector. The predicted efficiency agrees with the measurements within the systematic uncertainty for both the Maryland and the SDSM\&T detection chambers, which have a near-identical design. 

A second electrostatic PIN diode radon counting system is operated at Maryland in flow-through mode to measure the elution curves of charcoal samples in helium carrier gas. A piece of uranium ore acts as an uncalibrated radon source for these measurements. Once the elution curve is determined, an appropriate radon harvesting time can be chosen for each charcoal sample during its subsequent radon emanation measurement. 

\subsubsection{UCL}

The UCL system's custom-made electrostatic detector was originally developed for high sensitivity radon measurements for the ELEGANT V and Super-Kamiokande experiments~\cite{Mitsuda:2003eu}. The detector consists of a \SI{70}{\litre} electropolished steel chamber with a silicon PIN-diode located at the top, operating under the same principles as described above. By the use of a calibration source of known activity (a \SI{1.32}{\kilo\becquerel} ``flow-through" \ratts{} source by Pylon Electronics, RN-1025), the detection efficiencies for \potof{} and \potoe{} are determined to be \SI{31.6\pm1.6}{\percent} and \SI{27.1\pm1.4}{\percent}, respectively, with helium as the transfer gas. The system operates two \SI{2.7}{\litre} stainless steel chambers as the emanation media. The larger detector volume and the small emanation volumes allow a single step transfer process, where helium gas is flushed through the emanation chambers, directly into the detector. To eliminate the contamination of background radon from the carrier gas, the gas is initially allowed to flow through an activated carbon trap stored in an ultra-low temperate freezer (\SI{193}{\kelvin}) and the entire system is purged to remove accumulated radon emanating out of the transfer lines.

A second mode of operation for the system uses \SI{57}{\g} of activated carbon (a synthetic charcoal sourced from Carbo-Act International \cite{Pushkin:2018wdl}) as a radon collection trap. In larger emanation volumes, the radon is initially absorbed into the cooled trap while the carrier gas passes through. The trap is then heated to release the radon and the carrier gas is then used to transfer the concentrated radon into the detector volume. The trapping efficiency for this setup has been measured to be \SI{\approx 93}{\percent} at \SI{248}{\kelvin}. The cold trap was not generally necessary for the results reported in \tableWord~\ref{tab:RnBig}.  

A second facility with sensitivity to low-temperature emanation is under construction and will be operated at the Rutherford Appleton Laboratory. 

\subsubsection{University of Alabama}

The radon emanation facility at Alabama operates in a similar principle to those of the other three detectors, with the exception of their detection technique. Two \SI{2.6}{\litre} electropolished emanation chambers, of the same design as those utilized by the UCL group, are used to accumulate the radon outgassed from samples of interest.
Boil-off nitrogen, selected for its low intrinsic radon content, serves as a carrier to transfer the radon into about \SI{150}{\milli\litre} of organic liquid scintillator. The carrier gas is flowed for \SI{48}{\minute} at a rate of \SI{20}{\milli\litre\per\minute}. Experiments with a calibrated Pylon RN-1025 radon source showed that longer purge times and higher flow rates result in more effective radon removal from the collection vessel but, on the other hand, lead to radon loss in the liquid scintillator due to the dissolved radon being washed out. The chosen parameters constitute the optimal compromise between these two mechanisms. Measurements with cascaded scintillator collectors showed that under these conditions about \SI{70}{\percent} of the radon arriving at the scintillator-gas interface dissolves in the scintillator. Use of the calibrated radon source yields an overall radon transfer efficiency of \SI{34.3}{\percent}.
The radon-loaded scintillator is transferred into a small acrylic counting cell, equipped with one low-activity \SI{76}{\milli\meter} (3 inch) Hamamatsu R-1307 PMT. The measurement of delayed \bitof{}-\potof{} coincidences allows the determination of the decay rate with low background. The analysis of the delayed-coincidence data sets uses cuts on the $\beta$- and $\alpha$-like energy deposits. The distribution of time differences between $\beta$- and $\alpha$-like events is fit to an exponential plus a constant, with the correlation time frozen to the known \potof{} mean lifetime. The exponential component of the fit determines the decay rate; the constant term unfolds the random background. The efficiency of these cuts has been determined, by means of loading radon derived from the calibrated source, to be \SI{35.9}{\percent}.
However, because of the need to transfer both the carrier gas and the scintillator, the limiting factor is not the detector background but the blank (radon introduced during transfer and handling). Repeated measures of the handling blank allowed for estimation of the blank subtraction uncertainty. A further source of background is steady state leakage of radon into the sealed counting cells. Counting continued after allowing the radon to decay, thus, quantifying this background directly. This leakage varied between counting cells.


\section{Surface Contamination}
\label{S:4}
\subsection{Origins of Surface Contamination}
Radio-pure detector materials and components selected through the LZ screening campaign may be contaminated during the assembly process. Indeed, exposure to airborne radon at any stage of the assembly process results in the contamination of detector materials by radon daughters (mainly the long-lived \pbtoz{}, $\tau_{1/2} =$ \SI{22.3}{\year}) that plate-out onto surfaces~\cite{jacobi1972activity}. Environmental dust also deposits on detector material surfaces, and later, radon emanates from these surfaces and could yield daughter decays in the LXe volume during the LZ data-taking period. 

Of particular concern is \pbtof{}, which will emit naked betas leading to a continuous ER background down to the WIMP energy window as described in \sectionWord~\ref{S:3_1}. Also, \pbtoz{} will subsequently decay, with its granddaughter ${^{210}}$Po releasing \SI{5.3}{\mega\electronvolt} alphas. This induces neutrons via ($\alpha,n$) reactions on low-Z nuclei in various detector materials, leading to NR backgrounds. Stable \pbtos{} from the decay of ${^{210}}$Po  on material surfaces may recoil into the LXe volume producing a complicated wall background (0 to \SI{103}{\kilo\electronvolt} in energy), which, despite fiducialization, could yield leakage nuclear recoil events in the region of interest due to poor position reconstruction because of their high radius (near wall) and low energy~\cite{akerib2018position}. Surface contamination by radon daughters and dust thus contributes to LZ internal ER and NR backgrounds (at the expected level of about \SI{3}{\percent} and \SI{38}{\percent}, respectively) 
and should therefore be carefully controlled to meet the low level background requirements of the LZ experiment~\cite{Akerib:2018lyp}. To this effect, a stringent cleanliness campaign was developed to monitor and mitigate this contamination during sub-systems assemblies, mainly the TPC detector assembly in the SAL.

This section discusses the estimation of the TPC surface contamination from both Rn and dust accumulated during the different phases of the TPC construction, along with the cleanliness measures and procedures undertaken to ensure the contamination levels remain below LZ requirements, and to minimize the internal backgrounds.

\subsection{Dust Deposition}
\label{dp_intro}
The ambient dust in the SAL class 1000 cleanroom comes from two main sources: dust from outdoor air flowing through the air filtration system, and dust carried in and generated by personnel and material. The air filtration system consists of a series of six high-efficiency particulate air (HEPA) filters with fiber glass membranes
that remove dust particulates from outdoor air fed into the cleanroom with an efficiency \SI{\geq 99.97}{\percent} for particulates \SI{\geq 0.3} {\micro\m} \cite{us1997specification}.
Recorded dust level (from two Met One GT-526S particle counters installed at different locations) within the cleanroom with and without personnel demonstrates that outdoor dust passing through the air filtration system has a negligible contribution to the dust level within the cleanroom.

The main contribution to dust therefore comes from personnel who bring in dust from their bodies, cleanroom garbs, or equipment they bring into the cleanroom, as well as the dust they generate while working in the cleanroom. While continuous air recirculation within the cleanroom takes part of this dust back out, a portion of it does deposit onto open surfaces within the room, including those of the TPC, and should therefore be carefully quantified and controlled. 
Two technical probes were developed to estimate dust deposition on the TPC detector components surfaces, and a dust fallout model was developed for the same purpose. 

\subsubsection{Technical Probes: Witness Coupons}
\label{s_wc}

Witness coupons are small samples ideally made of the same materials as the detector component that is being assembled. Since dust deposits are expected to accumulate at the same rate on the detector material, the coupons are then used to infer the dust deposition on the component. For the LZ detector, the coupons are mostly PTFE and glass, since these make up a vast majority of the most background-sensitive surfaces of the TPC. 

\begin{figure}[!htb]
\includegraphics[width=\linewidth]{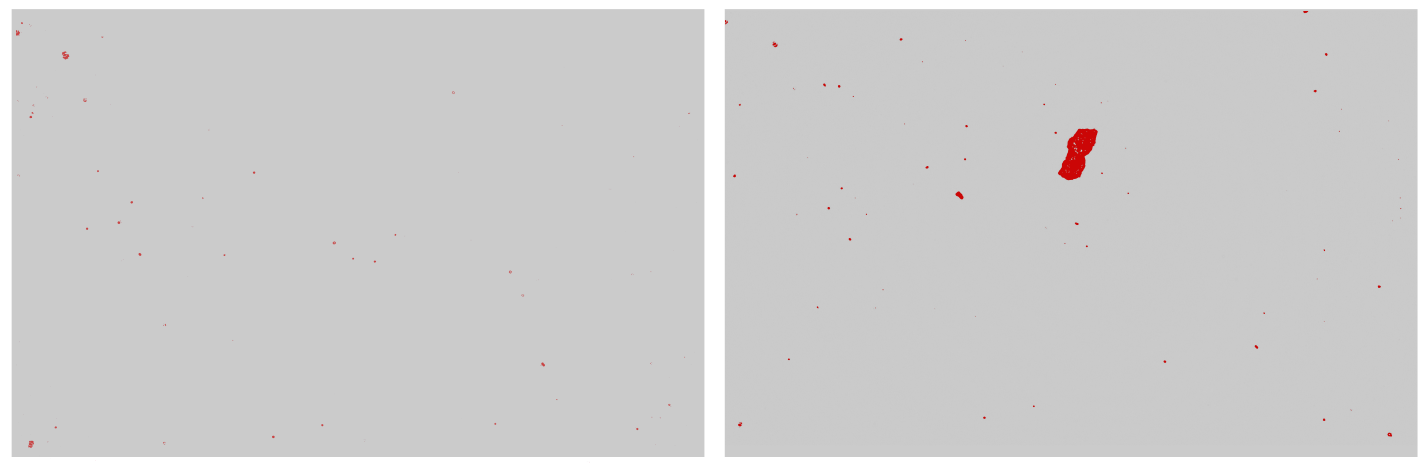}
\caption{Optical images of clean (Left) and dirty (Right) witness coupons, post optical processing. The feature common to them are dust from the optics, and serves as a calibration check.}
\label{fig_sizedistribution}
\end{figure}

Although coupon surfaces should preferably be flat, PMT cable coupons with curved surfaces were also deployed in an effort to thoroughly probe dust deposition. 
All coupons are carefully cleaned with isopropyl alcohol (IPA) soaked non-shedding wipes and then deployed in pairs, as close as possible to the detector, to collect dust under similar conditions as the detector surfaces. They are typically harvested after a couple of weeks, which, based on the cleanroom level and the sensitivity of the assay technique, is enough time to collect the required amount of dust to make an assay possible.  
The coupons are then assayed via optical or fluorescence microscopy depending on material type in order to enable a contrast between the dust particulates and the rest of the coupon surface. For the PTFE coupons, since PTFE does not fluoresce but many dust particulates do, ultraviolet light is shone on the exposed side of the samples. Glass coupons, on the other hand, may be imaged under visible light.
Both fluorescent or optical images are then processed via software written in for ImageJ~\cite{DBLP:journals/corr/RuedenSHDWE17} for further contrast to reveal smaller dust particulates down to \SI{0.5}{\micro\meter}, and accurately determine the size distribution of these particulates and their contribution to the dust density accumulated on the coupons' surfaces. Some dust from the optics, which shows up in the same place on each image (of a clean or dirty coupon as seen on \figureWord~\ref{fig_sizedistribution}),  needs to be removed from the analysis and provides some minor calibration which ensures a consistent imageJ threshold is set for all images.

\begin{figure}[!htb]
\minipage{0.5\textwidth}
  \includegraphics[width=\linewidth]{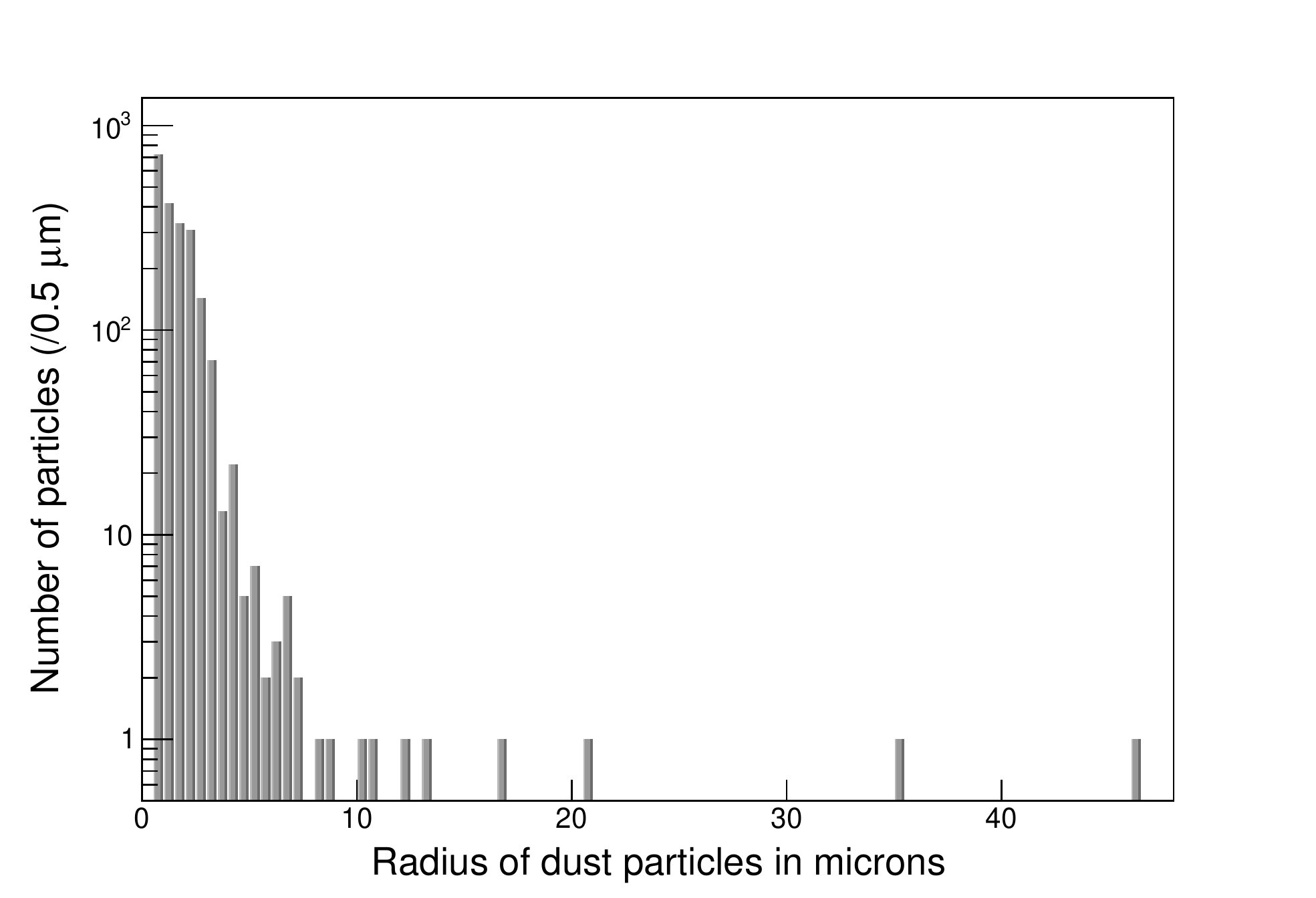}
\endminipage\hfill
\minipage{0.5\textwidth}%
  \includegraphics[width=\linewidth]{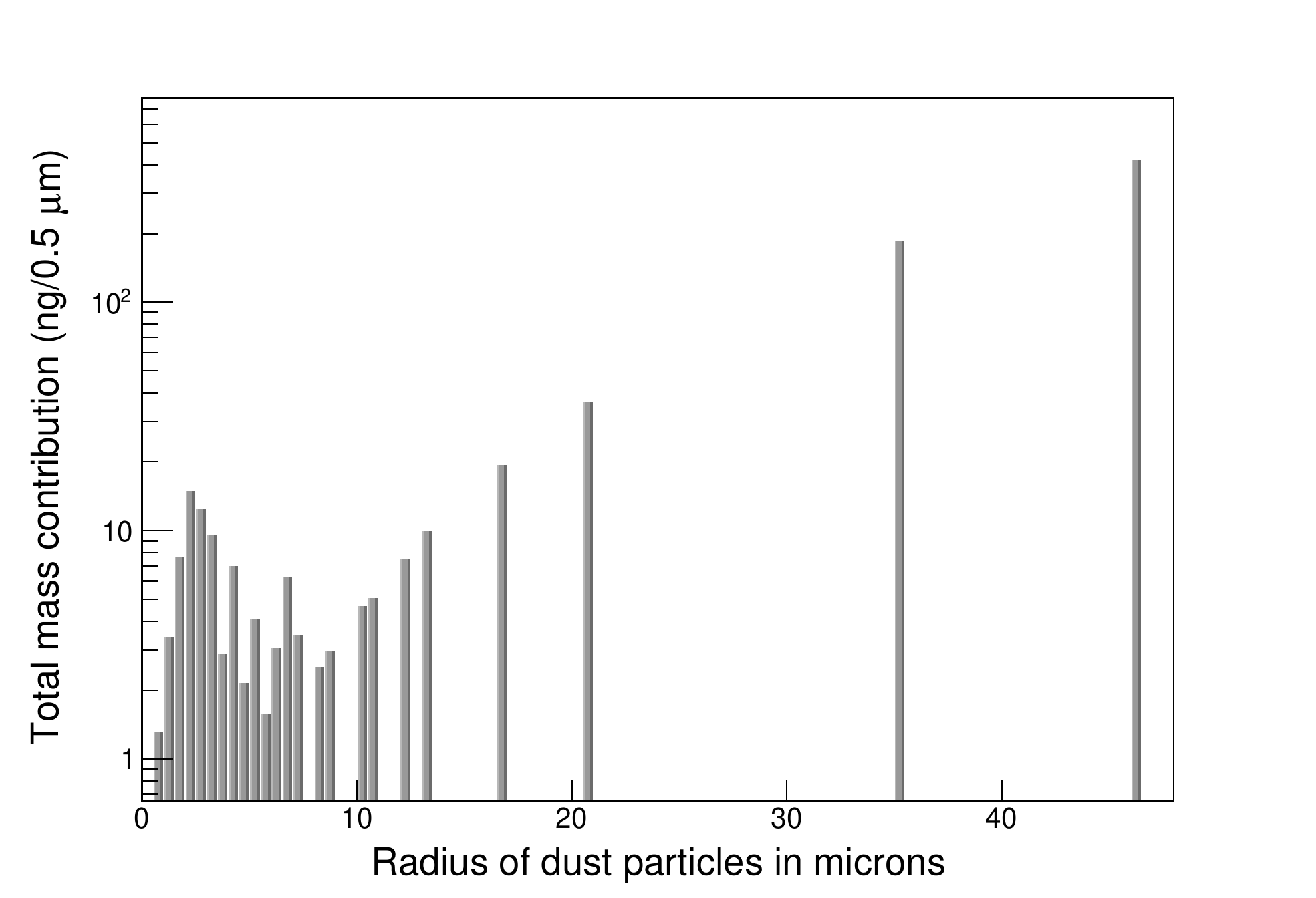} 
\endminipage
\caption{Left: Dust particulate size distribution from fluorescent image analysis of a witness plate. Particulates of size \SI{>50}{\micro\m} are rarely recorded. Right: Dust particulate mass distribution of the same witness coupon showing contribution of particulate size to mass. Although most particulates are small, most of the mass on the coupon is from a small number of larger particulates. }
\label{fig_size_mass_distribution}
\end{figure}

The number of particulates on the coupons decreases with increasing particulate size~\cite{borson2005iest}, such that large particles (\SI{\geq 50}{\micro\meter}) are relatively uncommon but they may dominate the total mass, as seen in \figureWord~\ref{fig_size_mass_distribution}.
Once the dust particulates size distribution is determined, the dust density accumulated on the coupon surface (in \si{\nano\g\per\cm\squared}) is calculated by dividing the accumulated mass (assuming particulates are spherical in shape with density of 1g/cm${^3}$ by the surface area of the coupons.

To obtain the dust fallout rate, witness coupons are assayed both before and after their exposure.  The pre-exposure dust concentration is subtracted from the post-exposure dust concentration before dividing by the coupon exposure time to determine the dust fallout rate.  Occasionally, mishandling of coupons led to their results being discarded.

\subsubsection{Technical Probes: Tape lifts}
\label{s_tl}
Since rough, fluorescing materials cannot be imaged accurately, and because witness coupons never have exactly the history of the material itself, another, more direct, probe of dust deposition was conducted. These so-called tape lifts are simple pieces of acetate or carbon tapes that are stuck to an assembly piece, and are lifted to remove dust for assay. Each tape lift can only be utilized on one spot and only lifted once. To get better statistics more tape lifts on different spots had to be taken. The choice of tape material depends on the surface texture/roughness of the component: acetate tape works better on smooth surfaces, like PTFE, while, for rougher surfaces like titanium, carbon tapes were found to perform better. The tape lifts are assayed using the same fluorescence microscopy technique utilized for the witness coupons. Unlike the coupons that collect dust throughout the assembly process, the tape lift assessment is conducted at the end of a main assembly. In addition to providing a more direct probe, tape lifts also mitigate against improper use or mishandling of coupons. For example, while the witness coupons are supposed to be wiped each time an assembly, sub-assembly or parts are being wiped, this step can be overlooked, in which case the coupons will collect more dust than the actual assembly. The tape lifts instead give confirmation of the dust level on a final assembly. However, tape lifts cannot be taken on particularly sensitive parts of the TPC, and therefore, they do not negate the need for coupons, but rather complement them. Both tape lifts and coupons are necessary for a full history of the dust deposition on every component during the assembly process.  
In addition, having two probes for dust deposition provided additional opportunities for the calibration and validation of the dust deposition models used by the LZ collaboration. 

\subsubsection{Dust Deposition Modeling}
As discussed above, the dust density inside the assembly cleanroom depends on the influx and outflux of dust particulates in the room. Its value per unit particulate size $D$ could be described by Eq.(\ref{E:fallout_int}) originally developed by the SNO  collaboration~\cite{Boger:1999bb,stokstad:1991}:
\begin{equation}
n(D) = R_{in}(D)\left[R_{AE}+v(D)\frac{S}{V}\right]^{-1}
\label{E:fallout_int}
\end{equation}
where ${R_{AE}}$ is the cleanroom air exchange rate defined as the volumetric air circulation rate in the cleanroom divided by the volume  $V$ of the cleanroom, $v(D)$ is the Stokes settling velocity which is dependent on the particulate  size~\cite{barber1999control}, ${R_{in}(D)}$ is the volume-normalized dust-particulate carry-in rate mostly from personnel, and  $S$ is the area of the projection of the cleanroom volume onto the horizontal plane.

The fallout rate per particulate size in mass per unit area and time could then be deduced using the continuity equation as shown in Eq.(\ref{E:fallout})
\begin{equation}
m(D) = \frac{\pi}{6}n(D)\rho D^3v(D)
\label{E:fallout}
\end{equation}
where ${\frac{\pi}{6}D^3}$ is the particulate volume assuming they are spherical in shape and ${\rho}$ is their mass density (${\equiv}$ 1g/cm${^3}$).   
The total fallout rate is then obtained by integrating over all particulate sizes (${m =\int m(D)~dD }$); which for LZ ranges between \SIrange{0.5}{100}{\micro\m} as determined by the witness coupons assay results. 

However, the carry-in rate per particulate size ${R_{in}(D)}$ was not measured by LZ. Only the air class within the cleanroom was measured using particulate counters. Consequently, a model of air particulate-size distribution from dust carry-in  was assumed based on ISO-14644-1 and the measured air class (which generally averages less than 10 particles \SI{\geq 0.5}{\micro\meter} per cubic foot due to our developed  cleanliness protocols) was used to constrain the model. Also, a scaling factor ${\eta}$ was added to the fallout rate in Eq.(\ref{E:fallout}) and was then calibrated using results of fallout rate measurements from witness coupon assays. Its average value was estimated to be $\eta=26.86\pm 5.09$. This modified fallout rate, written in  Eq.(\ref{E:modifiedfallout}) will be referred to as the modified SNO model. 

\begin{equation}
m =\int  \frac{\pi}{6}n(D)\rho D^3v(D)\eta ~~dD   
\label{E:modifiedfallout}
\end{equation}

The new modified SNO model was then compared with independent tape lift measurements to ensure agreement within uncertainty between the model and the measurements. \tableWord~\ref{tab:comparison_table} shows the agreement, hence validating the modified model. The $\eta$ factor may be influenced by several factors but the clearest observed correlation for an increase in particle fallout rate was with a decrease in the relative humidity in the cleanroom. It was observed that at the lowest relative humidity level of 25${\%}$, the fallout rate was the highest (${1.18 ^{+0.25}_{-0.19}}$ ng/cm$^{2}$/hour) while in normal relative humidity levels (35-45 ${\%}$), the fallout rate was as low as ${0.03 ^{+0.02}_{-0.01}}$ ng/cm$^{2}$/hour. This is  expected as, when the cleanroom air is drier, some surfaces like that of PTFE accumulate more static charges and attract more dust particulates.

\begin{table}[ht]
    \caption{Comparison of calibrated modified SNO model with tape lift assay results}
    \centering
    \resizebox{\textwidth}{!}{\begin{tabular}{|P{3cm}|P{3cm}|P{5cm}|P{6cm}|P{5cm}|}
    \hline
      \textbf{Tape lift date} & \textbf{Exposure time $T$ (days)} & \textbf{Air class (cts/m$^{3}$ \SI{\geq 0.5}{\micro\meter})} & \textbf{Calculated dust density $m*T$ from model (\si{\nano\gram\per\cm\squared})} &\textbf{Measured dust density from tape lifts (\si{\nano\gram\per\cm\squared})} \\
       \hline
       \hline
     09/27/2018 & 1  & 6074${\pm}$600 & 285 ${\pm}$53 & ${220^{+50} _{-20}}$ \\
       11/29/2018 & 64 & 283${\pm}$35  & 829 ${\pm}$ 154 & 885 ${\pm}$ 197  \\
       \hline
    \end{tabular}}
    \label{tab:comparison_table}
\end{table}

It is important to note that, while \tableWord~\ref{tab:comparison_table} shows the validation of the modified SNO model, dust densities recorded there are indicative of accumulated dust on some detector components at a particular period in time. These densities therefore cannot be used to infer the final dust density on the TPC surface since accumulated dust reduced to ${\sim}$ \SI{20}{\percent} of its value every time cleaning was performed on an assembly. Indeed, tape lift results taken before and after cleaning have verified that cleaning protocols developed by LZ ($e.g.$ wiping surfaces under UV light with IPA soaked non-shedding wipes) consistently reduce dust to the stated level, and serve as stringent mitigation procedures against surface dust contamination. This is discussed in more detail in \sectionWord~\ref{mitigation_Rn_dust}.

\subsubsection{Dust Fallout Calculation for the TPC}
\label{sc_calculations}
The dust deposition rate in mass per unit area per unit time on the various TPC components was estimated using the modified SNO model.  
Once individual estimates are obtained for different components for each daily work shift, the overall deposited dust density for the ${i^{th}}$ surface ($M_{i}$) is obtained by taking into account the exposure time ${T}$ of that given surface and the mass of dust deposited per unit area and time ($m$), and is then given by Eq.(\ref{e:mass}):

\begin{equation}
    M_i = Tm
    \label{e:mass}
\end{equation}

Since different areas of the detector are exposed for different times during the assembly process, one must also take into account the exposed area versus total detector area. Therefore, the overall dust density ($M_O$) in \si{\nano\g\per\cm\squared} for the entire TPC is: 

\begin{equation}
    M_o = \frac{\sum A_{exposed}^{i} M_i}{\sum A_{total}^{i}}
    \label{e:tpc}
\end{equation}

The surface areas, ${A_{exposed}^{i}}$ and ${A_{total}^{i}}$ are obtained from a sophisticated information repository developed by LZ to smoothly manage and track detector parts, their surface areas, and their exposure to ambient cleanroom air during TPC construction. In addition, every instance of detector surface cleaning is recorded and taken into account, as described above, which allows for an accurate estimation of the dust contamination on TPC surfaces. 
The final estimation amounts to a total of \SI{0.64 \pm 0.05}{\g} of dust accumulated on the entire TPC for a dust density of \SI{214 \pm 22}{\nano\g\per\cm\squared}, below LZ requirement of \SI{500}{\nano\g\per\cm\squared}. 

\subsection{Rn Progeny Plate-out}
\label{s_radon}

In addition to dust, another source of background in the experiment comes from the environmental radon-laden air itself, with radon daughters plating out onto the surfaces of materials \cite{Leung:2005gtk} during assembly. To limit plate-out, most assemblies are done in a radon-reduced cleanroom (RCR) at the SAL.  The radon-reduced system used in the RCR  was built by ATEKO, and is a continuous filtration system constantly circulating air through a cold carbon column to filter out the radon at an overall reduction factor of 2200 leading to an ambient radon level averaging \SI{<0.5}{\becquerel\per\metre\cubed} as measured by LZ. The RCR high recirculation rate of 8500 cubic feet per minute is enough to mostly sweep out radon daughters (particularly \potoe{}) before they plate-out onto detector surfaces. Absolute plate-out prevention is however not possible, and the remaining \potoe{} that plates out is problematic due to its long-lived \pbtoz{} daughter which will decay over time in the detector. Therefore, the plate-out rates on assemblies must be calculated. 

Plate-out rates onto materials are often estimated using the Jacobi model \cite{knutson1988modeling,jacobi1972activity}
, which, similar to the original SNO model, describes particle deposition from a balance of influx and outflux of particles in the room assuming that the room contents are well-mixed. This Jacobi model can also be modified to reflect a cleanroom setting, as is the case here \cite{Morrison:2017xul}. 
In its original version, the Jacobi model assumes that all surfaces within a given enclosure or a room are equivalent, with radon daughters ending up on all of them in equal concentrations. Under that assumption, the area-normalized plate-out rate (surface activity) depends on the conditions of the enclosure (air circulation rate, Rn concentration, volume, and surface area) within which the material surfaces are exposed. 
The \pbtoz\ plate-out rate expressed as the decay rate per unit area and unit time, $R_p$(Bq/m${^2}$/s), is thus described by Eq.(\ref{jac_eq}):
\begin{equation}
    R_p=C_{Rn}\lambda_{Pb_{210}}\frac{\Lambda_d}{\Lambda_d +\Lambda_v}\frac{V}{A}
    \label{jac_eq}
\end{equation}
where ${C_{Rn}}$ is the Rn concentration in the cleanroom (obtained from Durridge Rad7 radon monitors with the monitors' blank rates subtracted off), ${\lambda_{Pb_{210}}}$ is the \pbtoz{} decay rate, $V$ is the volume of the cleanroom, $A$ is the surface area within the cleanroom, ${{\Lambda_d =v\frac{A}{V}}}$ is the deposition rate that depends on the diffusion velocity $v$ of radon daughters measured to be between 5-15 m/h \cite{knutson1988modeling}, ${{\Lambda_v = \frac{R}{V}}}$ is the air ventilation rate obtained by dividing the recirculation rate $R$ of the cleanroom HEPA filters by its volume. The ratio ${\frac{\Lambda_d}{\Lambda_d +\Lambda_v}}$ corresponds to the probability that a Rn daughter will plate-out before being carried out by the ventilation; which, for the RCR, was around 0.17.

It is worth noting that the Jacobi model in Eq.(\ref{jac_eq}) is a direct analog of the SNO dust deposition model, as seen by expanding ${n(D)}$ from Eq.(\ref{E:fallout_int}) into Eq.(\ref{E:fallout}) and making a number of associations of variables.  While the SNO model describes deposition of dust onto horizontal surfaces, the Jacobi model describes deposition of Rn daughters on all surfaces, both controlled by a characteristic velocity in a similar filtration environment.  In particular we can identify the particle deposition rate per unit area per unit time in Eq.(\ref{jac_eq}) as
${R_p/\lambda_{Pb_{210}}}$,  whereas in Eq.(\ref{E:fallout}) it is expressed as ${ 6m/\pi \rho D^3}$.  The volume-normalized influx ${R_{in}}$  of dust particulates in Eq.(\ref{E:fallout_int})  is analogous to the volume-normalized rate of production of Rn daughters in Eq.(\ref{jac_eq}), that is,  the Rn activity per unit volume, ${C_{Rn}}$.  The volume-normalized filtration rate ${ R_{AE}}$ from Eq.(\ref{E:fallout_int}) is directly associated with ${\lambda_v}$ from Eq.(\ref{jac_eq}). The fallout area $S$ from Eq.(\ref{E:fallout_int}) is associated with the available Rn daughter plate-out area $A$ from Eq.(\ref{jac_eq}). Finally, the stokes velocity ${v(D)}$ from Eq.(\ref{E:fallout_int}) is physically similar and mathematically analogous to the Rn daughter diffusion velocity $v$ in Eq.(\ref{jac_eq}).  

However, the assumption in the Jacobi model that plate-out does not depend on material type has been shown to be incorrect, particularly for materials at the bottom of the triboelectric series, such as PTFE \cite{zou:2019tde}, which could 
have a plate-out rate $M$ times higher than for neutral metallic materials. An experiment conducted by the SDSM\&T measured the $M$ factor to be between 50 and 100 \cite{Morrison:2017xul}.
So for LZ, plate-out rate estimations using the Jacobi model are thus corrected with a multiplicative factor $M$ which has a value of ${M=1}$ for plate-out onto metals, and its highest value ${M=100}$ for plate-out rate on PTFE material surfaces which are naturally charged.

Much of the inner TPC is made from PTFE such that it is essential to mitigate against this high Rn plate-out. This is achieved by neutralizing the PTFE by using air deionizing fan units. 
These units are ISO 10 cleanroom compatible Simco 4008630 - 3 Fan Cleanroom Overhead Air Ionizer units which produce continuous ionized air through corona discharge, thus neutralizing the otherwise negatively charged PTFE material.
Usage of the fans was taken into account in the plate-out estimations by reducing the correction factor $M$ to the value of 1 for PTFE surfaces, after the fans' deployment. Indeed, electrostatic field measurements taken at regular time intervals between metallic surfaces and PTFE surfaces placed under air deionizing fan units showed a consistent reading of \SI{0}{\kilo\volt\per cm} within the uncertainty of the measurement device, while similar measurement for PTFE material not placed under these fans read ${\sim}$ \SI{0.6}{\kilo\volt\per cm}, thus demonstrating the successful neutralization of PTFE under these deionizer units. 

The weighted plate-out rate $R_w$ on a given TPC assembly for a single work shift time period (exposure time $T$) is thus given by Eq.(\ref{e:plate-out}) where $M$ is the plate-out rate multiplicative factor described above, $A$ is the surface area of the individual parts making up the assembly, and ${R_p}$ is the Jacobi plate-out rate per unit area and time obtained from Eq.(\ref{jac_eq}):

\begin{equation}
    R_w = \frac{\sum A_{exposed}^i (M R_p)T}{\sum A_{total}^i}
    \label{e:plate-out}
\end{equation}

The overall plate-out accumulated for all the work shift time periods for that assembly is obtained by combining all the weighted rates as was done previously for the dust estimation. 

Plate-out rate estimations are drastically different during and outside working hours (overnight and during weekends). As expected, the plate-out rates outside working hours were found to be negligible, of the order of less than \SI{1}{\percent}.
Overall, the average plate-out for the inner TPC PTFE surfaces in contact with the LXe is $R_{avg}=$ \SI{158\pm 13}{\micro \becquerel\per\square\meter}, which is below the LZ requirement for the TPC of \SI{500}{\micro\becquerel\per\square\meter}. After construction at the SAL, the TPC was sealed in the ICV before being transported underground where it was kept under N2 purge making its underground surface contamination negligible.

\subsection{Cleanliness Protocols to Mitigate against Dust and Rn-progeny Contamination}
\label{mitigation_Rn_dust}
Following manufacture, most detector components were sent to be cleaned at AstroPak Inc, a certified professional precision cleaning company. 
After cleaning, detector components were sealed in multiple Rn barrier bags under N$_{2}$ purge. The redundancy in the bags also provided layers to shed, thereby helping to reduce carry-in dust when components were brought into the cleanroom after transportation.

Both aluminized mylar and nylon bags have been shown to be very efficient against Rn penetration, with reduction factors of ${2500 \pm 1042}$ and ${130 \pm 3}$ respectively \cite{Meng:2019ker}. Once properly bagged, the detector components were shipped to the SAL facility where they were assembled in the RCR to mitigate, as previously mentioned, against surface contamination during assembly. The cleanliness protocols implemented within the cleanroom allowed its air class to always be measured at a significantly lower level (on average less than 10 particles with a size \SI{\geq 0.5}{\micro\meter} per cubic foot) than its class 1000 classification.

Upon arrival at the SAL, the outer layer of the shipping bags was removed before sealed components were brought inside the RCR. Inside the cleanroom, each component was un-bagged under deionizing fans to remove static charges on polymer-like materials such as PTFE. They were then inspected under UV light to check for dust particulates, which were cleaned off using  Abgenics Essence Gold  non-shedding mono-filament wipes saturated with \SI{99}{\percent} pure IPA as a basic cleaning method. Other cleaning methods involved IPA spray or bath followed by blow drying with filtered N$_{2}$, ultra-sonic or high-pressure cleaning using deionized water or IPA, and CO$_{2}$ blasting depending on the material type and the amount and type of particulates to be removed. In general, small hardware like screws and bolts were ultrasonically cleaned in deionized water and IPA bath. Smooth surface components, like PTFE parts,  were wiped down using IPA saturated non-shedding wipes, but this cleaning method could not be used on rough surfaces, like titanium, because of shedding residues. Those were instead cleaned with high pressure deionized water. CO$_{2}$ blasting was also used to clean the titanium surfaces to remove chemical residues from Astropak Inc cleaning. As for the PMT cables, wiping them with IPA saturated wipes was inefficient at removing dust. Instead, they were first sprayed with IPA within the droplets of which the dust accumulated. The cable was then gently blow-dried with N$_{2}$ thus removing the IPA droplets and the dust contained within.

\begin{figure}[!htb]
\minipage{0.32\textwidth}
  \includegraphics[width=\linewidth]{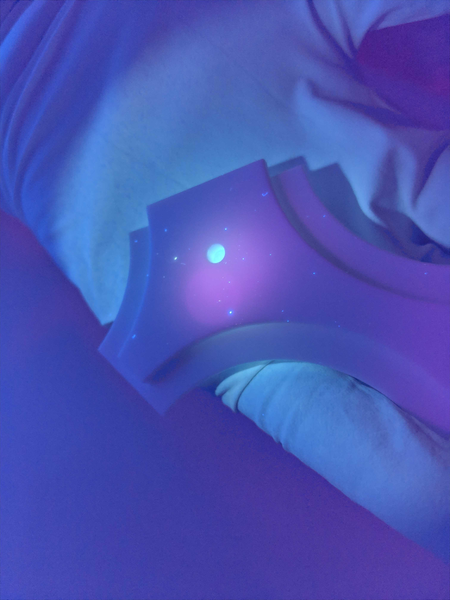}
\endminipage\hfill
\minipage{0.32\textwidth}
  \includegraphics[width=\linewidth]{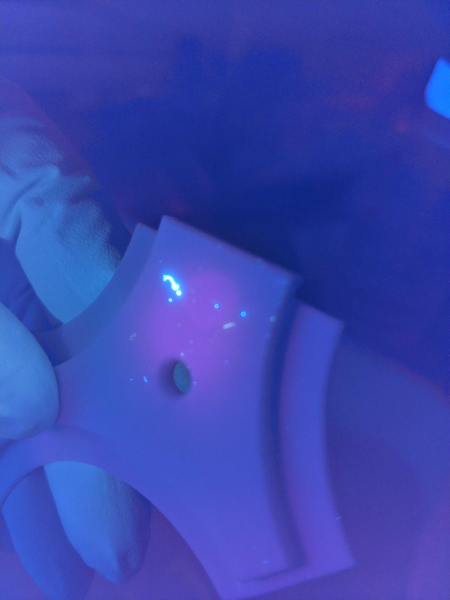}
\endminipage\hfill
\minipage{0.32\textwidth}%
  \includegraphics[width=\linewidth]{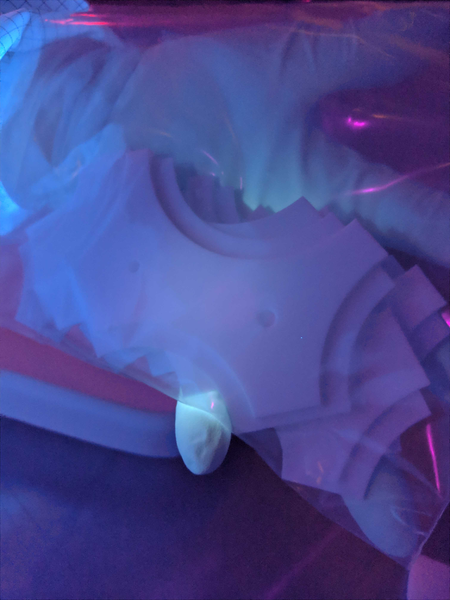} 
\endminipage
\caption{Pictures of some PTFE components (used on the PMT arrays) under UV light before (left and center) and after (right) cleaning with IPA saturated non-shedding mono-filament wipes. Fluorescent specks are dust particulates which are removed after cleaning the pieces in an IPA bath as seen in the right picture.}\label{fig:cleaning}
\end{figure}

All these methods have been visually investigated for efficiency before usage, but only the IPA cleaning was quantified, since it was the most used. \figureWord~\ref{fig:cleaning} shows images before and after cleaning of PTFE pieces under UV light, and visually and qualitatively demonstrates the effectiveness of the cleaning.
The quantitative estimates of the IPA cleaning protocol were done using tape lifts, as described in \sectionWord~\ref{s_tl}. The tape lifts were taken on sample coupons before and after cleaning for the various cleaning methods. The average dust removal efficiency was found to be about \SI{80}{\percent}. For instance, tape lift on sample PTFE before cleaning was of  ${700^{+700}_{-100}}$ ng/cm$^{2}$ and after cleaning was ${100^{+20}_{-9}}$ ng/cm$^{2}$ yielding ${\sim}$ \SI{90}{\percent} efficiency. 

During assembly, several deionizing fans were used to surround the assembly area, and ensured complete neutralization of materials, thereby reducing plate-out as discussed in \sectionWord~\ref{s_radon}. Usage of these deionizing fans during assembly was particularly important as the assembly process involved extensive manipulation and rubbing which would have otherwise cause charging, increasing the dust and radon-daughter plate-out, particularly on PTFE surfaces.

In addition to the deionizing fans, UV light inspections were conducted at regular time intervals to evaluate dust accumulation on assembly surfaces, and their cleaning was done every day at the beginning of the work shift. When judged necessary, or at the end of a sub-assembly, remedial cleaning of the surfaces was conducted (IPA wiping, IPA bath, vacuuming) with the appropriate cleaning methods, thus removing most of the accumulated dust.  Note that, as mentioned previously, remedial cleanings did not remove all the dust, which is why the witness coupons and tape lifts mentioned in \sectionWord~\ref{s_wc} were still invaluable in assessing the actual amount of dust collected on the assemblies.

Finally, cleanroom garbs worn by personnel working on the assembly were changed after every work shift to reduce carry-in dust, and wiped off with a lint roller multiple times during work to remove particulates that could deposit onto detector surfaces. 

At the end of all daily shifts, smaller detector components were sealed in double nylon bags which prevented plate-out since the components were then no longer in contact with the cleanroom 
air. Larger components like the PMT arrays, on the other hand, had bespoke airtight enclosures, allowing them to be kept under filtered and ventilated air from an extra HEPA filter unit or under N$_{2}$ purge, allowing further mitigation against Rn plate-out. 

All of the described procedures have ensured that exposure to Rn and dust during the assembly process was minimal, surpassing LZ requirements, as stated, for both dust and plate-out of radon progeny.


\section{Selected Highlights from the Assay Program}
\label{S:5}

The LZ design sensitivity imposes limits to radioactivity from particular components, principally items such as PMTs which are close to or in contact with the fiducial volume, thus demanding dedicated fixed contamination screening campaigns to meet requirements. The assay program also included dedicated QC and QA elements to ensure radioactivity and cleanliness compliance throughout all manufacturing processes, such as through the construction of the Ti cryostat. Dedicated radon emanation measurement campaigns were performed on complete purification and recirculation sub-systems that may contribute to radon, as well as \emph{in-situ} measurements of the inner cryostat with the TPC sealed within. The assay program also included construction and deployment of a dedicated detector system to determine radioactivity in the GdLS for the LZ OD system. 
In the following sub-sections, we present these dedicated campaigns in order to illustrate through specific examples the deployment of fixed contaminant assays, QA, radon assays, and detector construction and deployment.

\subsection{Hamamatsu R11410 PMT HPGe Assay Program}

LZ employs three different models of PMT. The main active volume will be viewed by 494 \SI{76}{\milli\meter} (3-inch) Hamamatsu R11410 PMTs. Other regions containing xenon are instrumented using a combination of \SI{30}{\milli\meter} Hamamatsu R8520 and \SI{56}{\milli\meter} Hamamatsu R8778 PMTs with the latter having been repurposed having been used in the LUX dark matter experiment~\cite{akerib:2012ys}. Additionally, the LZ OD system is instrumented using \SI{202}{\milli\meter} (8-inch) Hamamatsu R5912 PMTs.

The radioactive background of the R11410 PMTs is of particular importance as these sit adjacent to the main active volume of LZ. In order to ensure that these met the intrinsic radioactivity requirements of 3.0/3.0/30/2.5\,mBq/PMT for \utTe{}/\thtTt{}/\kfz{}/\cosz{}, respectively, a comprehensive screening program was conducted. Initially, a subset of the raw material used for the construction of the tubes was screened across detectors both at SURF and at Boulby, and from these measurements, a bottom-up prediction of the intrinsic radioactivity of the final tubes was calculated. This calculation is discussed in detail in~\cite{Mount:2017qzi}.

\begin{figure}[!htb]
\minipage{0.5\textwidth}
  \includegraphics[width=\linewidth]{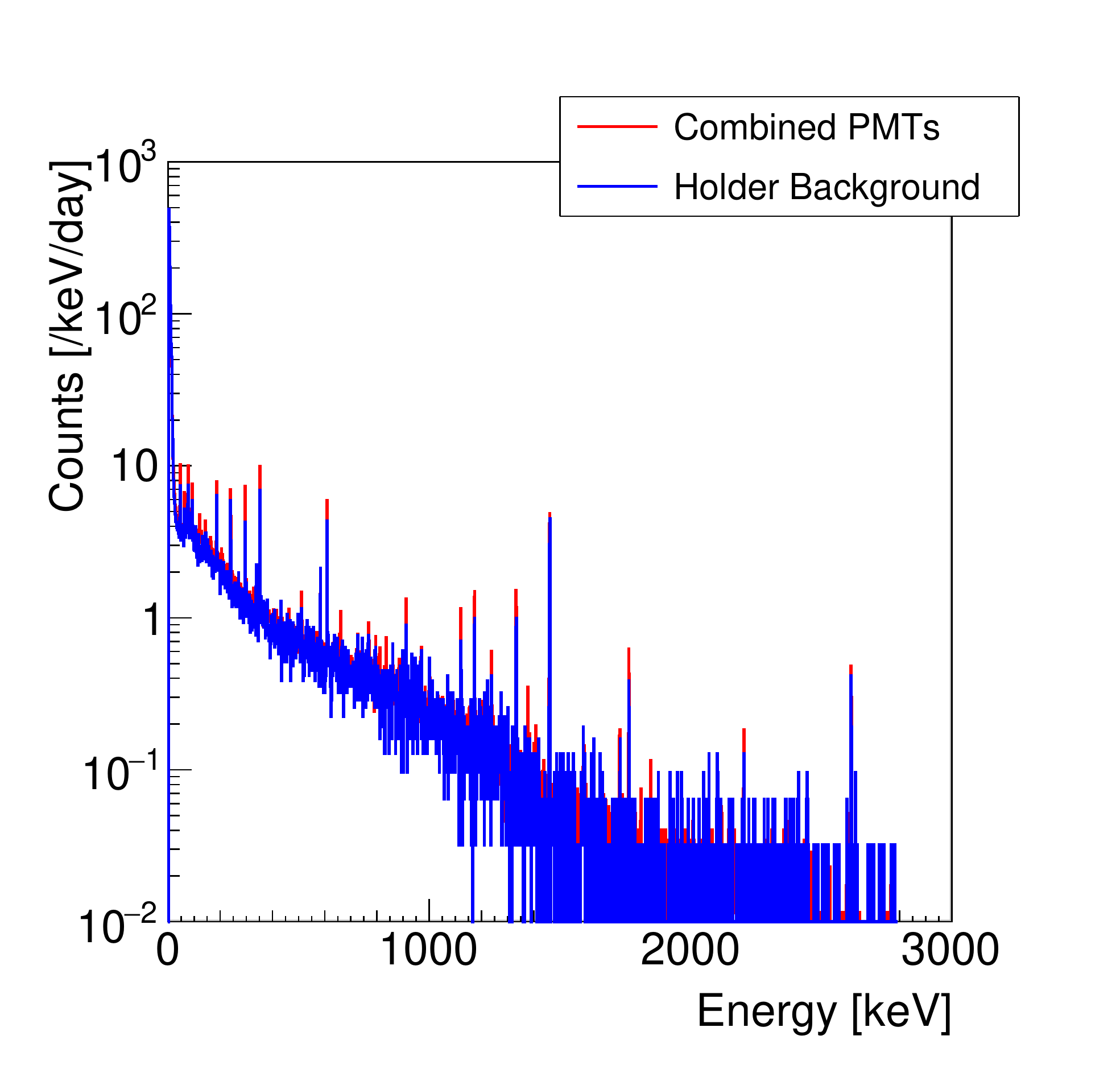}
\endminipage
\minipage{0.5\textwidth}
  \includegraphics[width=\linewidth]{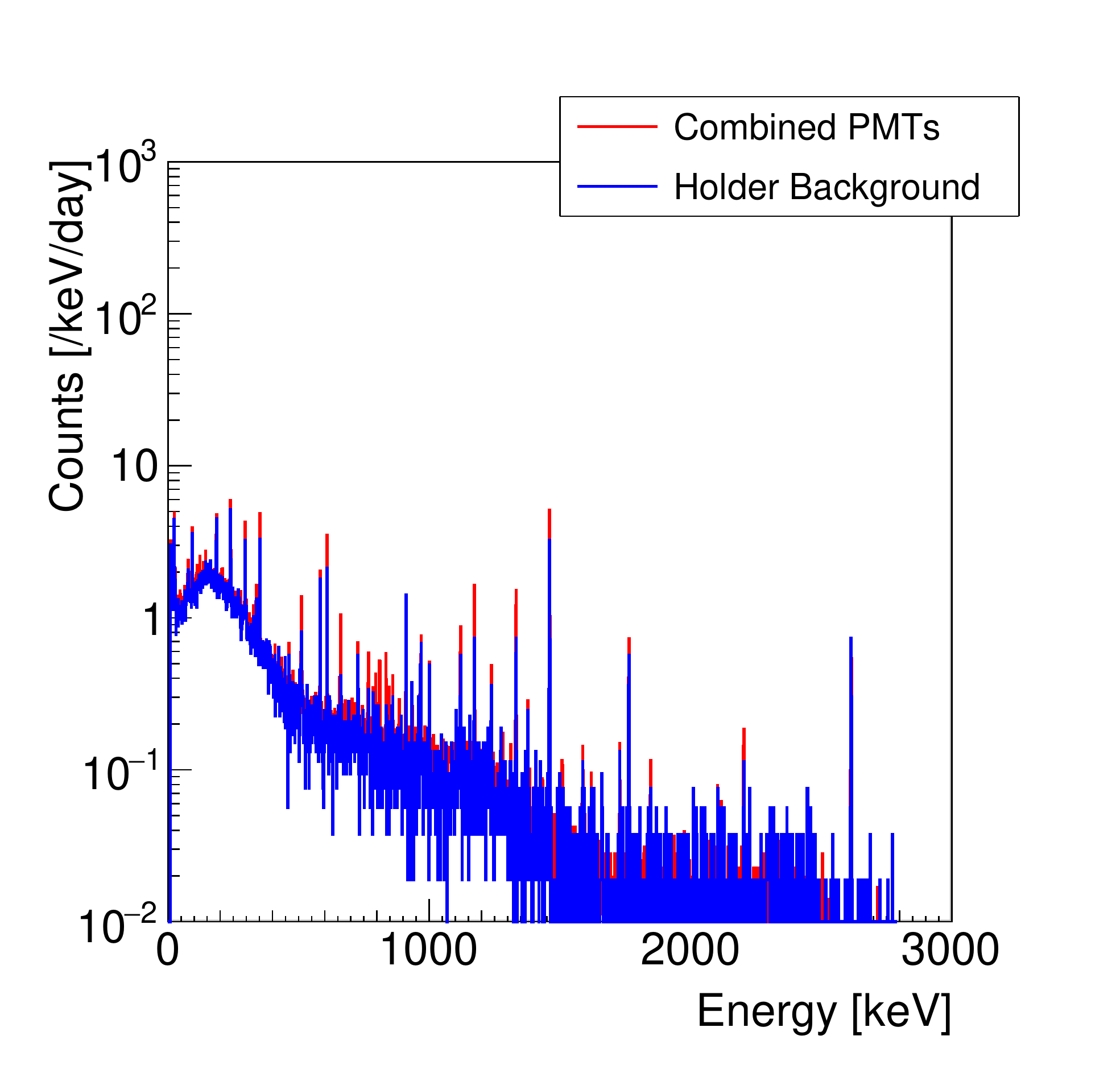}
\endminipage
\caption{Combined spectra for all R11410 PMT runs using Chaloner (left) and Lunehead (right) at the Boulby Underground Germanium Suite. In both cases, the background spectrum includes the holders used to secure PMTs in place in the detector castles. \label{fig:PMTsUK}}
 \end{figure}

Following the raw material screening program, the collaboration was satisfied that PMTs of the required radiopurity could be produced. However, even with knowledge of the radiopurity of the raw materials, it was important to repeat the screening process for the final tubes. This required substantial gamma-ray spectroscopy detector time both at SURF and at Boulby. The program began with an initial batch of tubes being screened at Boulby between August 2016 and February 2017. Over 32 weeks, 102 PMTs were screened and a substantial amount of background data was acquired. \figureWord~\ref{fig:PMTsUK} shows comparisons between a combination of all PMT runs on the detectors used in this campaign. For protection and cleanliness purposes, the R11410 PMTs assayed at Boulby were screened with PTFE tape wrapped around their body and with protective foam on the pins at the base of the PMT. Subsequently, the tape and foam were screened and their contribution (along with the contribution from the holder and detector setup backgrounds) subtracted to give final contamination levels for the tubes presented here. Screening of tubes at SURF was performed following performance testing at Brown University for which the foam and PTFE tape needed to be removed. No additional correction was required for this data. The full PMT screening program assayed 229 of the 494 R11410 tubes used in the LZ detector.

The tests of individual components showed that the radioactive contaminants were not uniformly distributed. Not only is this the case, but the relative levels of \utTe{}, \thtTt{} and \kfz{} were different in each component. Ordinarily, when calculating a geometric efficiency for an assayed material or component (using GEANT4~\cite{agostinelli:2002hh} in this case), gamma-rays are fired uniformly from the component being studied. In order to allow for non-uniform distribution of radioactivity when determining specific activities for the assayed PMTs, geometric efficiency curves had to be calculated for individual isotopes. The distribution of simulated gamma-rays was determined using the expected contamination distribution. This is detailed in \tableWord~\ref{table:UTh} where the three largest components (both by mass and, in the case of the ceramic stem, by radioactive content) are used to represent the whole PMT. In the case of all other isotopes (detailed in Tables~\ref{table:combined} and \ref{tab:unusual}), the contamination was assumed to be distributed uniformly.

The expected distributions of contamination detailed in \tableWord~\ref{table:UTh} do not take into account one important unknown factor: the distribution of \kfz{} as the process of forming a PMT photocathode requires the evaporation of potassium onto the inside of the quartz window face of the PMT. This adds a substantial systematic uncertainty to the final measured values of \kfz{} in this study. As an approximation, the distribution of \kfz{} in the PMTs was modified in order to give a reasonable systematic error on each setup. The systematic error is set assuming a distribution of 0.1, 0.45 and 0.45 for the ceramic, the Kovar and the quartz face, respectively. In the case of the Chaloner detector, where PMTs are placed so that the pins are the closest part to the front face of the detector and the quartz window the furthest, a systematic error of \SI{125}{\percent} is calculated. In the case of SOLO, where PMTs uniformly surround the detector, there is no substantial systematic error as the geometric efficiency has a very weak dependence on the distribution of \kfz{} in the PMTs. The systematic errors for \kfz{} are captured in \tableWord~\ref{table:UTh}. In all other isotopes, it was assumed that the distribution of radiocontaminants follows expectation so only statistical errors are presented. 

\setlength{\extrarowheight}{3pt}
\begin{table}[!htb]
\caption{Fractional contamination levels for \utTe{}, \thtTt{} and \kfz{} from individual R11410 component assay. The PMT is split into major components and the expected distribution of  activity is calculated using individual component measurements. This distribution acts as weighting factor when determining the contribution of each individual component to the overall efficiency. In the \utTe{} and \thtTt{}, secular equilibrium is assumed.}
\centering
\begin{tabular}{c c c c c}
\hline\hline 
Isotope & Ceramic & Kovar & Quartz\\ 
 & Stem & Bulb & Face\\ [0.5ex]
\hline
\utTe\ & 0.79 & 0.17 & 0.04 \\
\thtTt\ & 0.45 & 0.45 & 0.10\\
\kfz\ & 0.71 & 0.23 & 0.06\\[1ex]
\hline
\end{tabular}
\label{table:UTh}
\end{table}

Overall average values are calculated for assayed tubes and these are presented in \tableWord~\ref{table:combined}. Even allowing for the systematic error in the measurement, it is clear that there is substantial variation in the levels of \kfz{} found in the tubes. This is likely due to variability in the manufacturing process as uniformity in \kfz{} values is seen across batches of PMTs produced at similar times. Additionally, some significant variation in \cosz{} can be seen in the various batches of tubes.
The combined values were calculated using two methods. The first ignores data where only upper limits are obtained and relies solely on measured values whereas the second incorporates measured \SI{90}{\percent} CL upper limits as measured values. For \thtTt{} and \utTF{}, the incorporation of upper limits substantially reduces the combined value.

\begin{table}[!htb]
\caption{Combined activities for all runs in the LZ PMT screening program. In total, 233 PMTs were screened across six detectors in batches of varying sizes. For each detector, all acquired spectra are summed and the resultant peaks fitted for use in the calculation. For the combined values, detection errors are combined in quadrature. Two combined values are presented. One $(MO)$ uses  measured values only, ignoring upper limits and the other $(UL)$ incorporates upper limits as measured values. The component average comes from~\cite{Mount:2017qzi}. We include for comparison the results from the equivalent assay campaign of an earlier model of Hamamatsu R11410 PMT by \textsc{Xenon1T}, taken from~\cite{Aprile:2015lha}. \label{table:combined}}
\begin{adjustbox}{width=\textwidth,center}
    \tabcolsep=4pt
    \centering
    \begin{tabular}{ lcc|cccccccc| }
    \hline
    \multicolumn{1}{|c}{\multirow{2}{*}{\textbf{Detector}}} & 
    \textbf{\# PMTs}  & 
    \textbf{Total Live} &  
    \multirow{2}{*}{\textbf{\utTeE}}  & 
    \multirow{2}{*}{\textbf{\utTeL}}   &  
    \multirow{2}{*}{\textbf{\thtTtE}}  &  
    \multirow{2}{*}{\textbf{\thtTtL}} & 
    \multirow{2}{*}{\textbf{\kfz}} & 
    \multirow{2}{*}{\textbf{\cosz}} & 
    \multirow{2}{*}{\textbf{\utTF}} & 
    \multirow{2}{*}{\textbf{\pbtoz}} \\  
    \multicolumn{1}{|c}{}& 
    \rule{0pt}{3ex} \textbf{Assayed}  & 
    \textbf{Time (d)} &  
    \multicolumn{8}{c|}{\textbf{all values in mBq/PMT}} \\
    \hline
    \hline
    \multicolumn{1}{|c}{Chaloner} & 48 & 160 & 
    $4.8\pm1.3$ &   
    $1.5\pm0.3$ &   
    $<0.3$ &        
    $<0.1$ &        
    $5.4^{+9.9}_{-3.1}$ &   
    $0.6\pm0.3$ &   
    $0.7\pm0.4$ &   
    $7.1\pm2.9$ \\  
    \multicolumn{1}{|c}{Lunehead} & 54 & 144 & 
    $<4.1$ &        
    $0.9\pm0.3$ &   
    $<0.2$ &        
    $<0.2$ &        
    $14.0^{+3.4}_{-2.0}$ &  
    $0.7\pm0.1$ &   
    $0.6\pm0.6$ &   
    - \\            
    \multicolumn{1}{|c}{SOLO} & 90 & 121 & 
    $<7.1$ &        
    $0.9\pm0.1$ &   
    $<0.1$ &        
    $0.5\pm0.2$ &   
    $13.0\pm0.2$ &  
    $1.0\pm0.1$ &   
    $<0.1$ &        
    - \\            
    \multicolumn{1}{|c}{\textsc{Mordred}} & 27 & 45 & 
    $5.6\pm4.9$ &    
    $<0.2$ &        
    $0.9\pm0.4$ &   
    $0.9\pm0.2$ &   
    $6.4^{+3.0}_{-2.8}$ &   
    $3.0\pm0.2$ &   
    $1.4\pm0.9$ &   
    $<28$ \\        
    \multicolumn{1}{|c}{\textsc{Maeve}} & 10 & 28.7 & 
    $<10$ &         
    $0.4\pm0.3$ &   
    $1.9\pm1.5$ &   
    $0.9\pm0.3$ &   
    $11.0^{+3.0}_{-4.2}$ &  
    $6.2\pm0.3$ &   
    $<0.3$ &        
    - \\            
    \multicolumn{1}{|c}{\textsc{Morgan}} & 4 & 17.4 & 
    $<15$ &         
    $<0.8$ &        
    $3.0\pm1.9$ &   
    $<0.8$ &        
    $6.3^{+6.6}_{-7.1}$ &   
    $6.0\pm0.7$ &   
    $<0.4$ &        
    - \\            
    \hline
    & \multicolumn{2}{|r|}{\textbf{Combined (MO)}} &
    $5.0\pm2.1$ &         
    $0.9\pm0.2$ &         
    $1.4\pm0.8$ &         
    $0.8\pm0.2$ &         
    \multirow{2}{*}{$12.2^{+1.1}_{-0.9}$} &  
    \multirow{2}{*}{$1.9\pm0.2$} &         
    $0.8\pm0.6$ &         
    $7.1\pm2.9$ \\        
    & \multicolumn{2}{|r|}{\textbf{Combined (UL)}} &
    $4.3\pm2.9$ &         
    $0.6\pm0.2$ &         
    $0.20\pm0.15$ &       
    $0.3\pm0.1$ &         
    & 
    & 
    $0.2\pm0.2$ &         
    $8.3\pm4.8$ \\        
    \cline{2-11}
    & \multicolumn{2}{|r|}{Component Average} & $<13.3$ & $<0.6$ & $<0.6$ & $<0.6$ & $<2.9$ & $<0.5$ & - & $<0.1$ \\
    & \multicolumn{2}{|r|}{LZ Requirement} & - & $<3.0$ & $<3.0$ & $<3.0$ & $<30$ & $<2.5$ & - & - \\
    & \multicolumn{2}{|r|}{\textsc{Xenon1T}} & $<12.9$ & $0.5\pm0.1$ & $0.5\pm0.1$ & $0.4\pm0.1$ & $12\pm2$ & $0.7\pm0.1$ & $0.4\pm0.1$ & - \\
    \cline{2-11}
\end{tabular}
\end{adjustbox}
\end{table}

In addition to values for U (\utTe{}, \utTF{}, \pbtoz{}), \thtTt{}, \kfz{} and \cosz{}, small levels of contamination from other isotopes are observed. This includes isotopes of cobalt (other than \cosz{}), \csoTS{} and the meta-stable \agooz{} isotope as reported in~\cite{Aprile:2015lha}. Results from measurements of these isotopes made with the Chaloner and Lunehead detectors at Boulby are shown in \tableWord~\ref{tab:unusual}. 

\begin{table}[!htb]
\caption{Combined contaminations from the Chaloner (BEGe) and Lunehead (p-type) detectors at Boulby for the less common isotopes detected when combining all PMT runs together. As we have no prior knowledge of the distribution of contamination for these isotopes we include a systematic uncertainty due to potential variations in geometric efficiency.\label{tab:unusual}}

    \centering
    \begin{tabular}{ |l|cccc| }
    \hline
    \multirow{2}{*}{\textbf{Detector}} & 
    \textbf{\csoTS}  & 
    \textbf{\agooz}  &  
    \textbf{\coFS}  &  
    \textbf{\coFe} \\ 
    
    & 
    \multicolumn{4}{c|}{\textbf{all values in mBq/PMT}} \\
    
    \hline
    \hline
    Chaloner & 
    $0.21 ^{+0.48}_{-0.12}$ & 
    $0.07 ^{+0.19}_{-0.05}$ & 
    $0.25 ^{+0.64}_{-0.11}$ & 
    $0.20 ^{+0.48}_{-0.10}$ \\
    Lunehead & 
    $0.51 \pm 0.03$ & 
    $0.13 \pm 0.02$ & 
    $0.25 \pm 0.03$ & 
    $0.37 \pm 0.03$ \\
    \hline
\end{tabular}
\end{table}

\begin{figure}[!htb]
\noindent\makebox[\textwidth][c]{%
\minipage{0.5\textwidth}
  \includegraphics[width=\linewidth]{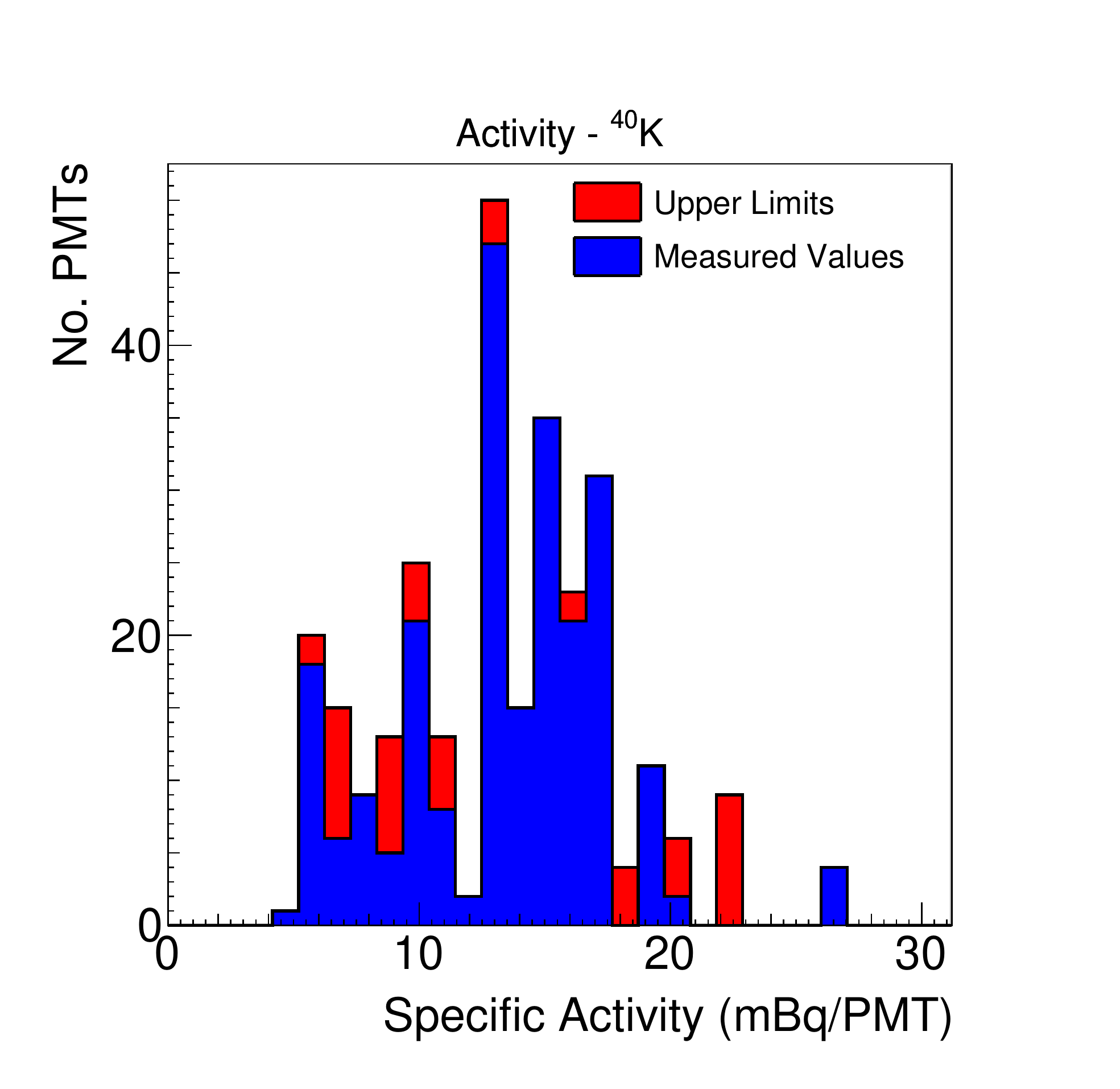}
\endminipage
}
\hfill\allowbreak%
\minipage{0.5\textwidth}
  \includegraphics[width=\linewidth]{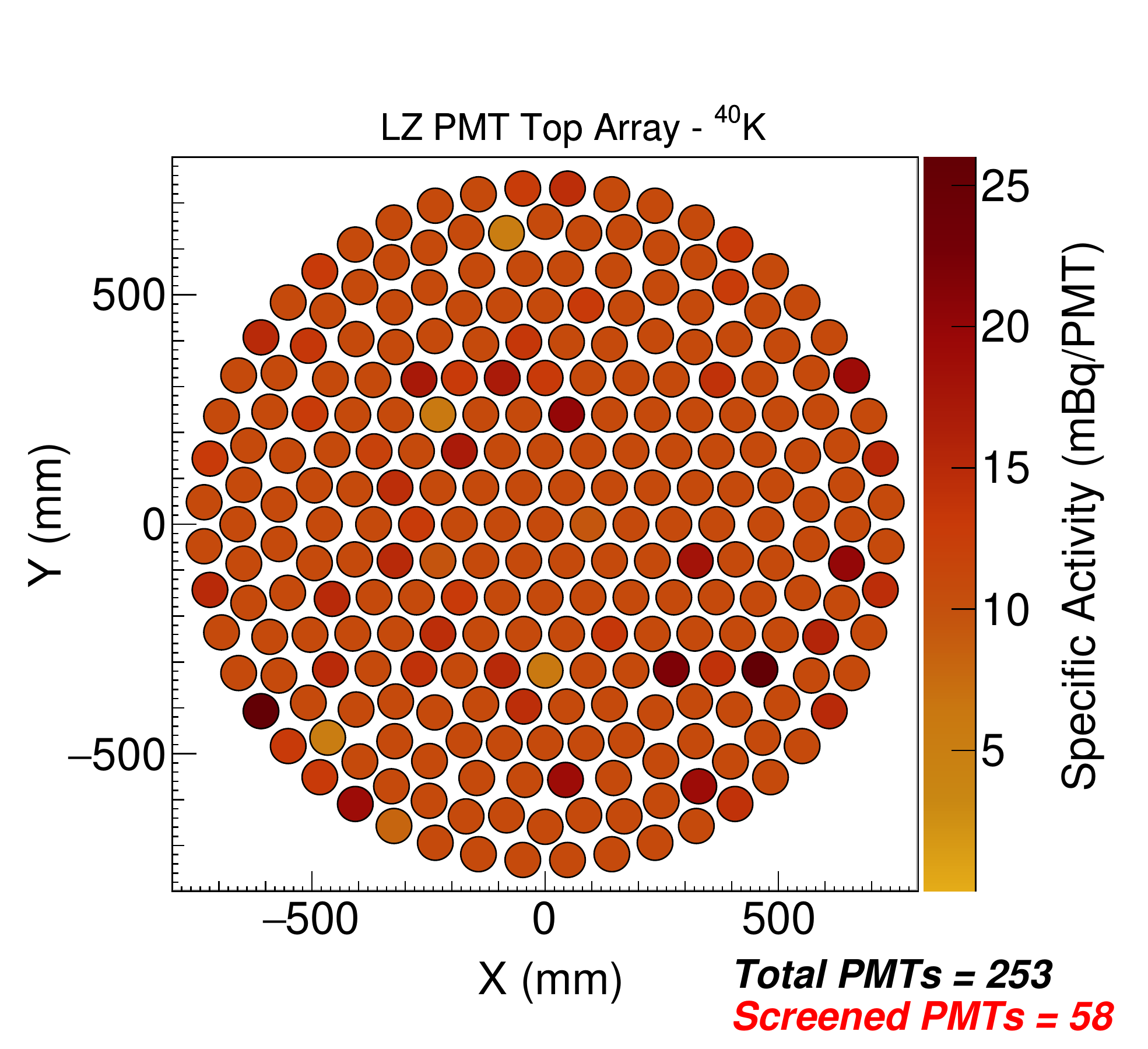}
\endminipage
\minipage{0.5\textwidth}
  \includegraphics[width=\linewidth]{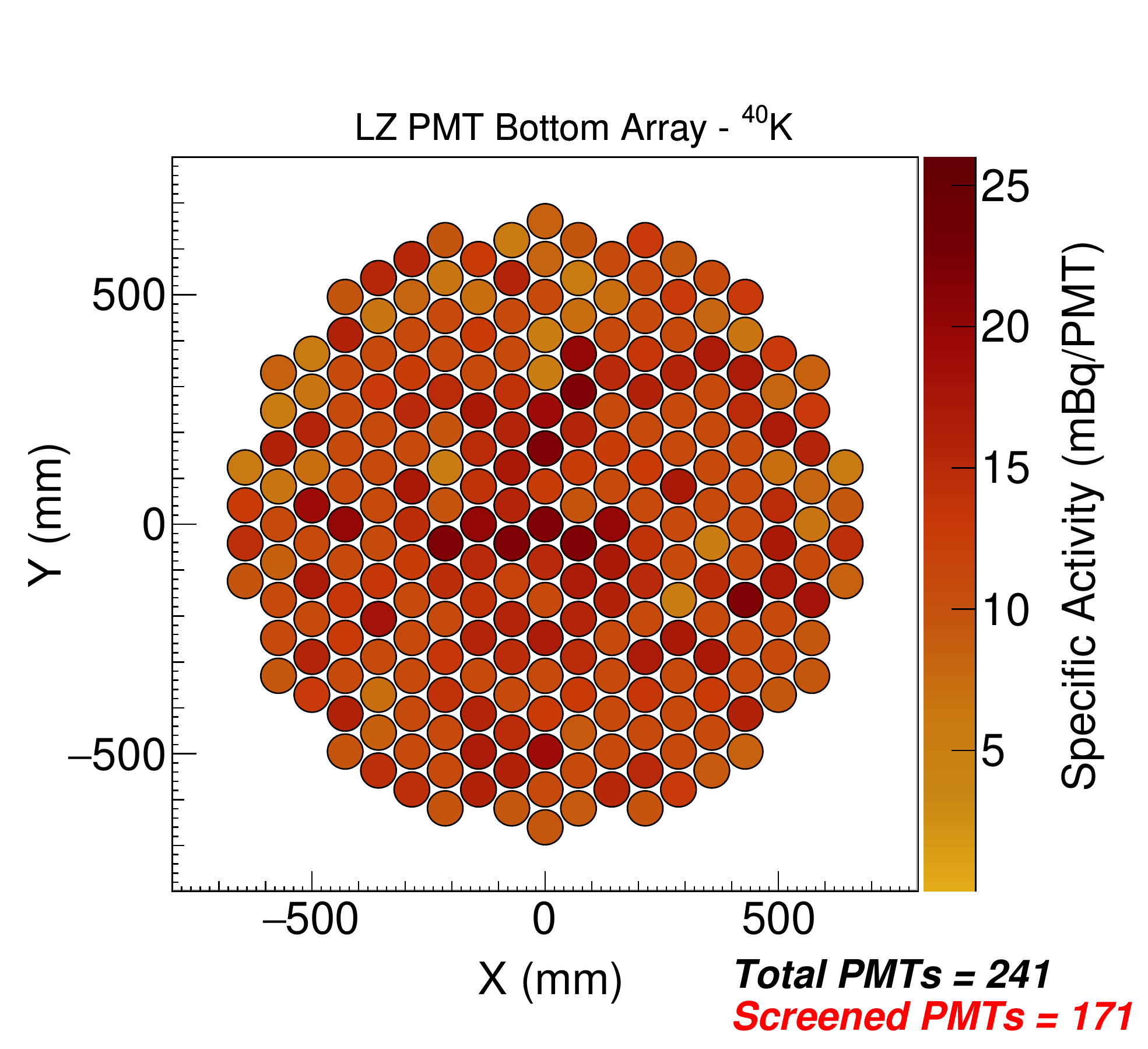}
\endminipage
\caption{Top: Distribution of measured value for \kfz{} in the LZ PMT assay program. Measured values are plotted in blue and upper limits are plotted in red. Bottom, left and right: Distribution of \kfz{} contamination for the top and bottom arrays, respectively. For tubes which were not measured, the calculated average value of $12.2^{+1.1}_{-0.9}$~mBq/PMT, as given in \tableWord~\ref{table:combined} is used.\label{fig:pmtK40}}
 \end{figure}

As an example, the distribution of measured values of \kfz{} in the assayed PMTs along with their distribution in the upper and lower PMT arrays of LZ are shown in \figureWord~\ref{fig:pmtK40}. For those PMTs not assayed, the average value of $12.2^{+1.1}_{-0.9}$~mBq/PMT, as given in \tableWord~\ref{table:combined} is used. Also shown in \figureWord~\ref{fig:pmtK40} is the distribution of measured \kfz{} values. This figure incorporates both measured values and upper limits.

\subsection{ICP-MS Titanium QC and QA assays}
\label{sec:NSTi}

In the early stages of the LZ experiment, an extensive R\&D campaign was conducted to source and produce enough titanium for the cryostat vessels of the detector. The inner and the outer cryostat vessels (ICV and OCV, respectively), containing the TPC and the 10 tonnes of LXe, make up a significant bulk of the LZ detector. Due to their scale and proximity to the TPC, it was necessary to ensure ultra-low levels of radiopurity for the \utTe{} and \thtTt{} isotopes as well as \kfz{} and \cosz{}. A detailed analysis using ICP-MS and gamma-ray spectroscopy of 22 different titanium samples was conducted, and the sample of the HN3469 product manufactured by TIMET was found to have the lowest background. The measured activities for \utTe{}, \thtTt{}, \cosz{} and \kfz{} from the sample are significantly lower than requirements and were the lowest reported to date~\cite{Akerib:2017iwt}. 

The titanium R\&D campaign was followed by a radiopurity screening program to monitor and mitigate the risk of radioactive contamination during the construction process of the vessels. Although the bulk titanium was found to be ultra-pure, the manufacturing stage of the vessels possessed a large risk for contamination. In assessing the radiopurity of the welding process a welding sample issued to the project as one of the many regular samples for assessment in the LZ QC and QA program, was measured to have 6 ppb of thorium---equating to roughly 100 times higher than the concentration initially measured in the TIMET HN3469. Material used for the test sample was from the LZ titanium stock and the welding was performed with an automatic tungsten inert gas (TIG) machine purportedly using the lanthanated tungsten electrodes commonly used by the manufacturer (Loterios). 

The high levels of thorium found in the sample prompted a suspension of the cryostat production and the start of a screening campaign to identify the source of the contamination. Over a period of two months, the campaign performed 24 radioassays with HPGe detectors and 67 ICP-MS measurements, screening representative samples from the titanium stock, welding wires and electrodes used by Loterios. 	 	 	 		
Assays of two welding electrodes used by Loterios and removed from the TIG and plasma machines located in the cryostat production area showed a very high level of thorium. The \thtTtE{} levels, as measured by the ICP-MS system, resulted in \SI{3.20\pm0.16}{ppb} of Th, in comparison to \SI{0.069\pm0.007}{ppb} measured for the TIMET HN3469 stock. This indicated that the unexpected excess in Th activity was due to an isolated and erroneous use of a small number of thoriated electrodes, as opposed to the pre-agreed lanthanated tungsten electrodes throughout.

Additionally, it was found that the standard lanthanated tungsten electrodes used do contain a small amount of thorium which, in the welding process, gets into the weld at the level of \SI{0.7}{ppb}. Although this is not ideal, it would not have explained the observed increase of thorium in the welded sample which prompted this contamination study.

The rapid turnaround for uranium and thorium assays at sub-ppb levels made possible by the ICP-MS system allowed LZ to arrest production, thoroughly investigate and precisely identify the contaminant introduced by the manufacturing process.
Upon these findings, appropriate control measures were implemented in the cryostat manufacturing process to avoid additional inadvertent use of thoriated welding electrodes. This is a key example of how the LZ QA program identified a potentially serious manufacturing error which otherwise would have had the potential to substantially increase the overall thorium background in LZ.

\subsection{Radon Emanation Studies}

\begin{figure}[htp!]
\centering
\hspace*{-0.2cm}
\includegraphics[scale=0.29]{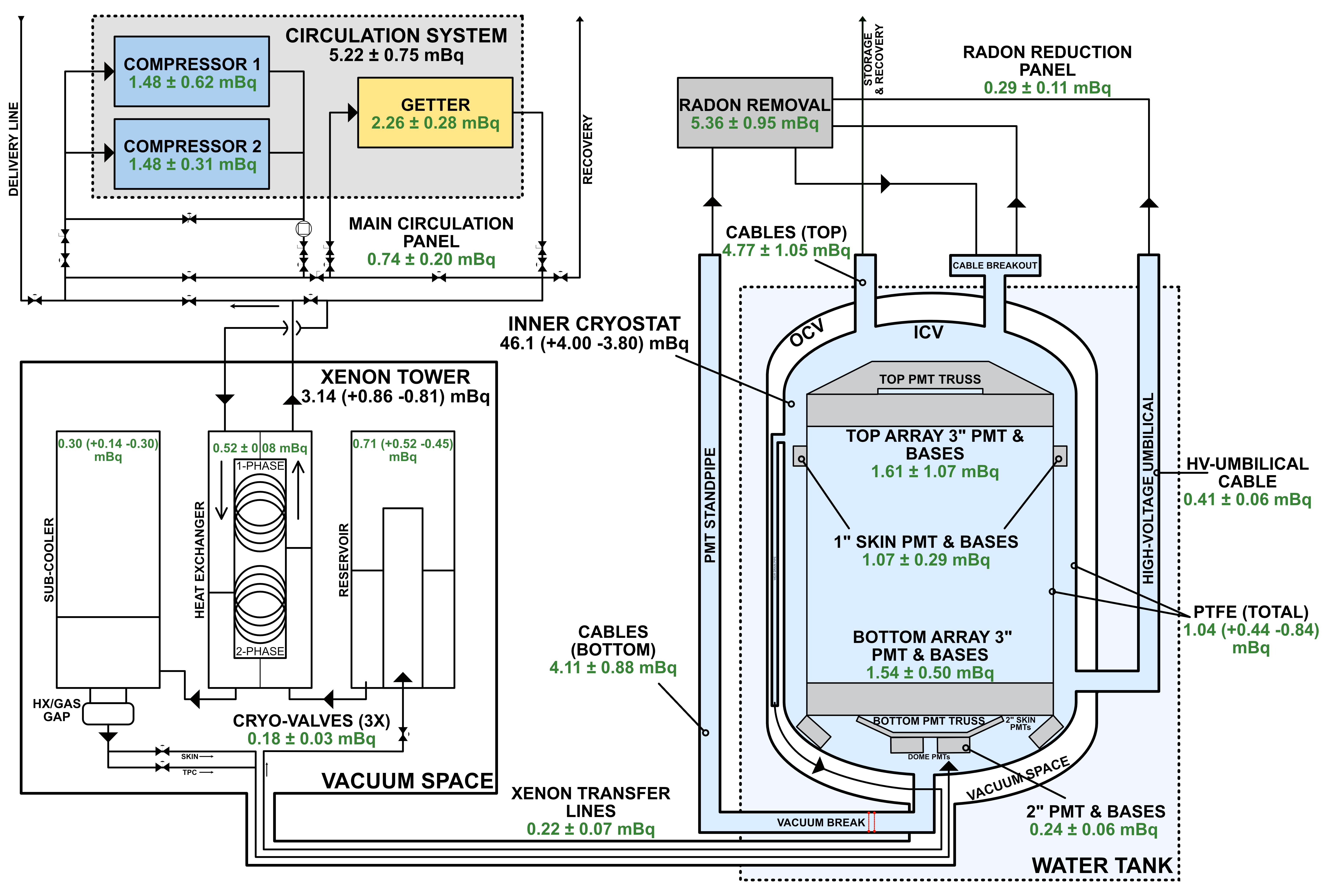}
\caption{Approximate schematic of LZ highlighting key sub-systems and xenon circulation paths in and out of the ICV. The diagram condenses some of the key radon emanation measurements that will contribute to the overall radon budget of LZ. The results presented in green text are directly from measurements and those in black show summed results for that particular component. All of the results shown are measurements made at room temperature and do not include any radon removal or cold temperature suppression expected under operational conditions. \label{LZ_radon_diagram_paper}}
\end{figure}

Sensitivity projections for LZ presented in~\cite{Akerib:2018lyp} include the effect of the online charcoal-based radon-removal system, operating continuously to scrub gaseous Xe \cite{Mount:2017qzi, Pushkin:2018wdl}. Projections also assume the expected suppression of radon diffusion in certain materials at low temperatures. In this article, \tableWord~\ref{tab:RnBig} presents the as-measured results from room temperature assays of individual materials and components. Room temperature measurements from fully assembled LZ systems and contributions from individually screened components within those systems are highlighted in \figureWord~\ref{LZ_radon_diagram_paper}. Measurements for three of these assemblies are discussed in the following sub-sections, with radon emanation results.

\subsubsection{Inner Cryostat Vessel}

Radon emanation from the ICV was measured several times during various integration stages of the construction of the skin veto region and the TPC installation. The final assay was made following after the ICV was fully complete and sealed. The cryostat at this stage housed both the top and the bottom PMT arrays for the TPC and the skin veto regions, and their corresponding PMT bases and cables. Furthermore, the entire field cage, PTFE coating, various sensors, and conduit volumes of the cables were a part of these measurement.

A portable radon trapping system was deployed underground at SURF with minimal plumbing due to space constraints. After leak-checking and purging, the trapping system was opened to the ICV and the emanated gas was harvested over a 6.3 hour period---equivalent to 18.25\% of the gas within the ICV. After the harvest, the trap was carefully disconnected and transported to SDSM\&T radon facility for screening. The radon trap also captures outgassing molecular species that would serve as neutralisers of positively charged radon-daughters, leading to a drop in detection efficiency. An in-house procedure was followed to separate out these species by transferring the sample from a cold trap held at -109$^{\circ{}}$C to one at -196$^{\circ{}}$C with sufficient flow to effectively transfer all of the radon atoms while leaving most of the contaminants behind. This process was repeated until measurements with a residual gas analyzer indicated no further reduction in contaminant concentration, after which the sample was transferred to the detection chamber via a secondary small cold trap.

Results indicate a room-temperature emanation rate of $46.1^{+4.0}_{-3.8}\,$mBq under the assumption of an even sampling of the radon within the ICV.

\subsubsection{Xenon Circulation System}

The xenon gas circulation system brings together multiple components and surfaces that are potential radon emitters. The system consists of two gas compressors, a heated zirconium getter, and a main valve and instrumentation panel. The compressors (model A2-5/15 from Fluitron) have two heads, each enclosing a flexible all-metal diaphragm sealed with copper plating. Check valves, accumulation bottles, and associated plumbing and instrumentation are also included in the compressor assemblies. The compressors operate in parallel to achieve a gas flow rate of 500 standard liters per minute. Much of the system was fabricated at the University of Wisconsin's Physical Sciences Laboratory. Whilst there, a portable radon trapping system was used to harvest emanation samples that were then shipped to shipped to the U. Maryland radon facility for counting.
Initial radon emanation measurements of compressor 2 found that the heads emanated \SI{< 1}{\milli\becquerel} each; however, the integrated compressor skid assembly presented $\sim$\SI{17}{\milli\becquerel}. After replacing most of the welded stainless steel plumbing and etching the accumulation bottles in citric acid, the rate was reduced to $1.48\pm0.31$\,mBq. A similar treatment was applied to compressor 1 but this compressor was not radon emanated and hence is assumed to have the same rate as compressor 2. The main circulation panel contains most of the valves and instrumentation exposed to the xenon in gas phase, and it was found to contribute \SI{0.74}{\milli\becquerel} of radon. The fully loaded getter (model PS5-MGT50-R-535 from SAES) was emanated at its operational temperature of $400 ^{\circ}$C using helium carrier gas and its emanation rate was determined to be \SI{2.26}{\milli\becquerel}. The entire circulation system amounted to a total emanation rate of $5.22\pm0.75$\,mBq.

\subsubsection{Xenon Tower}

The xenon tower is a cryogenic system that thermally couples the gaseous and LXe portions of the purification circuit for efficient heat transfer, serving to vaporize and re-condense the liquid for continuous purification. It consists of a two-phase heat exchanger (supplied from Standard Xchange), three cryogenic valves (manufactured by WAKE), a sub-cooler/phase-separator vessel to hold LXe returning to the detector, a reservoir vessel to hold liquid exiting the detector, two liquid xenon purity monitors, and several custom liquid xenon heat exchangers. The tower can be viewed as having two sides: the heat exchanger assembly on one side and the weir reservoir, sub-cooler and purity monitor on the other. Radon emanation from sub-components was measured prior to full integration of the xenon tower and was found to contribute a total of \SI{<1}{\milli\becquerel}. 

A preliminary measurement of the tower after integration found a very high radon activity in the reservoir side, possibly due to a leak into the system from laboratory air. As a precautionary measure, the reservoir vessel was flushed with a concentrated solution designed for removing radioactive contamination (Radiacwash\textsuperscript{TM}) and rinsed with deionized water. The portable radon trap was then deployed underground to measure the two sides of the complete xenon tower prior to the installation of the purity monitor and found a total emanation rate of $3.14^{+0.86}_{-0.81}$\,mBq.

\subsection{GdLS screener for the Outer Detector (OD)}

Achieving a low level of radioimpurities in the \SI{17.3}{\tonne} of GdLS used in the LZ OD is crucial for its performance as both an effective monitor of the experiment's backgrounds and as an efficient veto for coincident events in the TPC. The admissible radioactive impurity levels in the GdLS are derived from the LZ requirement for a false veto rate of \SI{<5}{\percent} assuming an energy threshold of \SI{200}{\keV} and a maximum coincidence window of \SI{500}{\micro\second} between an interaction in the TPC and the OD. This restricts the event rate in the full \SI{17.3}{\tonne} of GdLS to be \SI{<100}{\Hz}. Approximately \SI{39}{\Hz} is expected from a combination of radioactivity in LZ construction materials and the flux of gamma-rays in the experimental hall which penetrate the water shield to reach the OD~\cite{Akerib:2019sek}. This allows for at most \SI{61}{\Hz} of rate to arise from impurities dissolved in the GdLS.

GdLS generally features a combination of common radioimpurities and more exotic isotopes which enter during the Gd-loading process. Common radioisotopes include \cof{}, \kref{}, \kfz{}, and those in the \utTe{} and \thtTt{} chains, while the more unusual isotopes include \laoTe{}, \luoSs{}, \smofS{}, \gdoFt{}, and those in the \utTF{} decay chain. Chemical processing and purification of the components of GdLS alters the relative abundance of these isotopes, often resulting in out-of-equilibrium activities within decay chains~\cite{Yeh:2010zz}.

Before it is added to LS, the raw gadolinium product (in the form of GdCl$_3$) is purified via pH-controlled partial hydrolysis. During this process, certain actinides including uranium and thorium are precipitated out of solution, but others such as actinium itself tend to remain in solution~\cite{Yeh:2010zz}. The purified Gd is then combined with the chelating ligand TMHA, or 3,5,5-trimethylhexanoic acid. The resulting $\mathrm{Gd(TMHA)}_{\mathrm{3}}$ compound is dissolved in the non-polar LS solvent, linear alkylbenzene, to achieve a final concentration of \SI{0.1}{\percent} Gd by mass.

Measurements of the radioactive impurity concentrations in both Gd-loaded and unloaded scintillator samples were made using a dedicated liquid scintillation counter known as the ``LS Screener". The LS Screener was comprised of an acrylic vessel containing approximately 23~kg of scintillator and three R11410 PMTs. For each sample measurement, the detector was lowered into the ultra-low background environment at the center of the filled LZ water tank. A detailed description of that detector and the results from those measurement campaigns are published in~\cite{Haselschwardt:2018vmp}. Two results from that work yielded results particularly significant for the OD's performance. First, a fit using the \cof{} $\beta$ shape ($Q=$ \SI{156}{\kilo\electronvolt}) to the low-energy spectrum collected with the unloaded scintillator sample measured the \cof{}/\cotw{} atom fraction in the LS to be $(2.83 \pm 0.06 \textrm{(stat.)}\pm 0.01 \textrm{(sys.)}) \times 10^{-17}$. This concentration of \cof{} validates that hydrocarbons used for the GdLS chemicals are suitably derived from underground sources (as opposed to biogenic sources). As a result, the expected rates of single and pile-up pulses from \cof{} near the OD threshold are negligible. Second, pulse shape discrimination and delayed coincidence counting were used to measure the concentrations of $\alpha$ and $\beta$-emitting radioisotopes with relatively small uncertainties. In particular, a significant amount of activity from the \utTF{} series was measured by identifying the unique pulse pair from \rnton{}-\potoF{} in the GdLS sample. No such activity was observed in the unloaded LS sample, confirming that this activity is introduced with the Gd.

To further characterize the impact of loading with Gd, a \SI{0.307}{\kg} sample of the same purified $\mathrm{Gd(TMHA)}_{\mathrm{3}}$ compound used in the LS Screener was counted by HPGe at the BHUC. A large activity from \luoSs{} was easily detected above background confirming the presence of rare earth impurities. Additionally, the strong $\gamma$-line from \utTF{} was not detected, suggesting that equilibrium is broken during purification causing either \patTo{} or \acttS{} to serve as the head of the decay chain.

These findings motivated a modification to the Gd purification procedure in which a higher pH was used to more aggressively precipitate out impurities at the cost of a slightly lower Gd yield. To confirm the effectiveness of this procedure, a larger, \SI{1.44}{\kg} sample of the newly purified $\mathrm{Gd(TMHA)}_{\mathrm{3}}$ was counted using HPGe at the BHUC. \tableWord~\ref{tab:gdScreening} reports the results before and after the new purification as well as the reduction factor in each isotope's central value or quoted upper limit. Concentrations are given per kg of $\mathrm{Gd(TMHA)}_{\mathrm{3}}$ and per kg of LZ GdLS, doped to \SI{0.1}{\percent} Gd by mass. The observed reduction in the detected \luoSs{} concentration demonstrates the effectiveness of the purification on rare earth metals. No significant signatures from \acttS{} and its daughters in the newer sample were observed. By applying the reduction factors in \tableWord~\ref{tab:gdScreening} to the results in~\cite{Haselschwardt:2018vmp}, the total OD rate from GdLS radioimpurities above \SI{200}{\keV} is predicted to be \SI{<10}{\Hz}, well below the requirement.

\begin{table}[!htb]
\centering
\caption{Results from HPGe counting of two purified $\mathrm{Gd(TMHA)}_{3}$ powder samples given in \si{\mBqkg} of $\mathrm{Gd(TMHA)}_{\mathrm{3}}$. For GdLS, loaded to 0.1\% Gd by mass, the values listed can be reduced by a factor of 250. The reduction factors of the limits/central values for the improved purification results are also given. Limits are given at~$1\sigma$~confidence level. The late \utTF{} chain is defined as starting at \patTo{} as discussed in~\cite{Haselschwardt:2018vmp}.\label{tab:gdScreening}}
\begin{adjustbox}{width=\textwidth,center}
    \begin{tabular}{ l | c | c | c } \hline
    \textbf{Isotope or} & \multicolumn{1}{c|}{\textbf{0.307~kg, Initial Purification}} & \multicolumn{1}{c|}{\textbf{1.44~kg, Improved Purification}} & \multirow{2}{*}{\textbf{Reduction}} \\ 
     {\textbf{Subchain}} & \textbf{mBq/(kg $\textrm{Gd(TMHA)}_{3}$)} & \textbf{mBq/(kg $\textrm{Gd(TMHA)}_{3}$)} & \\ \hline
    \utTeE{}    & $<259$     & $<4.36$        & 59.4 \\
    \utTeL{}    & $23(5)$    & $2.6(9)$       & 8.8 \\
    \utTFE{}    & $<2.8$     & $<4.5$         & 0.6 \\
    \utTFL{}    & $26(10)$   & $<3.0$         & 8.7 \\
    \thtTtE{}   & $<6.7$     & $<0.89$        & 7.5 \\
    \thtTtL{}   & $<5.1$     & $<0.76$        & 6.7 \\
    \kfz{}      & $<56$      & $<2.29$        & 24.5 \\
    \laoTe{}    & $<1.4$     & $<0.42$        & 3.3 \\
    \luoSs{}    & $75(18)$   & $2.03(46)$     & 36.9 \\
    \hline
\end{tabular}
\end{adjustbox}
\end{table}

https://www.overleaf.com/project/5c5310397a978a1f50a01c1c

\section{Conclusion}

The LZ collaboration has concluded one of the most wide ranging and sensitive assay programs performed to-date for a rare event search experiment, successfully meeting requirements on fixed contaminants, radon emanation, and surface cleanliness. This program began in October 2013 and has continued for more than six years, building a database of over 1200 entries. The results obtained in the screening program have been used to select the materials used in the final construction of the LZ detector, to ensure radioactivity compliance through to detector assembly, and to inform the background model used to determine the predicted sensitivity of LZ.

In parallel to these assay efforts, a comprehensive set of cleanliness quality assurance protocols were developed and implemented. These sought to ensure that additional radiocontaminants were not introduced into the detector in the construction period, particularly during the time when the TPC was being constructed, up until the point where it was sealed within the inner cryostat vessel for transportation to the underground laboratory at SURF.

Comprehensive tables of results for the LZ assay program are presented in~\ref{app:tabs}, documenting the results from the LZ assays performed since 2013.


\section{Acknowledgements}

The LZ collaboration lost one of its dedicated researchers this year and the father of low background assay at Berkeley Lab, Al Smith.  For more than sixty years, he pursued low background assays and identification of materials in support of rare search experiments including CDMS, SNO, KamLAND, Daya Bay, KATRIN, Majorana Demonstrator, LUX, and LZ.  We also wish to acknowledge the contributions from Emma Meehan at the Boulby Underground Laboratory for her help. The research supporting this work took place in whole or in part at the Sanford Underground Research Facility (SURF) in Lead, South Dakota. Funding for this work is supported by the U.S. Department of Energy, Office of Science, Office of High Energy Physics under Contract Numbers DE-AC02-05CH11231, DE-SC0020216, DE-SC0012704, DE-SC0010010, DE-AC02-07CH11359, DE-SC0012161, DE-SC0014223, DE-FG02-13ER42020, DE-SC0009999, DE-NA0003180, DE-SC0011702, \\
DE-SC0010072, DE-SC0015708, DE-SC0006605, DE-FG02-10ER46709, \\
UW PRJ82AJ, DE-SC0013542, DE-AC02-76SF00515, DE-SC0019066, DE-AC52-07NA27344, \& DE-SC0012447.	This research was also supported by U.S. National Science Foundation (NSF); the U.K. Science \& Technology Facilities Council under award numbers, ST/M003655/1, ST/M003981/1, ST/M003744/1, ST/M003639/1, ST/M003604/1, and ST/M003469/1; Portuguese Foundation for Science and Technology (FCT) under award numbers PTDC/FIS-Â­PAR/28567/2017; the Institute for Basic Science, Korea (budget numbers IBS-R016-D1); University College London and Lawrence Berkeley National Laboratory thank the U.K. Royal Society for travel funds under the International Exchange Scheme (IE141517). We acknowledge additional support from the STFC Boulby Underground Laboratory in the U.K., the GridPP Collaboration \cite{Faulkner:2006px,Britton:2009ser}, in particular at Imperial College London and additional support by the University College London (UCL) Cosmoparticle Initiative. This research used resources of the National Energy Research Scientific Computing Center, a DOE Office of Science User Facility supported by the Office of Science of the U.S. Department of Energy under Contract No. DE-AC02-05CH11231. The University of Edinburgh is a charitable body, registered in Scotland, with the registration number SC005336. The assistance of SURF and its personnel in providing physical access and general logistical and technical support is acknowledged.
\bibliography{Support_Files/LZNew}

\appendix
\setcounter{table}{0}
\renewcommand{\thetable}{\Alph{section}\arabic{table}}

\section{Assay Results Tables}\label{app:tabs}

In this section, the results from the assay campaign of the LZ experiment are captured. The tables are split into the various techniques which were used. \tableWord~\ref{tab:HPGeBig} details all assays completed using the HPGe detectors available to the collaboration. For this table, there are a number of repeat measurements. Where this is the case, items are grouped with a single entry having multiple results. Each named item in the table is unique even if it shares the same name as another. \tableWord{}s~\ref{tab:NAABig},\ref{tab:ICPMSBig} and \ref{tab:GDMSBig} detail results from NAA, ICP-MS and GD-MS, respectively. Finally, \tableWord~\ref{tab:RnBig} details results from the radon emanation assay campaign.

For ease of interpretation, the tables are further subdivided to detail locations within the LZ experiment where specific items were either used or where there intended use would be. Those items which are constructed ``in-house'' by the LZ collaboration are highlighted as such. Not all isotopes are detailed for all items as the ability to do this may depend on the ability of the detector or method employed or if the isotope is of particular interest ($e.g.$ for \cosz{} in stainless steel components). The upper limits reported are generally at \SI{90}{\percent} confidence level but upper limits from the LBNL and BHUC detectors (\textsc{Merlin}, \textsc{Maeve}, \textsc{Morgan}, \textsc{Mordred}, and SOLO) are quoted at $1\sigma$.

\clearpage
\thispagestyle{empty}
\newgeometry{margin=3cm}

\begin{sidewaystable*}[tp]
\caption{HPGe Screening Results.\label{tab:HPGeBig}}
\begin{adjustbox}{width=\textwidth,center}
    \tabcolsep=4pt
    \centering
\begin{tabular}{ H L{9cm} L{7cm} H C{2.cm} C{2.cm} C{2.cm}
    C{2.5cm} C{2.5cm} C{2.5cm} C{2.5cm} C{2.5cm} C{2.5cm} C{2.5cm} }

    \toprule
    \\
    \textbf{Item} & 
    \textbf{Sample}  & 
    \textbf{Supplier} &  
    \textbf{LZ Location}  &  
    \textbf{Detector}  & 
    \textbf{Mass [kg]}   &  
    \textbf{Units}  &  
    \textbf{\utTeE} & 
    \textbf{\utTeL} & 
    \textbf{\thtTtE} & 
    \textbf{\thtTtL} & 
    \textbf{\kfz} & 
    \textbf{\pbtoz} & 
    \textbf{\cosz} 
    \\
    \hline
    \hline
    \\
\\
\multicolumn{14}{l}{\textbf{Calibration}}\\
\hline \\
0 & Tungsten Pig & Midwest Tungsten & Calibration & \textsc{Merlin} & 0.89 & mBq/kg & - & $<$1200 & $<$690 & $<$610 & $<$150 & - & - \\
1 & Tungsten Alloy & Midwest Tungsten & Calibration & \textsc{Merlin} & 0.98 & Bq/kg & - & 1.7(10) & 0.15(1) & 0.14(1) & 77(5) & - & - \\
\end{tabular}
\end{adjustbox}
\end{sidewaystable*}
\clearpage
\thispagestyle{empty}
\newgeometry{margin=3cm}

\begin{sidewaystable*}[tp]
\begin{adjustbox}{width=\textwidth,center}
    \tabcolsep=4pt
    \centering
    \begin{tabular}{ H L{9cm} L{7cm} H C{2.cm} C{2.cm} C{2.cm}
    C{2.5cm} C{2.5cm} C{2.5cm} C{2.5cm} C{2.5cm} C{2.5cm} C{2.5cm}}

    \toprule
    \\
    \textbf{Item} & 
    \textbf{Sample}  & 
    \textbf{Supplier} &  
    \textbf{LZ Location}  &  
    \textbf{Detector}  & 
    \textbf{Mass [kg]}   &  
    \textbf{Units}  &  
    \textbf{\utTeE} & 
    \textbf{\utTeL} & 
    \textbf{\thtTtE} & 
    \textbf{\thtTtL} & 
    \textbf{\kfz} & 
    \textbf{\pbtoz} & 
    \textbf{\cosz} 
    \\
    \hline
    \hline
    \\
    
30 & \multirow{2}{*}{Stainless Steel Sheet} & \multirow{2}{*}{Nironit} & Cryostat & \textsc{Maeve} & \multirow{2}{*}{10} & mBq/kg & 7.3(25) & 0.35(5) & 1.1(2) & 4.0(3) & $<$0.80 & - & 14.5(3) \\
31 &  &  & Cryostat & \textsc{Merlin} & & mBq/kg & - & $<$0.70 & - & 6.3(10) & $<$3.0 & - & 16(1) \\
32 & \multirow{2}{*}{Stainless Steel Sheet} & \multirow{2}{*}{Nironit} & Cryostat & \textsc{Maeve} & \multirow{2}{*}{10} & mBq/kg & $<$1.0 & 0.54(10) & 0.49(8) & 1.1(2) & $<$0.59 & - & 1.7(1) \\
33 &  &  & Cryostat & \textsc{Merlin} & & mBq/kg & - & $<$0.60 & - & $<$1.3 & $<$3.0 & - & 1.7(3) \\
34 & \multirow{2}{*}{Stainless Steel Sheet} & \multirow{2}{*}{Nironit} & Cryostat & \textsc{Merlin} & \multirow{2}{*}{11} & mBq/kg & - & $<$0.60 & - & 0.90(20) & $<$4.0 & - & 2.2(3) \\
\end{tabular}
\end{adjustbox}
\end{sidewaystable*}
\clearpage
\thispagestyle{empty}
\newgeometry{margin=3cm}

\begin{sidewaystable*}[tp]
\begin{adjustbox}{width=\textwidth,center}
    \tabcolsep=4pt
    \centering
    \begin{tabular}{ H L{9cm} L{7cm} H C{2.cm} C{2.cm} C{2.cm}
    C{2.5cm} C{2.5cm} C{2.5cm} C{2.5cm} C{2.5cm} C{2.5cm} C{2.5cm}}

    \toprule
    \\
    \textbf{Item} & 
    \textbf{Sample}  & 
    \textbf{Supplier} &  
    \textbf{LZ Location}  &  
    \textbf{Detector}  & 
    \textbf{Mass [kg]}   &  
    \textbf{Units}  &  
    \textbf{\utTeE} & 
    \textbf{\utTeL} & 
    \textbf{\thtTtE} & 
    \textbf{\thtTtL} & 
    \textbf{\kfz} & 
    \textbf{\pbtoz} & 
    \textbf{\cosz} 
    \\
    \hline
    \hline
    \\
60 & Stainless Steel Cryostat Flange & Italy \#422511 & Cryostat & \textsc{Merlin} & 1.0 & mBq/kg & 2.5(12) & 1.9(4) & 4.9(8) & 4.9(8) & $<$31 & - & 3.1(3) \\
61 & \multirow{2}{*}{Stainless Steel Round Stock} & \multirow{2}{*}{Italian} & Cryostat & \textsc{Morgan} & \multirow{2}{*}{3.4} & mBq/kg & $<$0.99 & $<$0.86 & 3.3(4) & 4.9(4) & $<$1.0 & - & 6.7(4) \\
62 &  &  & Cryostat & \textsc{Merlin} & & mBq/kg & 37(12) & 12(1) & 3.7(12) & 3.7(12) & $<$31 & - & 9.0(40) \\
63 & Stainless Steel Plate & Coulter & Cryostat & \textsc{Merlin} & 13 & mBq/kg & $<$6.2 & $<$2.5 & $<$2.0 & $<$2.0 & $<$31 & - & 12.1(4) \\
64 & 3.2 mm Titanium Welding Wire & Timet & Cryostat & \textsc{Merlin} & 0.95 & mBq/kg & $<$123 & $<$12 & $<$8.1 & $<$8.1 & $<$31 & - & - \\
\end{tabular}
\end{adjustbox}
\end{sidewaystable*}
\clearpage
\thispagestyle{empty}
\newgeometry{margin=3cm}

\begin{sidewaystable*}[tp]
\begin{adjustbox}{width=\textwidth,center}
    \tabcolsep=4pt
    \centering
    \begin{tabular}{ H L{9cm} L{7cm} H C{2.cm} C{2.cm} C{2.cm}
    C{2.5cm} C{2.5cm} C{2.5cm} C{2.5cm} C{2.5cm} C{2.5cm} C{2.5cm}}

    \toprule
    \\
    \textbf{Item} & 
    \textbf{Sample}  & 
    \textbf{Supplier} &  
    \textbf{LZ Location}  &  
    \textbf{Detector}  & 
    \textbf{Mass [kg]}   &  
    \textbf{Units}  &  
    \textbf{\utTeE} & 
    \textbf{\utTeL} & 
    \textbf{\thtTtE} & 
    \textbf{\thtTtL} & 
    \textbf{\kfz} & 
    \textbf{\pbtoz} & 
    \textbf{\cosz} 
    \\
    \hline
    \hline
    \\
90 & 1.6 mm Titanium Welding Wire & Loterios & Cryostat & \textsc{Morgan} & 1.1 & mBq/kg & $<$6.0 & $<$0.80 & $<$0.90 & $<$2.7 & $<$4.3 & - & - \\
91 & PEEK Rods 3/8 \& 1/2 & McMaster Carr & Cryostat & \textsc{Merlin} & 0.60 & mBq/kg & $<$38 & 21(5) & $<$3.2 & $<$2.9 & $<$19 & - & - \\
92 & PEEK Rods 1/4 & McMaster Carr & Cryostat & \textsc{Merlin} & 0.23 & mBq/kg & $<$18 & $<$15 & $<$9.0 & $<$6.0 & $<$51 & - & - \\
93 & PEEK 1/4 Threaded Rod & McMaster Carr & Cryostat & \textsc{Merlin} & 0.12 & mBq/kg & $<$25 & $<$12 & $<$16 & $<$11 & $<$94 & - & - \\
94 & UMHW Polyethylene Blocks & Direct Plastics & Cryostat & Chaloner & 1.5 & mBq/kg & $<$24 & $<$1.0 & $<$1.0 & $<$0.50 & $<$24 & - & - \\
\end{tabular}
\end{adjustbox}
\end{sidewaystable*}
\clearpage
\thispagestyle{empty}
\newgeometry{margin=3cm}

\begin{sidewaystable*}[tp]
\begin{adjustbox}{width=\textwidth,center}
    \tabcolsep=4pt
    \centering
    \begin{tabular}{ H L{9cm} L{7cm} H C{2.cm} C{2.cm} C{2.cm}
    C{2.5cm} C{2.5cm} C{2.5cm} C{2.5cm} C{2.5cm} C{2.5cm} C{2.5cm}}

    \toprule
    \\
    \textbf{Item} & 
    \textbf{Sample}  & 
    \textbf{Supplier} &  
    \textbf{LZ Location}  &  
    \textbf{Detector}  & 
    \textbf{Mass [kg]}   &  
    \textbf{Units}  &  
    \textbf{\utTeE} & 
    \textbf{\utTeL} & 
    \textbf{\thtTtE} & 
    \textbf{\thtTtL} & 
    \textbf{\kfz} & 
    \textbf{\pbtoz} & 
    \textbf{\cosz} 
    \\
    \hline
    \hline
    \\
120 & PTFE Screws & Product Components Corporation & Cryostat & \textsc{Merlin} & 0.08 & mBq/kg & $<$145 & $<$44 & $<$52 & $<$31 & $<$188 & - & - \\
121 & Bellows & Palatine Precision Ltd & Cryostat & Roseberry & 0.54 & mBq/kg & 14(7) & 8.0(20) & 12(4) & 15(1) & $<$8.0 & 140(51) & 10(2) \\
122 & Viton O-Rings & Dichtomatik & Cryostat & Lunehead & 0.37 & mBq/kg & 171(26) & 62(8) & 19(7) & 15(3) & 3267(105) & - & $<$3.0 \\
123 & Elastomer Cord & Dichtomatik & Cryostat & Chaloner & 0.35 & mBq/kg & $<$19 & 572(20) & 1776(45) & 1747(15) & 7895(172) & - & $<$2.0 \\
124 & \multirow{2}{*}{PEEK Monofilament} & \multirow{2}{*}{Accu-Glass Products Inc} & Cryostat & \textsc{Merlin} & \multirow{2}{*}{0.02} & mBq/kg & $<$1060 & $<$320 & $<$710 & $<$760 & $<$2030 & - & - \\
\end{tabular}
\end{adjustbox}
\end{sidewaystable*}
\clearpage
\thispagestyle{empty}
\newgeometry{margin=3cm}

\begin{sidewaystable*}[tp]
\begin{adjustbox}{width=\textwidth,center}
    \tabcolsep=4pt
    \centering
    \begin{tabular}{ H L{9cm} L{7cm} H C{2.cm} C{2.cm} C{2.cm}
    C{2.5cm} C{2.5cm} C{2.5cm} C{2.5cm} C{2.5cm} C{2.5cm} C{2.5cm}}

    \toprule
    \\
    \textbf{Item} & 
    \textbf{Sample}  & 
    \textbf{Supplier} &  
    \textbf{LZ Location}  &  
    \textbf{Detector}  & 
    \textbf{Mass [kg]}   &  
    \textbf{Units}  &  
    \textbf{\utTeE} & 
    \textbf{\utTeL} & 
    \textbf{\thtTtE} & 
    \textbf{\thtTtL} & 
    \textbf{\kfz} & 
    \textbf{\pbtoz} & 
    \textbf{\cosz} 
    \\
    \hline
    \hline
    \\
150 & Copper Block & Silicon Valley MFG & Cryostat & \textsc{Merlin} & 0.35 & mBq/kg & $<$24 & $<$8.6 & $<$9.3 & $<$8.5 & $<$28 & - & - \\
151 & Terminal Ring Connectors & Digikey/TE Connectivity & Cryostat & \textsc{Merlin} & 0.04 & mBq/kg & $<$343 & $<$120 & $<$110 & $<$121 & $<$359 & - & - \\
152 & Vacuum RTD Screws & Made in house & Cryostat & Roseberry & 0.08 & mBq/kg & 58(14) & 29(6) & 6.0(40) & 14(3) & 30(18) & - & 18(4) \\
153 & Stainless Steel Tubes & Tube Supply Intl. & Cryostat & Lunehead & 0.92 & mBq/kg & $<$25 & $<$4.0 & $<$4.0 & $<$3.0 & $<$18 & - & 172(3) \\
154 & Stainless Steel Tubes & Tube Supply Intl. & Cryostat & Lunehead & 1.5 & mBq/kg & 40(20) & $<$3.0 & $<$4.0 & $<$5.0 & $<$17 & - & 15(1) \\
\end{tabular}
\end{adjustbox}
\end{sidewaystable*}
\clearpage
\thispagestyle{empty}
\newgeometry{margin=3cm}

\begin{sidewaystable*}[tp]
\begin{adjustbox}{width=\textwidth,center}
    \tabcolsep=4pt
    \centering
    \begin{tabular}{ H L{9cm} L{7cm} H C{2.cm} C{2.cm} C{2.cm}
    C{2.5cm} C{2.5cm} C{2.5cm} C{2.5cm} C{2.5cm} C{2.5cm} C{2.5cm}}

    \toprule
    \\
    \textbf{Item} & 
    \textbf{Sample}  & 
    \textbf{Supplier} &  
    \textbf{LZ Location}  &  
    \textbf{Detector}  & 
    \textbf{Mass [kg]}   &  
    \textbf{Units}  &  
    \textbf{\utTeE} & 
    \textbf{\utTeL} & 
    \textbf{\thtTtE} & 
    \textbf{\thtTtL} & 
    \textbf{\kfz} & 
    \textbf{\pbtoz} & 
    \textbf{\cosz} 
    \\
    \hline
    \hline
    \\
180 & Stainless Steel Mount & McMaster Carr & High Voltage & \textsc{Morgan} & 0.32 & mBq/kg & $<$19 & $<$3.6 & $<$6.0 & $<$4.1 & $<$20 & - & 24(7) \\
181 & Pure and Ceriated Tungsten Welding Electrodes & McMaster Carr & High Voltage & \textsc{Merlin} & 0.30 & mBq/kg & $<$52 & 156(35) & 506(107) & 303(34) & $<$102 & - & - \\
182 & Stainless Steel Welding Wire & McMaster Carr & High Voltage & \textsc{Merlin} & 0.49 & mBq/kg & $<$28 & $<$12 & $<$13 & 34(15) & $<$36 & - & - \\
183 & PEEK Cylinder 2-1/2 Diameter & McMaster Carr & High Voltage & \textsc{Merlin} & 0.63 & mBq/kg & $<$34 & $<$17 & $<$17 & $<$13 & $<$45 & - & - \\
184 & FEP Compression To Conflat Adaptor & McMaster Carr & High Voltage & \textsc{Merlin} & - & mBq/Unit & $<$64 & $<$3.0 & 89(9) & 101(6) & $<$10 & - & - \\
\end{tabular}
\end{adjustbox}
\end{sidewaystable*}
\clearpage
\thispagestyle{empty}
\newgeometry{margin=3cm}

\begin{sidewaystable*}[tp]
\begin{adjustbox}{width=\textwidth,center}
    \tabcolsep=4pt
    \centering
    \begin{tabular}{ H L{9cm} L{7cm} H C{2.cm} C{2.cm} C{2.cm}
    C{2.5cm} C{2.5cm} C{2.5cm} C{2.5cm} C{2.5cm} C{2.5cm} C{2.5cm}}

    \toprule
    \\
    \textbf{Item} & 
    \textbf{Sample}  & 
    \textbf{Supplier} &  
    \textbf{LZ Location}  &  
    \textbf{Detector}  & 
    \textbf{Mass [kg]}   &  
    \textbf{Units}  &  
    \textbf{\utTeE} & 
    \textbf{\utTeL} & 
    \textbf{\thtTtE} & 
    \textbf{\thtTtL} & 
    \textbf{\kfz} & 
    \textbf{\pbtoz} & 
    \textbf{\cosz} 
    \\
    \hline
    \hline
    \\
210 & OFHC Copper & City Special Metals & Misc & \textsc{Merlin} & 2.3 & mBq/kg & $<$15 & $<$11 & $<$10 & $<$8.0 & $<$28 & $<$11 & - \\
211 & Deionised Water From SURF Faucet & N/A & Misc & Roseberry & 1.2 & mBq/kg & 2.0(10) & 3.0(10) & 2.0(10) & 1.1(20) & 9.0(60) & 19(6) & - \\
212 & Superwash 2000 & Imperial College London Fabrication & Misc & Chaloner & 1.1 & mBq/kg & $<$7.0 & $<$2.0 & $<$3.0 & 2.0(10) & 1280(40) & $<$18 & - \\
213 & Citranox & Alconox & Misc & Chaloner & 1.1 & mBq/kg & $<$6.0 & 3.0(20) & $<$4.0 & $<$2.0 & 147(25) & $<$18 & - \\
214 & Deionised Water & Imperial College London Fabrication & Misc & Chaloner & 1.1 & mBq/kg & $<$6.0 & 3.0(20) & $<$3.0 & $<$2.0 & $<$25 & $<$16 & - \\
\end{tabular}
\end{adjustbox}
\end{sidewaystable*}
\clearpage
\thispagestyle{empty}
\newgeometry{margin=3cm}

\begin{sidewaystable*}[tp]
\begin{adjustbox}{width=\textwidth,center}
    \tabcolsep=4pt
    \centering
    \begin{tabular}{ H L{9cm} L{7cm} H C{2.cm} C{2.cm} C{2.cm}
    C{2.5cm} C{2.5cm} C{2.5cm} C{2.5cm} C{2.5cm} C{2.5cm} C{2.5cm}}

    \toprule
    \\
    \textbf{Item} & 
    \textbf{Sample}  & 
    \textbf{Supplier} &  
    \textbf{LZ Location}  &  
    \textbf{Detector}  & 
    \textbf{Mass [kg]}   &  
    \textbf{Units}  &  
    \textbf{\utTeE} & 
    \textbf{\utTeL} & 
    \textbf{\thtTtE} & 
    \textbf{\thtTtL} & 
    \textbf{\kfz} & 
    \textbf{\pbtoz} & 
    \textbf{\cosz} 
    \\
    \hline
    \hline
    \\
240 & Fiberoptic Cable & Super ESKA & Outer Detector & Lumpsey & 0.94 & mBq/kg & $<$74 & 22(6) & 14(12) & 13(6) & 411(28) & $<$210 & $<$1.5 \\
241 & Alconox Detegent-8 & Alconox & Outer Detector & Chaloner & 0.32 & mBq/kg & $<$15 & $<$5.0 & $<$7.0 & $<$2.0 & 86000(1300) & $<$46 & $<$2.0 \\
242 & Deionized Water & UCSB Fabrication & Outer Detector & Chaloner & 0.88 & mBq/kg & 8.0(110) & 2.0(40) & $<$2.0 & 6.0(20) & 21(31) & $<$11 & $<$0.50 \\
243 & Bonding Syrup & Reynolds & Outer Detector & Chaloner & 0.41 & mBq/kg & 12(21) & 11(8) & 13(11) & 6.0(30) & 510(70) & $<$18 & 4.0(30) \\
244 & Machining Water & UCSB Fabrication & Outer Detector & Chaloner & 0.82 & mBq/kg & 7.0(90) & 3.0(40) & 5.0(50) & 3.0(20) & 100(30) & $<$9.0 & $<$0.70 \\
\end{tabular}
\end{adjustbox}
\end{sidewaystable*}
\clearpage
\thispagestyle{empty}
\newgeometry{margin=3cm}

\begin{sidewaystable*}[tp]
\begin{adjustbox}{width=\textwidth,center}
    \tabcolsep=4pt
    \centering
    \begin{tabular}{ H L{9cm} L{7cm} H C{2.cm} C{2.cm} C{2.cm}
    C{2.5cm} C{2.5cm} C{2.5cm} C{2.5cm} C{2.5cm} C{2.5cm} C{2.5cm}}

    \toprule
    \\
    \textbf{Item} & 
    \textbf{Sample}  & 
    \textbf{Supplier} &  
    \textbf{LZ Location}  &  
    \textbf{Detector}  & 
    \textbf{Mass [kg]}   &  
    \textbf{Units}  &  
    \textbf{\utTeE} & 
    \textbf{\utTeL} & 
    \textbf{\thtTtE} & 
    \textbf{\thtTtL} & 
    \textbf{\kfz} & 
    \textbf{\pbtoz} & 
    \textbf{\cosz} 
    \\
    \hline
    \hline
    \\
270 & Resistors VISHAY 7M5 - 0805 & Farnell & PMT & Chaloner & 0.01 & mBq/kg & 1097(102) & 389(47) & 106(22) & 106(22) & 3082(608) & 41600(830) & $<$16 \\
271 & PMT Pin Receptacles & TE Connectivity & PMT & Lunehead & 0.21 & mBq/kg & 1706(770) & $<$15 & $<$21 & $<$21 & 159(240) & - & $<$5.1 \\
272 & Receptacles For PMT HV & Mill-Max & PMT & Chaloner & 0.02 & mBq/kg & 1252(220) & $<$24 & $<$17 & $<$17 & $<$107 & 34971(1283) & 18(8) \\
274 & R11410-20 Insulator Plate (2013) & Hamamatsu & PMT & SOLO & 0.21 & mBq/kg & $<$306 & $<$10 & $<$16 & $<$16 & $<$104 & - & - \\
275 & R11410-20 Metal Bulb (2013) & Hamamatsu & PMT & SOLO & 0.20 & mBq/kg & $<$389 & $<$7.7 & $<$10 & $<$10 & $<$72 & - & $<$2.1 \\
\end{tabular}
\end{adjustbox}
\end{sidewaystable*}
\clearpage
\thispagestyle{empty}
\newgeometry{margin=3cm}

\begin{sidewaystable*}[tp]
\begin{adjustbox}{width=\textwidth,center}
    \tabcolsep=4pt
    \centering
    \begin{tabular}{ H L{9cm} L{7cm} H C{2.cm} C{2.cm} C{2.cm}
    C{2.5cm} C{2.5cm} C{2.5cm} C{2.5cm} C{2.5cm} C{2.5cm} C{2.5cm}}

    \toprule
    \\
    \textbf{Item} & 
    \textbf{Sample}  & 
    \textbf{Supplier} &  
    \textbf{LZ Location}  &  
    \textbf{Detector}  & 
    \textbf{Mass [kg]}   &  
    \textbf{Units}  &  
    \textbf{\utTeE} & 
    \textbf{\utTeL} & 
    \textbf{\thtTtE} & 
    \textbf{\thtTtL} & 
    \textbf{\kfz} & 
    \textbf{\pbtoz} & 
    \textbf{\cosz} 
    \\
    \hline
    \hline
    \\
300 & R11410-20 Faceplate (2015) & Hamamatsu & PMT & \textsc{Maeve} & 1.2 & mBq/kg & $<$11 & $<$0.67 & $<$1.0 & $<$0.80 & $<$4.0 & - & - \\
301 & R11410-20 Insulator Plate (2015) & Hamamatsu & PMT & \textsc{Maeve} & 0.84 & mBq/kg & $<$21 & $<$1.0 & $<$1.6 & $<$1.2 & $<$6.3 & - & - \\
302 & R11410-20 Electrode Disk (2015) & Hamamatsu & PMT & SOLO & 0.52 & mBq/kg & $<$203 & 9.5(36) & $<$4.3 & 14(4) & $<$9.6 & - & 8.5(21) \\
303 & R11410-20 Faceplate Flange (2015) & Hamamatsu & PMT & SOLO & 0.53 & mBq/kg & $<$162 & $<$2.8 & $<$3.8 & $<$4.2 & $<$14 & - & 13(3) \\
304 & R11410-20 Ceramic Stem (2015) & Hamamatsu & PMT & SOLO & 0.76 & mBq/kg & 247(175) & 20(4) & 18(5) & 11(4) & 96(21) & - & - \\
\end{tabular}
\end{adjustbox}
\end{sidewaystable*}
\clearpage
\thispagestyle{empty}
\newgeometry{margin=3cm}

\begin{sidewaystable*}[tp]
\begin{adjustbox}{width=\textwidth,center}
    \tabcolsep=4pt
    \centering
    \begin{tabular}{ H L{9cm} L{7cm} H C{2.cm} C{2.cm} C{2.cm}
    C{2.5cm} C{2.5cm} C{2.5cm} C{2.5cm} C{2.5cm} C{2.5cm} C{2.5cm}}

    \toprule
    \\
    \textbf{Item} & 
    \textbf{Sample}  & 
    \textbf{Supplier} &  
    \textbf{LZ Location}  &  
    \textbf{Detector}  & 
    \textbf{Mass [kg]}   &  
    \textbf{Units}  &  
    \textbf{\utTeE} & 
    \textbf{\utTeL} & 
    \textbf{\thtTtE} & 
    \textbf{\thtTtL} & 
    \textbf{\kfz} & 
    \textbf{\pbtoz} & 
    \textbf{\cosz} 
    \\
    \hline
    \hline
    \\
330 & PEEK Fasteners & McMaster Carr & PMT & \textsc{Merlin} & 0.48 & mBq/kg & $<$19 & 17(8) & $<$10 & $<$6.0 & $<$45 & - & - \\
331 & Set Screws & RediVac & PMT & \textsc{Merlin} & 0.07 & mBq/kg & $<$388 & $<$19 & $<$24 & $<$10 & $<$186 & - & - \\
332 & Teflon Screws & Craftech & PMT & \textsc{Morgan} & 0.94 & mBq/kg & $<$18 & 3.9(13) & 2.9(16) & 9.8(33) & 839(24) & - & - \\
333 & Kovar KV-2 & Hamamatsu & PMT & \textsc{Maeve} & 0.58 & mBq/kg & $<$36 & 7.0(20) & $<$7.4 & $<$3.1 & $<$15 & - & - \\
334 & Co-Free Kovar YFE42 & Hamamatsu & PMT & \textsc{Morgan} & 0.51 & mBq/kg & $<$48 & $<$2.2 & $<$4.2 & $<$2.2 & 64(18) & - & - \\
\end{tabular}
\end{adjustbox}
\end{sidewaystable*}
\clearpage
\thispagestyle{empty}
\newgeometry{margin=3cm}

\begin{sidewaystable*}[tp]
\begin{adjustbox}{width=\textwidth,center}
    \tabcolsep=4pt
    \centering
    \begin{tabular}{ H L{9cm} L{7cm} H C{2.cm} C{2.cm} C{2.cm}
    C{2.5cm} C{2.5cm} C{2.5cm} C{2.5cm} C{2.5cm} C{2.5cm} C{2.5cm}}

    \toprule
    \\
    \textbf{Item} & 
    \textbf{Sample}  & 
    \textbf{Supplier} &  
    \textbf{LZ Location}  &  
    \textbf{Detector}  & 
    \textbf{Mass [kg]}   &  
    \textbf{Units}  &  
    \textbf{\utTeE} & 
    \textbf{\utTeL} & 
    \textbf{\thtTtE} & 
    \textbf{\thtTtL} & 
    \textbf{\kfz} & 
    \textbf{\pbtoz} & 
    \textbf{\cosz} 
    \\
    \hline
    \hline
    \\
360 & PTFE Screws & MMC & PMT & \textsc{Merlin} & 0.06 & mBq/kg & $<$194 & $<$83 & $<$100 & 134(117) & $<$306 & - & - \\
361 & PEEK Nuts and Washers & Solid Spot & PMT & \textsc{Merlin} & 0.53 & mBq/kg & $<$45 & $<$10 & $<$12 & 37(17) & $<$49 & - & - \\
362 & PEEK Hex Nuts & McMaster Carr & PMT & \textsc{Merlin} & 0.01 & mBq/kg & 12000(5000) & 9000(700) & 6200(1000) & 8200(700) & 13000(5000) & - & - \\
363 & 1-inch PMT Batch 1 & Hamamatsu & PMT & Roseberry & - & mBq/Unit & 0.34(11) & $<$0.12 & $<$0.06 & 0.08(2) & 6.4(5) & $<$0.60 & 0.30(3) \\
364 & 1-inch PMT Batch 2 & Hamamatsu & PMT & Roseberry & - & mBq/Unit & 0.16(6) & $<$0.08 & $<$0.05 & 0.05(1) & 6.5(4) & 0.90(20) & 0.33(2) \\
\end{tabular}
\end{adjustbox}
\end{sidewaystable*}
\clearpage
\thispagestyle{empty}
\newgeometry{margin=3cm}

\begin{sidewaystable*}[tp]
\begin{adjustbox}{width=\textwidth,center}
    \tabcolsep=4pt
    \centering
    \begin{tabular}{ H L{9cm} L{7cm} H C{2.cm} C{2.cm} C{2.cm}
    C{2.5cm} C{2.5cm} C{2.5cm} C{2.5cm} C{2.5cm} C{2.5cm} C{2.5cm}}

    \toprule
    \\
    \textbf{Item} & 
    \textbf{Sample}  & 
    \textbf{Supplier} &  
    \textbf{LZ Location}  &  
    \textbf{Detector}  & 
    \textbf{Mass [kg]}   &  
    \textbf{Units}  &  
    \textbf{\utTeE} & 
    \textbf{\utTeL} & 
    \textbf{\thtTtE} & 
    \textbf{\thtTtL} & 
    \textbf{\kfz} & 
    \textbf{\pbtoz} & 
    \textbf{\cosz} 
    \\
    \hline
    \hline
    \\
390 & LZ PMT Cable Offcuts (1) & Axon & PMT Cabling & Chaloner & - & mBq/m & $<$0.15 & $<$0.15 & $<$0.10 & $<$0.03 & $<$1.2 & $<$0.25 & $<$0.06 \\
391 & LZ PMT Cable Offcuts (2) & Axon & PMT Cabling & Chaloner & - & mBq/m & $<$0.21 & $<$0.20 & $<$0.15 & $<$0.04 & $<$0.98 & $<$0.58 & $<$0.06 \\
392 & LZ PMT Cable Offcuts (3) & Axon & PMT Cabling & Chaloner & - & mBq/m & $<$0.22 & $<$0.20 & $<$0.12 & $<$0.04 & $<$0.91 & $<$0.45 & $<$0.04 \\
393 & LZ PMT Cable Offcuts (4) & Axon & PMT Cabling & Chaloner & - & mBq/m & $<$0.12 & $<$0.10 & $<$0.10 & $<$0.05 & $<$0.59 & $<$0.36 & 0.03(1) \\
394 & FEP Pellets & Axon & PMT Cabling & Roseberry & 0.67 & mBq/kg & $<$0.50 & 6.6(90) & $<$0.80 & 0.40(10) & 6.6(3) & $<$2.6 & $<$0.30 \\
\end{tabular}
\end{adjustbox}
\end{sidewaystable*}
\clearpage
\thispagestyle{empty}
\newgeometry{margin=3cm}

\begin{sidewaystable*}[tp]
\begin{adjustbox}{width=\textwidth,center}
    \tabcolsep=4pt
    \centering
    \begin{tabular}{ H L{9cm} L{7cm} H C{2.cm} C{2.cm} C{2.cm}
    C{2.5cm} C{2.5cm} C{2.5cm} C{2.5cm} C{2.5cm} C{2.5cm} C{2.5cm}}

    \toprule
    \\
    \textbf{Item} & 
    \textbf{Sample}  & 
    \textbf{Supplier} &  
    \textbf{LZ Location}  &  
    \textbf{Detector}  & 
    \textbf{Mass [kg]}   &  
    \textbf{Units}  &  
    \textbf{\utTeE} & 
    \textbf{\utTeL} & 
    \textbf{\thtTtE} & 
    \textbf{\thtTtL} & 
    \textbf{\kfz} & 
    \textbf{\pbtoz} & 
    \textbf{\cosz} 
    \\
    \hline
    \hline
    \\
420 & PEEK Stock Material & Boedeker Plastics Inc & TPC & \textsc{Merlin} & 1.0 & mBq/kg & $<$12 & $<$12 & $<$4.1 & $<$4.1 & $<$31 & - & - \\
421 & PEEK For TPC & TECAPEEK & TPC & \textsc{Merlin} & 0.15 & mBq/kg & 741(247) & 198(37) & $<$41 & $<$41 & 402(124) & - & - \\
422 & PEEK Flat Stock Material & Boedeker Plastics Inc & TPC & SOLO & 2.4 & mBq/kg & $<$14 & 4.6(7) & 5.0(10) & 3.0(3) & 19(5) & - & $<$0.20 \\
423 & Silicon Dioxide Powder & American Elements & TPC & \textsc{Merlin} & 0.33 & mBq/kg & 617(123) & 49(12) & $<$4.1 & 28(16) & $<$31 & - & - \\
424 & Synthetic Alumina Powder & Taimei & TPC & \textsc{Morgan} & 0.33 & mBq/kg & $<$63 & $<$35 & 4.8(41) & $<$2.5 & 61(26) & - & - \\
\end{tabular}
\end{adjustbox}
\end{sidewaystable*}
\clearpage
\thispagestyle{empty}
\newgeometry{margin=3cm}

\begin{sidewaystable*}[tp]
\begin{adjustbox}{width=\textwidth,center}
    \tabcolsep=4pt
    \centering
    \begin{tabular}{ H L{9cm} L{7cm} H C{2.cm} C{2.cm} C{2.cm}
    C{2.5cm} C{2.5cm} C{2.5cm} C{2.5cm} C{2.5cm} C{2.5cm} C{2.5cm}}

    \toprule
    \\
    \textbf{Item} & 
    \textbf{Sample}  & 
    \textbf{Supplier} &  
    \textbf{LZ Location}  &  
    \textbf{Detector}  & 
    \textbf{Mass [kg]}   &  
    \textbf{Units}  &  
    \textbf{\utTeE} & 
    \textbf{\utTeL} & 
    \textbf{\thtTtE} & 
    \textbf{\thtTtL} & 
    \textbf{\kfz} & 
    \textbf{\pbtoz} & 
    \textbf{\cosz} 
    \\
    \hline
    \hline
    \\
450 & Grading Resistors & Riedon & TPC & \textsc{Merlin} & 0.09 & mBq/kg & 4100(400) & 3700(200) & 5300(200) & 5900(200) & 4500(700) & - & - \\
451 & PTFE Washers & Craftech & TPC & \textsc{Merlin} & 0.42 & mBq/kg & $<$50 & $<$19 & $<$22 & $<$12 & 820(115) & - & - \\
452 & PEEK Virgin Natural Sheet 1.000 & Boedeker Plastics Inc & TPC & \textsc{Merlin} & 2.4 & mBq/kg & $<$10 & 16(3) & 18(6) & 13(3) & 39(6) & 274(37) & - \\
453 & PEEK Virgin Natural Sheet 1.500 & Boedeker Plastics Inc & TPC & \textsc{Merlin} & 2.1 & mBq/kg & $<$10 & $<$5.0 & $<$6.0 & $<$6.0 & $<$31 & - & - \\
454 & PTFE Powder & Dupont & TPC & Roseberry & 0.07 & mBq/kg & 8.0(50) & 13(5) & 4.0(50) & 5.0(20) & 45(25) & $<$22 & $<$3.0 \\
\end{tabular}
\end{adjustbox}
\end{sidewaystable*}
\clearpage
\thispagestyle{empty}
\newgeometry{margin=3cm}

\begin{sidewaystable*}[tp]
\begin{adjustbox}{width=\textwidth,center}
    \tabcolsep=4pt
    \centering
    \begin{tabular}{ H L{9cm} L{7cm} H C{2.cm} C{2.cm} C{2.cm}
    C{2.5cm} C{2.5cm} C{2.5cm} C{2.5cm} C{2.5cm} C{2.5cm} C{2.5cm}}

    \toprule
    \\
    \textbf{Item} & 
    \textbf{Sample}  & 
    \textbf{Supplier} &  
    \textbf{LZ Location}  &  
    \textbf{Detector}  & 
    \textbf{Mass [kg]}   &  
    \textbf{Units}  &  
    \textbf{\utTeE} & 
    \textbf{\utTeL} & 
    \textbf{\thtTtE} & 
    \textbf{\thtTtL} & 
    \textbf{\kfz} & 
    \textbf{\pbtoz} & 
    \textbf{\cosz} 
    \\
    \hline
    \hline
    \\
480 & \multirow{2}{*}{Harwin Pins} & \multirow{2}{*}{Digikey} & TPC & Lunehead & \multirow{2}{*}{0.01} & mBq/kg & 450(220) & 200(90) & $<$100 & 80(25) & $<$630 & - & $<$40 \\
481 & & & TPC & Merrybent & & mBq/kg & 420(260) & 730(210) & $<$40 & 90(30) & $<$580 & - & $<$17 \\
482 & Stainless Steel Fasteners & McMaster Carr & TPC & \textsc{Merlin} & 2.5 & mBq/kg & 26(19) & $<$2.5 & 2.4(50) & 4.0(30) & $<$9.3 & $<$36 & 6.5(5) \\
483 & PEEK Fasteners & Solid Spot & TPC & \textsc{Merlin} & 0.53 & mBq/kg & $<$45 & $<$10 & $<$12 & 37(17) & $<$49 & - & - \\
484 & PEEK Resistor Tap & LBNL Fabrication & TPC & \textsc{Merlin} & - & mBq/Unit & $<$2.3 & $<$0.60 & $<$1.3 & $<$0.90 & $<$4.1 & - & - \\
\end{tabular}
\end{adjustbox}
\end{sidewaystable*}
\clearpage
\thispagestyle{empty}
\newgeometry{margin=3cm}

\begin{sidewaystable*}[tp]
\begin{adjustbox}{width=\textwidth,center}
    \tabcolsep=4pt
    \centering
    \begin{tabular}{ H L{9cm} L{7cm} H C{2.cm} C{2.cm} C{2.cm}
    C{2.5cm} C{2.5cm} C{2.5cm} C{2.5cm} C{2.5cm} C{2.5cm} C{2.5cm}}

    \toprule
    \\
    \textbf{Item} & 
    \textbf{Sample}  & 
    \textbf{Supplier} &  
    \textbf{LZ Location}  &  
    \textbf{Detector}  & 
    \textbf{Mass [kg]}   &  
    \textbf{Units}  &  
    \textbf{\utTeE} & 
    \textbf{\utTeL} & 
    \textbf{\thtTtE} & 
    \textbf{\thtTtL} & 
    \textbf{\kfz} & 
    \textbf{\pbtoz} & 
    \textbf{\cosz} 
    \\
    \hline
    \hline
    \\
510 & Position Sensors & University of Oxford Fabrication & TPC & \textsc{Maeve} & - & mBq/Unit & $<$3.1 & 5.3(3) & $<$1.0 & $<$0.64 & $<$1.9 & - & - \\
511 & Weir Precision Sensors (WPS) & University of Oxford Fabrication & TPC & \textsc{Maeve} & - & mBq/Unit & $<$1.9 & 3.7(5) & 2.3(4) & 2.9(3) & 19(2) & - & - \\
512 & PEEK Grid Spacer Replacement Parts & Boedeker & TPC & \textsc{Merlin} & - & mBq/Unit & $<$0.80 & $<$0.30 & $<$0.50 & $<$0.50 & $<$1.4 & - & - \\
513 & Acoustic Sensors & Measurement Specialties & TPC & \textsc{Maeve} & 0.18 & mBq/kg & $<$433 & $<$58 & $<$78 & 71(67) & $<$216 & - & - \\
514 & Hardware For Vent Tube Assembly & McMaster Carr & TPC & \textsc{Morgan} & 0.48 & mBq/kg & $<$39 & $<$4.9 & $<$6.5 & $<$4.4 & $<$28 & - & 27(8) \\
\hline

\end{tabular}
\end{adjustbox}
\end{sidewaystable*}

\clearpage
\thispagestyle{empty}
\newgeometry{margin=3cm}

\begin{sidewaystable*}[tp]
\caption{ICP-MS Screening Results.\label{tab:ICPMSBig}}
\begin{adjustbox}{width=\textwidth,center}
    \tabcolsep=4pt
    \centering
    \begin{tabular}{ H L{12cm} L{7cm} H C{7.cm} H C{2.cm}
    C{2.5cm} H C{2.5cm} H H H H H H }
    \toprule
    \\
    \textbf{Item} & 
    \textbf{Sample}  & 
    \textbf{Supplier} &  
    \textbf{LZ Location}  &  
    \textbf{Detector}  & 
    \textbf{Mass [kg]}   &  
    \textbf{Units}  &  
    \textbf{\utTe} & 
    \textbf{\utTeL} & 
    \textbf{\thtTt} & 
    \textbf{\thtTtL} & 
    \textbf{\kfz} & 
    \textbf{\pbtoz} & 
    \textbf{\cosz} & 
    \textbf{\csoTS}& 
    \textbf{\utTF} 
    \\
    \hline
    \hline
    \\
\\
\multicolumn{15}{l}{\textbf{Cryostat}}\\
\hline \\
0 & Titanium Welding Wire 1.6 mm & Timet (LZ Stock) & Cryostat & UCL & - & mBq/kg & 2.6(4) & - & 0.63(6) & - & - & - & - & - & - \\
1 & Titanium Welding Wire 1.6 mm & Timet (LZ Stock) & Cryostat & UCL & - & mBq/kg & 66(1) & - & 13(1) & - & - & - & - & - & - \\
\end{tabular}
\end{adjustbox}
\end{sidewaystable*}
\clearpage
\thispagestyle{empty}
\newgeometry{margin=3cm}

\begin{sidewaystable*}[tp]
\begin{adjustbox}{width=\textwidth,center}
    \tabcolsep=4pt
    \centering
    \begin{tabular}{ H L{12cm} L{7cm} H C{7.cm} H C{2.cm}
    C{2.5cm} H C{2.5cm} H H H H H H }
    \toprule
    \\
    \textbf{Item} & 
    \textbf{Sample}  & 
    \textbf{Supplier} &  
    \textbf{LZ Location}  &  
    \textbf{Detector}  & 
    \textbf{Mass [kg]}   &  
    \textbf{Units}  &  
    \textbf{\utTe} & 
    \textbf{\utTeL} & 
    \textbf{\thtTt} & 
    \textbf{\thtTtL} & 
    \textbf{\kfz} & 
    \textbf{\pbtoz} & 
    \textbf{\cosz} & 
    \textbf{\csoTS}& 
    \textbf{\utTF} 
    \\
    \hline
    \hline
    \\
20 & Ti Cryostat Weld Chips; TIG Welds; Uncleaned & Timet (LZ Stock) & Cryostat & UCL & - & mBq/kg & 4.1(3) & - & 1.5(2) & - & - & - & - & - & - \\
21 & Welding Electrode; 4.0 mm W-La From Wolfram; Etched & Timet (LZ Stock) & Cryostat & UCL & - & mBq/kg & 2.1(2) & - & 4.7(7) & - & - & - & - & - & - \\
22 & Welding Electrode; 4.0 mm W-La From Wolfram; Cleaned & Timet (LZ Stock) & Cryostat & UCL & - & mBq/kg & 2.1(2) & - & 5.9(3) & - & - & - & - & - & - \\
23 & Welding Electrode; 4.0 mm W-La From Wolfram; Uncleaned & Timet (LZ Stock) & Cryostat & UCL & - & mBq/kg & 2.1(2) & - & 6.2(2) & - & - & - & - & - & - \\
24 & Welding Electrode; Pure W From Plansee; Etched & Timet (LZ Stock) & Cryostat & UCL & - & mBq/kg & 2.1(1) & - & 36(2) & - & - & - & - & - & - \\
\end{tabular}
\end{adjustbox}
\end{sidewaystable*}
\clearpage
\thispagestyle{empty}
\newgeometry{margin=3cm}

\begin{sidewaystable*}[tp]
\begin{adjustbox}{width=\textwidth,center}
    \tabcolsep=4pt
    \centering
    \begin{tabular}{ H L{12cm} L{7cm} H C{7.cm} H C{2.cm}
    C{2.5cm} H C{2.5cm} H H H H H H }
    \toprule
    \\
    \textbf{Item} & 
    \textbf{Sample}  & 
    \textbf{Supplier} &  
    \textbf{LZ Location}  &  
    \textbf{Detector}  & 
    \textbf{Mass [kg]}   &  
    \textbf{Units}  &  
    \textbf{\utTe} & 
    \textbf{\utTeL} & 
    \textbf{\thtTt} & 
    \textbf{\thtTtL} & 
    \textbf{\kfz} & 
    \textbf{\pbtoz} & 
    \textbf{\cosz} & 
    \textbf{\csoTS}& 
    \textbf{\utTF} 
    \\
    \hline
    \hline
    \\
40 & Polystyrene Foam & Styrodure & Cryostat & UCL & - & mBq/kg & 3.3(10) & - & 0.57(21) & - & - & - & - & - & - \\
41 & Forward Field Region Titanium Weld Sample & Timet & Cryostat & UCL & - & mBq/kg & 2.1(2) & - & 0.14(4) & - & - & - & - & - & - \\
42 & Forward Field Region Titanium Weld Sample & Timet & Cryostat & UCL & - & mBq/kg & 2.4(3) & - & 4.3(2) & - & - & - & - & - & - \\
43 & Ti weld sample from CHAMP; Material Adjacent To Weld & Timet & Cryostat & UCL & - & mBq/kg & 2.3(4) & - & 0.59(13) & - & - & - & - & - & - \\
44 & Ti weld sample from CHAMP; Weld Material & Timet & Cryostat & UCL & - & mBq/kg & 2.9(3) & - & 15(1) & - & - & - & - & - & - \\
\hline

\end{tabular}
\end{adjustbox}
\end{sidewaystable*}

\clearpage
\thispagestyle{empty}
\newgeometry{margin=3cm}

\begin{sidewaystable*}[tp]
\caption{GDMS Screening Results.\label{tab:GDMSBig}}
\begin{adjustbox}{width=\textwidth,center}
    \tabcolsep=4pt
    \centering
    \begin{tabular}{ H L{9cm} L{7cm} H C{2.cm} C{2.cm} C{2.cm}
    C{2.5cm} H C{2.5cm} H C{2.5cm} H H H H}

    \toprule
    \\
    \textbf{Item} & 
    \textbf{Sample}  & 
    \textbf{Supplier} &  
    \textbf{LZ Location}  &  
    \textbf{Detector}  & 
    \textbf{Mass [g]}   &  
    \textbf{Units}  &  
    \textbf{\utTeE} & 
    \textbf{\utTeL} & 
    \textbf{\thtTtE} & 
    \textbf{\thtTtL} & 
    \textbf{\kfz} & 
    \textbf{\pbtoz} & 
    \textbf{\cosz} & 
    \textbf{\csoTS}& 
    \textbf{\utTF} 
    \\
    \hline
    \hline
    \\
\\
\multicolumn{15}{l}{\textbf{Cryostat}}\\
\hline \\
0 & Ti Plate HN3469 & Timet & Cryostat & NRC & 0.35 & mBq/kg & $<$2.5 & - & $<$0.81 & - & $<$0.31 & - & - & - & - \\
1 & Ti Plate HN3469 & Timet & Cryostat & NRC & 0.35 & mBq/kg & $<$2.5 & - & $<$0.81 & - & $<$0.34 & - & - & - & - \\
\hline

\end{tabular}
\end{adjustbox}
\end{sidewaystable*}

\clearpage
\thispagestyle{empty}
\newgeometry{margin=3cm}

\begin{sidewaystable*}[tp]
\caption{NAA Screening Results.\label{tab:NAABig}}
\begin{adjustbox}{width=\textwidth,center}
    \tabcolsep=4pt
    \centering
    \begin{tabular}{ H L{9cm} L{7cm} H C{7cm} C{2.cm} C{2.cm}
    C{2.5cm} H C{2.5cm} H C{2.5cm} H H H H}

    \toprule
    \\
    \textbf{Item} & 
    \textbf{Sample}  & 
    \textbf{Supplier} &  
    \textbf{LZ Location}  &  
    \textbf{Detector}  & 
    \textbf{Mass [g]}   &  
    \textbf{Units}  &  
    \textbf{\utTeE} & 
    \textbf{\utTeL} & 
    \textbf{\thtTtE} & 
    \textbf{\thtTtL} & 
    \textbf{\kfz} & 
    \textbf{\pbtoz} & 
    \textbf{\cosz} & 
    \textbf{\csoTS}& 
    \textbf{\utTF} 
    \\
    \hline
    \hline
    \\
\\
\multicolumn{15}{l}{\textbf{PMT Cabling}}\\
\hline \\
0 & Raw FEP Pellets & Axon & PMT cabling & University of Alabama & 3.1 & mBq/kg & $<$0.017 & - & 0.036(2) & - & 0.168(4) & - & - & - \\
1 & Raw FEP Pellets & Axon & PMT cabling & University of Alabama & 3.1 & mBq/kg & $<$0.005 & - & 0.037(2) & - & 0.175(4) & - & - & - \\
2 & FEP Inner Cable Jacket & Axon & PMT cabling & University of Alabama & 2.6 & mBq/kg & $<$0.057 & - & 0.083(8) & - & 5.053(78) & - & - & - \\
3 & FEP Outer Cable Jacket & Axon & PMT cabling & University of Alabama & 2.6 & mBq/kg & $<$0.024 & - & 0.096(6) & - & 10.7(20) & - & - & - \\
\\
\multicolumn{15}{l}{\textbf{TPC}}\\
\hline \\
4 & PTFE 8764 For Skin Region & Boedeker Plastics & TPC & University of Alabama & 8.0 & mBq/kg & 0.018(40) & - & 0.029(1) & - & 0.076(1) & - & - & - & - \\
5 & PTFE FLON008 & Flontech & TPC & University of Alabama & 2.8 & mBq/kg & $<$0.027 & - & 0.051(3) & - & 0.33(2) & - & - & - & - \\
6 & Teflon NXT85 & DuPont & TPC & University of Alabama & 2.9 & mBq/kg & $<$0.021 & - & 0.028(2) & - & 0.122(10) & - & - & - \\
7 & Teflon 807NX & DuPont & TPC & University of Alabama & 2.8 & mBq/kg & 0.038(10) & - & 0.029(2) & - & 0.096(5) & - & - & - \\
8 & PTFE 8764 & Boedeker Plastics & TPC & University of Alabama & 2.8 & mBq/kg & $<$0.043 & - & 0.013(3) & - & 0.066(4) & - & - & - \\
\hline

\end{tabular}
\end{adjustbox}
\end{sidewaystable*}

\clearpage
\thispagestyle{empty}
\newgeometry{margin=3cm}

\begin{sidewaystable*}[tp]
\caption{Radon Emanation Results.\label{tab:RnBig}}
\begin{adjustbox}{width=\textwidth,center}
    \tabcolsep=4pt
    \centering
    \begin{tabular}{H L{9cm} L{7cm} H C{7.cm} C{4.cm} C{2.cm} C{3.5cm}}
    \toprule
    \\
    \textbf{ID}  & 
    \textbf{Sample}  & 
    \textbf{Supplier} &  
    \textbf{LZ Location}  &  
    \textbf{Detector}  & 
    \textbf{Quantity Screened}   &  
    \textbf{Units}  &  
    \textbf{\rnttt{}} 
    \\
    \hline
    \hline
    \\
\\
\multicolumn{8}{l}{\textbf{Calibration}}\\
\hline \\
0 & Welding Rods & HTP from USAWeld & Calibration & SDSM\&T & 1 unit & \si{\micro\becquerel}/unit & $185^{+104}_{-81}$  \\
\\
\end{tabular}
\end{adjustbox}
\end{sidewaystable*}
\clearpage
\thispagestyle{empty}
\newgeometry{margin=3cm}

\begin{sidewaystable*}[tp]
\begin{adjustbox}{width=\textwidth,center}
    \tabcolsep=4pt
    \centering
    \begin{tabular}{H L{9cm} L{7cm} H C{7.cm} C{4.cm} C{2.cm} C{3.5cm}}
    \toprule
    \\
    \textbf{ID}  & 
    \textbf{Sample}  & 
    \textbf{Supplier} &  
    \textbf{LZ Location}  &  
    \textbf{Detector}  & 
    \textbf{Quantity Screened}   &  
    \textbf{Units}  &  
    \textbf{\rnttt{}} 
    \\
    \hline
    \hline
    \\
15 & Delrin (Polyoxymethylene) & DuPont & High Voltage & UCL & 0.54 m$^{2}$ & \si{\micro\becquerel}/m$^{2}$ & $440\pm116$ \\
16 & Viton O-Rings & Dichtomatik & High Voltage & UCL & 90 unit & \si{\micro\becquerel}/unit & $17\pm2.3$ \\
17 & HV Cathode System Components & Various & High Voltage & UCL & 1 unit & \si{\micro\becquerel}/batch & $369\pm85$ \\
18 & Umbilical Cables & Dielectric Sciences & High Voltage & SDSM\&T & 25 m & \si{\micro\becquerel}/m & $52^{+8.1}_{-12}$ \\
19 & HV Umbilical Epoxy & Stycast & High Voltage & University of Alabama & 0.11 m$^{2}$ & \si{\micro\becquerel}/m$^{2}$ & $5010\pm2630$ \\
\end{tabular}
\end{adjustbox}
\end{sidewaystable*}
\clearpage
\thispagestyle{empty}
\newgeometry{margin=3cm}

\begin{sidewaystable*}[tp]
\begin{adjustbox}{width=\textwidth,center}
    \tabcolsep=4pt
    \centering
    \begin{tabular}{H L{9cm} L{7cm} H C{7.cm} C{4.cm} C{2.cm} C{3.5cm}}
    \toprule
    \\
    \textbf{ID}  & 
    \textbf{Sample}  & 
    \textbf{Supplier} &  
    \textbf{LZ Location}  &  
    \textbf{Detector}  & 
    \textbf{Quantity Screened}   &  
    \textbf{Units}  &  
    \textbf{\rnttt{}} 
    \\
    \hline
    \hline
    \\
30 & AFRL-02 Dust & Powder Technology & Others & University of Alabama & 295.7 g & \si{\micro\becquerel}/g & $330\pm110$ \\
31 & AFRL-03 Dust & Powder Technology & Others & University of Alabama & 323.6 g & \si{\micro\becquerel}/g & $190\pm110$ \\
32 & Dust From SURF 4850L & N/A & Others & University of Alabama & 0.2172 kg & \si{\micro\becquerel}/g & $739\pm125$ \\
33 & Dust From SURF 4850L & N/A & Others & University of Alabama & 0.2172 kg & \si{\micro\becquerel}/g & $900\pm153$ \\
34 & TIG Welded Titanium Block (untreated) & Loterios  & Others & UCL & 325 cm$^{2}$ & \si{\micro\becquerel}/cm$^{2}$ & $1.7\pm0.3$ \\
\end{tabular}
\end{adjustbox}
\end{sidewaystable*}
\clearpage
\thispagestyle{empty}
\newgeometry{margin=3cm}

\begin{sidewaystable*}[tp]
\begin{adjustbox}{width=\textwidth,center}
    \tabcolsep=4pt
    \centering
    \begin{tabular}{H L{9cm} L{7cm} H C{7.cm} C{4.cm} C{2.cm} C{3.5cm}}
    \toprule
    \\
    \textbf{ID}  & 
    \textbf{Sample}  & 
    \textbf{Supplier} &  
    \textbf{LZ Location}  &  
    \textbf{Detector}  & 
    \textbf{Quantity Screened}   &  
    \textbf{Units}  &  
    \textbf{\rnttt{}} 
    \\
    \hline
    \hline
    \\
45 & PMT Bases (3-inch) & Imperial College London Fabrication & PMT & UCL & 124 unit & \si{\micro\becquerel}/unit & $5\pm0.87$ \\
46 & PCB/Cirlex  & Cirlex & PMT & UCL & 421 g & \si{\micro\becquerel}/g & $1.4\pm0.35$ \\
47 & PMT Capacitors (Before Cooling) & KEMET & PMT & University of Alabama & 0.5 kg & \si{\micro\becquerel}/kg & $3510\pm961$ \\
48 & PMT HV Feedthrough (After Potting) & Ceramaseal & PMT & University of Maryland & 21 unit & \si{\micro\becquerel}/unit & $5.2\pm2.4$ \\
49 & PMTs (3-inch) & Hamamatsu & PMT & SDSM\&T & 93 unit & \si{\micro\becquerel}/unit & $1.9^{+1.7}_{-1.9}$ \\
\end{tabular}
\end{adjustbox}
\end{sidewaystable*}
\clearpage
\thispagestyle{empty}
\newgeometry{margin=3cm}

\begin{sidewaystable*}[tp]
\begin{adjustbox}{width=\textwidth,center}
    \tabcolsep=4pt
    \centering
    \begin{tabular}{H L{9cm} L{7cm} H C{7.cm} C{4.cm} C{2.cm} C{3.5cm}}
    \toprule
    \\
    \textbf{ID}  & 
    \textbf{Sample}  & 
    \textbf{Supplier} &  
    \textbf{LZ Location}  &  
    \textbf{Detector}  & 
    \textbf{Quantity Screened}   &  
    \textbf{Units}  &  
    \textbf{\rnttt{}} 
    \\
    \hline
    \hline
    \\
60 & Kapton & Fralock (McMaster Carr) & TPC & SDSM\&T & 25000 cm$^{2}$ & \si{\micro\becquerel}/cm$^{2}$ & $56^{+104}_{-56}$ \\
\\
\multicolumn{8}{l}{\textbf{Xenon Circulation System}}\\
\hline \\
61 & Heat Exchanger & FNAL Fabrication & Xe & SDSM\&T & 1 unit & \si{\micro\becquerel}/unit & $197^{+185}_{-197}$\\
\end{tabular}
\end{adjustbox}
\end{sidewaystable*}
\clearpage
\thispagestyle{empty}
\newgeometry{margin=3cm}

\begin{sidewaystable*}[tp]
\begin{adjustbox}{width=\textwidth,center}
    \tabcolsep=4pt
    \centering
    \begin{tabular}{H L{9cm} L{7cm} H C{7.cm} C{4.cm} C{2.cm} C{3.5cm}}
    \toprule
    \\
    \textbf{ID}  & 
    \textbf{Sample}  & 
    \textbf{Supplier} &  
    \textbf{LZ Location}  &  
    \textbf{Detector}  & 
    \textbf{Quantity Screened}   &  
    \textbf{Units}  &  
    \textbf{\rnttt{}} 
    \\
    \hline
    \hline
    \\
75 & Shirasagi Charcoal & Osaka Gas Chemicals & Xe & University of Maryland & 0.512 kg & \si{\micro\becquerel}/kg & $101000\pm8100$ \\
76 & Cryogenic Valve & WEKA & Xe & University of Maryland & 1 unit & \si{\micro\becquerel}/unit & $247\pm56$  \\
77 & Circulation Compressor \#1 heads & Fluitron & Xe & University of Maryland & 2 unit & \si{\micro\becquerel}/unit & $375\pm50$  \\
78 & Circulation Compressor \#2 heads & Fluitron & Xe & University of Maryland & 2 unit & \si{\micro\becquerel}/unit & $391\pm56$ \\
79 & Full Circulation Compressor \#2 & Fluitron & Xe & University of Maryland & 1 unit & \si{\micro\becquerel}/unit & $17000\pm500$  \\
80 & Xe Transfer Lines & U. Maryland Fabrication & Xe & University of Maryland & 1 unit & \si{\micro\becquerel}/unit & $223\pm72$ \\
81 & Vespel Gas Valve Seats & DuPont & Xe & University of Alabama & 36 unit & \si{\micro\becquerel}/unit & $15\pm9.3$ \\
82 & Radon Reduction System Panel & U. Michigan Fabrication & Xe & University of Maryland & 1 unit & \si{\micro\becquerel}/unit & $286^{+110}_{-91}$ \\
83 & Radon Reduction System Plumbing & U. Michigan Fabrication & Xe & University of Maryland & 1 unit & \si{\micro\becquerel}/unit & $160\pm57$  \\
84 & TIG Welds On 0.5" SS Tubing & U. Maryland Fabrication & Xe & University of Maryland & 1 m & \si{\micro\becquerel}/m & $34\pm21$  \\
85 & Main Circulation Panel & U. Maryland Fabrication & Xe & University of Maryland & 1 unit & \si{\micro\becquerel}/unit & $744\pm198$  \\
\hline

\end{tabular}
\end{adjustbox}
\end{sidewaystable*}

\end{document}